\documentclass[
  journal=pasa,
  manuscript=research-paper, 
  year=2020,
  volume=37,
]{cup-journal}

\usepackage{amsmath}
\usepackage{amssymb,microtype,siunitx,booktabs}
\title{EMU/GAMA: Radio detected galaxies are more obscured than optically selected galaxies}

\author{U. T. Ahmed}
\affiliation{Australian Astronomical Optics, Macquarie University, 105 Delhi Rd, North Ryde, NSW 2113, Australia}
\alsoaffiliation{Centre for Astrophysics, University of Southern Queensland, 37 Sinnathamby Boulevard, Springfield Central, QLD 4300, Australia} 
\email[U. T. Ahmed]{ummeetania.ahmed@mq.edu.au}

\author{A. M. Hopkins}
\affiliation{Australian Astronomical Optics, Macquarie University, 105 Delhi Rd, North Ryde, NSW 2113, Australia}

\author{J. Ware}
\affiliation{Australian Astronomical Optics, Macquarie University, 105 Delhi Rd, North Ryde, NSW 2113, Australia}

\author{Y. A. Gordon}
\affiliation{Department of Physics, University of Wisconsin-Madison, Madison, WI 53706-1390, USA}

\author{M. Bilicki}
\affiliation{Center for Theoretical Physics, Polish Academy of Sciences, Al. Lotników 32/46, 02-668 Warsaw, Poland}

\author{M. J. I. Brown}
\affiliation{School of Physics \& Astronomy, Monash University, Clayton, VIC 3800, Australia}

\author{M. Cluver}
\affiliation{Centre for Astrophysics and Supercomputing, Swinburne University of Technology, John Street, Hawthorn 3122, Australia}

\author{G. G{\"u}rkan}
\affiliation{Th{\"u}ringer Landessternwarte, Sternwarte 5, D-07778 Tautenburg, Germany}
\alsoaffiliation{CSIRO Space and Astronomy, ATNF, PO Box 1130, Bentley WA 6102, Australia}

\author{\'A. R. L\'opez-S\'anchez}
\affiliation{Australian Astronomical Optics, Macquarie University, 105 Delhi Rd, North Ryde, NSW 2113, Australia}
\alsoaffiliation{Macquarie University Research Centre for Astronomy, Astrophysics \& Astrophotonics, Sydney, NSW 2109, Australia}
\alsoaffiliation{ARC Centre of Excellence for All Sky Astrophysics in 3 Dimensions (ASTRO-3D)}

\author{D. A. Leahy}
\affiliation{Department of Physics and Astronomy, University of Calgary, Calgary, Alberta, T2N 1N4, Canada}

\author{L. Marchetti}
\affiliation{Department of Astronomy, University of Cape Town, Private Bag X3, Rondebosch 7701, South Africa}
\alsoaffiliation{INAF - Istituto di Radioastronomia, via Gobetti 101, I-40129 Bologna, Italy}

\author{S. Phillipps}
\affiliation{Astrophysics Group, School of Physics, University of Bristol, Tyndall Avenue, Bristol BS8 1TL, UK}

\author{I. Prandoni}
\affiliation{INAF - Istituto di Radioastronomia, via Gobetti 101, I-40129 Bologna, Italy}

\author{N. Seymour}
\affiliation{International Centre for Radio Astronomy Research, Curtin University, 1 Turner Avenue, Bentley, WA 6102, Australia}

\author{E. N. Taylor}
\affiliation{Centre for Astrophysics and Supercomputing, Swinburne University of Technology, John Street, Hawthorn 3122, Australia}

\author{E. Vardoulaki}
\affiliation{Th{\"u}ringer Landessternwarte, Sternwarte 5, D-07778 Tautenburg, Germany}

\doi{10.1017/pasa.2020.32}

\received {dd Mmm YYYY}
\revised  {dd Mmm YYYY}
\accepted {dd Mmm YYYY}
\published{22 September 2020}


\begin{document}

\begin{abstract}
We demonstrate the importance of  radio selection in probing heavily obscured galaxy populations. We combine Evolutionary Map of the Universe (EMU) Early Science data in the Galaxy and Mass Assembly (GAMA) G23 field with the GAMA data, providing optical photometry and spectral line measurements, together with Wide-field Infrared Survey Explorer (WISE) infrared (IR) photometry, providing IR luminosities and colours. We investigate the degree of obscuration in star forming galaxies, based on the Balmer decrement (BD), and explore how this trend varies, over a redshift range of $0<z<0.345$. We demonstrate that the radio detected population has on average higher levels of obscuration than the parent optical sample, arising through missing the lowest BD and lowest mass galaxies, which are also the lower star formation rate (SFR) and metallicity systems. We discuss possible explanations for this result, including speculation around whether it might arise from steeper stellar initial mass functions in low mass, low SFR galaxies.
\end{abstract}

\section{INTRODUCTION }
\label{sec:int}

All galaxies with star formation contain dust, which has long been recognised as a key element of a wide variety of astrophysical processes \citep{2003ARA&A..41..241D}. The dust acts to obscure emission preferentially at bluer wavelengths, and can hamper analyses of star formation rate (SFR) and other galaxy properties unless suitable corrections are applied  \citep[e.g.,][]{2000ApJ...533..682C,2001PASP..113.1449C, 2001AJ....122..288H,2010MNRAS.409..421G,2016ApJ...818...13B}.
Dust obscuration is one of the largest uncertainties in measuring the SFR in galaxies, when using traditional optical tracers such as the H$\alpha$ emission line.
One primary concern of large galaxy surveys is to be able to make reliable corrections for dust obscuration when estimating properties such as SFR, stellar mass, or stellar population age \citep{1998ARA&A..36..189K}, and conventionally the emission line ratio H$\alpha$/H$\beta$, the Balmer decrement \citep[BD,][]{1989agna.book.....O}, a dust sensitive parameter, is used for making such corrections.

Many other approaches, however, are used in probing the dusty universe, including surveys at infrared wavelengths \citep[e.g.,][]{2010AJ....140.1868W, 2014MNRAS.442.3361N} which are efficient at detecting dusty galaxies that would be missed by blind optical surveys. Surveys at radio wavelengths too, which are unaffected by dust obscuration, have been used to explore star formation in galaxies at high redshifts \citep[e.g.,][]{1999ApJ...525..609S,2003NewAR..47..357W,2007ApJ...654..764B,2008MNRAS.386.1695S,2017A&A...602A...5N}, and have been shown to be sensitive to more heavily obscured galaxies than those sampled at optical wavelengths \citep[e.g.,][] {2003ApJ...597..269A,2010ApJ...714.1305S}. This effect has been incorporated into simulations developed for the SKA \citep[e.g.,][]{2010MNRAS.405..447W}.


Investigations of the faint radio source population \citep[e.g.,][]{1999ASPC..193...55W,2021A&A...648A...3K} have clearly established that very sensitive optical datasets are required in order to identify counterparts for the vast bulk of the radio population. This is a consequence of the fact that many faint radio sources are either at very high redshift, or are heavily obscured, or both \citep[e.g.,][]{2017A&A...602A...2S,2022MNRAS.516..245W}. It is natural that such work, in exploring the boundaries of the faint radio population, has pushed to maximise the numbers of optical counterparts from very deep complementary datasets. Despite these results, there has not been much work investigating systematically how  radio selected samples differ from those selected at other wavelengths in terms of their sensitivity to the obscured galaxy population.

Here we are interested in understanding the impact of dust on sample selection. We start, in contrast to earlier work, with a magnitude limited optically-selected sample, and then investigate the obscuration properties of their radio counterparts and how they compare to the parent optical sample. As a result, rather than investigating the sensitivity of a deep radio selected population to the obscured universe, our analysis looks at a somewhat different question: What fraction of a given magnitude-limited optical sample has radio counterparts, and are their obscuration properties similar or not?

Given the growth and variety of radio surveys being pursued using the Square Kilometre Array (SKA) pathfinder telescopes \citep{2013PASA...30...20N}, it is timely to revisit the utility of radio selected samples in probing star forming galaxy (SFG) populations, and to explore in more detail their sensitivity to obscured systems. We investigate this issue here by combining data from the Galaxy And Mass Assembly (GAMA) survey \citep{2011MNRAS.413..971D,2013MNRAS.430.2047H,2015MNRAS.452.2087L,2018MNRAS.474.3875B,2022MNRAS.513..439D} with early science data taken using the Australian SKA Pathfinder (ASKAP) telescope \citep{2021PASA...38....9H} for the Evolutionary Map of the Universe (EMU) survey \citep{2011PASA...28..215N}.

ASKAP is capable of quickly surveying wide sky regions with good resolution and sensitivity, exemplified by programs such as the Rapid ASKAP Continuum Survey \citep[RACS;][]{2020PASA...37...48M}, which is mapping the sky in three frequency bands (covering the full ASKAP range from 700-1800\,MHz) with a $\sigma \approx 0.2 - 0.4\,$mJy beam$^{-1}$ root-mean-square (rms) noise sensitivity.
The EMU survey \citep{2021PASA...38...46N} will provide a significantly deeper radio view of the sky, with $\sigma \approx 0.02\,$mJy beam$^{-1}$ at $943\,$MHz. EMU early science and pilot observations have covered a range of fields over the sky, including the GAMA G23 field, 
covering an area of 82.7 deg$^2$ centered around $\alpha = 23$\,h and $\delta = -32^{\circ}$ with $\sigma \approx 0.038\,$mJy beam$^{-1}$, at $887.5$\,MHz \citep{2019PASA...36...24L,2022MNRAS.512.6104G}. We use the latest EMU measurements of G23 \citep{2022MNRAS.512.6104G} in this work to investigate the dust obscuration properties of galaxies, measured using the BD from GAMA optical spectra. We also compare with the properties of counterparts identified from Wide-field Infrared Survey Explorer \citep[WISE,][]{2010AJ....140.1868W} data. This allows us to contrast the radio detected subset with an infrared detected subset.

Early work \citep{2003ApJ...599..971H,2003ApJ...597..269A} demonstrated that BDs in SFGs detected at radio wavelengths are systematically higher than those in the respective optically selected samples. With the combination of data from GAMA and EMU we are able to explore this trend over a redshift range of $0<z<0.345$, and probe its origin. 

The layout of this paper is as follows. We provide a brief introduction to the EMU, GAMA and WISE surveys, and describe the data and our sample selection in \S~\ref{sec:data}. In \S~\ref{sec:analysis} we present our findings, with details of the analysis of the obscuration properties of three samples, a parent optically selected sample, and two subsamples, detected at radio and infrared wavelengths from EMU and WISE. In \S~\ref{sec:discussion}, we summarise the key results and explore the possible origins for the differences we find in the radio detected subsample, before presenting our conclusions in \S~\ref{sec:conclusion}.
Throughout we assume cosmological parameters of $H_0=70\,$km\,s$^{-1}$\,Mpc$^{-1}$, $\Omega_M=0.3$,
$\Omega_\Lambda=0.7$ and $\Omega_{\rm{k}} = 0$.





\section{DATA}
\label{sec:data}

We use early science data taken with the  ASKAP telescope \citep{2021PASA...38....9H} for the EMU survey \citep{2011PASA...28..215N, 2021PASA...38...46N} in the GAMA G23 field \citep{2019PASA...36...24L,2022MNRAS.512.6104G}. We choose this region due to the wealth of complementary photometric and spectroscopic data.
We start with the combined photometry from GAMA, the Kilo-Degree survey \citep[KiDs;][]{2015A&A...582A..62D} and the VISTA\footnote{Visible Infrared Survey Telescope for Astronomy} Kilo-degree INfrared Galaxy survey \citep[VIKING;][]{2012sngi.confE..40S} datasets referred to as the GAMA-KiDS-VIKING catalogue \citep[GKV;][]{2020MNRAS.496.3235B}.
We cross-matched this with the radio catalogue from \citet{2022MNRAS.512.6104G} and the WISE catalogue constructed for the GAMA G23 sources by \citet{2020ApJ...903...91Y}. Our main emphasis is to explore the degree of obscuration in SFGs, based on the BD, and to investigate any differences in the radio and infrared detected sub-populations.

\begin{table}
\centering
\caption{The number of galaxies in G23 from cross-matched optical parent, radio and WISE samples.}
\label{tab:Table 1}
\begin{tabular}{lcccr} 
\hline
\hline
Cross-matched samples & Number of galaxies\\
\hline
GKV & $47\,735$\\
GKV ($nQ$ $\geq$ $3$) & $45\,020$\\
GKV (gold) & $3038$\\
GKV -- Radio & $8303$\\
GKV -- Radio ($nQ$ $\geq$ $3$) & $7999$\\
GKV -- Radio (gold) & $1207$\\
GKV -- WISE & $40\,774$\\
GKV -- WISE ($nQ$ $\geq$ $3$) & $38\,581$\\
GKV -- WISE (gold) & $2533$\\
GKV -- Radio -- WISE & $7336$\\
GKV -- Radio -- WISE ($nQ$ $\geq$ $3$) & $7059$\\
GKV -- Radio -- WISE (gold) & $1047$\\
Radio -- WISE & $7583$\\
\hline
\end{tabular}
\end{table}

\subsection{Optical Data: GAMA}
\label{subsec:gama}

GAMA is a multiwavelength photometric and spectroscopic survey, which covers an area of 286 deg$^2$ with 300\,000 galaxy spectra \citep{2022MNRAS.513..439D, 2020MNRAS.496.3235B, 2011MNRAS.413..971D} across five sky regions (G02, G09, G12, G15, and G23), with a limiting magnitude of $r$ $=$ $19.8$ for most of these fields \citep{2010MNRAS.404...86B, 2010PASA...27...76R, 2011MNRAS.413..971D, 2013MNRAS.430.2047H,2015MNRAS.452.2087L, 2018MNRAS.474.3875B}. The selection limit for G23 was $i<19.2$ \citep{2015MNRAS.452.2087L}. The latest summary and final data release for GAMA are presented in \citet{2022MNRAS.513..439D}.

We take photometry from the gkvScience-v02 (GKV) catalogue, which is now the standard UV-to-IR photometery for GAMA DR4, and use the pre-computed photometry-to-spectroscopy matches in that table.
We also require accurate emission line measurements to select our SFGs. We use the optical line fluxes from “GaussFitComplexv05” within the ``SpecLineSFR" data management unit \citep[DMU;][]{2017MNRAS.465.2671G}. This DMU provides the line flux and equivalent width measurements, and the redshifts from “SpecAllv27” within the SpecCat DMU \citep{2015MNRAS.452.2087L}.
We could have chosen to use the measurements from GaussFitSimplev05, and our initial work based on these measurements gives qualitatively similar results to what we present here. We chose to use the measurements from GaussFitComplexv05, which calculates line fluxes using several Gaussian components in order to improve the fits, using the Bayesian information criterion (BIC) to identify the optimal component number required, as described in \citet{2017MNRAS.465.2671G}. This is because the [N{\sc ii}] lines are blended with H$\alpha$ in the GAMA spectra, and the joint fit to these lines gives a more robust estimate of H$\alpha$. This is desirable in order to have the most accurate BD measurements.
The full GAMA sample includes 295\,853 spectra of main survey targets, of which 259\,720 galaxies have robustly measured redshifts, quantified through the redshift quality $nQ$ $\geq$ $3$ \citep{2015MNRAS.452.2087L}. We link this with the photometry from the GKV catalogue and restrict the sample to the G23 field, giving us a ``parent sample'' for the purposes of this current analysis of 47\,735 galaxies, of which 45\,020 galaxies have $nQ$ $\geq$ $3$ (Table~\ref{tab:Table 1}). Next, we cross-match this parent sample against radio and infrared samples, giving the numbers shown in Table~\ref{tab:Table 1}. Following these steps, we further limit each of our samples to those labelled as ``gold'' in Table \ref{tab:Table 1} by imposing quality limits on the redshifts and emission lines. These correspond to the redshift quality $nQ$ $\geq 3$, signal-to-noise limits on the Balmer lines, $S/N$ (H$\alpha$) $\geq 5$, and $S/N$ (H$\beta$) $\geq 5$, positive detections of the [O{\sc iii}] and [N{\sc ii}] emission lines, and line fitting flags ${\rm NPEG}=0$ for both the H$\alpha$ and H$\beta$ lines, indicating that none of the fitting parameters were ``pegged'' at their extreme parameter range during the fits. This last step ensures that our sample only includes galaxies with the most reliably measured spectral lines.

\begin{figure}[hbt!]
\centering
\includegraphics[width=\textwidth]{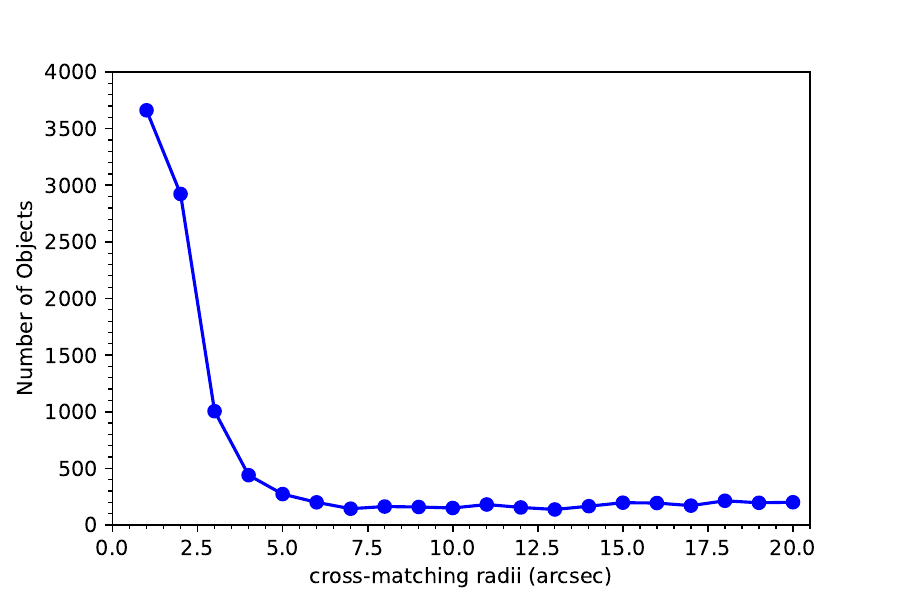}
\caption{The number of cross-matched GKV - Radio objects against the GKV sample as a function of different cross-matching radii. Our selection of a $5''$ matching radius corresponds to the distance at which the curve flattens.}
\label{Figure 1}

\end{figure}

\subsection{Radio Data: EMU}
\label{subsec:emu}

EMU is an ongoing deep radio continuum sky survey with ASKAP, covering the Southern Hemisphere up to $\delta < +30^{\circ}$ and with resolution of  $\sim 15"$ FWHM, which is expected to generate a catalogue of about 40 million galaxies at $943\,$MHz \citep{2021PASA...38...46N}. The characteristics of EMU are similar to earlier generations of sub-mJy surveys like the Phoenix Deep Field \citep[PDF;][]{1998MNRAS.296..839H, 2003AJ....125..465H}, and the Australia Telescope Large Area Survey \citep[ATLAS;][]{2006AJ....132.2409N, 2015MNRAS.453.4020F, 2014MNRAS.441.2555H}. ATLAS covered about 7\,deg$^2$ \citep{2012MNRAS.426.3334M} of the sky with the aim of producing the widest deep (10–15\,$\mu$Jy rms) radio survey at the time, whereas EMU is aiming to ultimately observe three-quarters of the sky (30\,000\,deg$^2$), with similar sensitivity ($20\,\mu$Jy) and resolution ($15''$). Even with just the data observed in early science and pilot phases for EMU, we have covered more than 300\,deg$^2$ \citep{2021PASA...38...46N}.

One challenge for EMU is that optical spectroscopy will not be available for the whole survey area, only selected patches, and typically only for relatively low redshifts (e.g., $z \lesssim 0.5$ for GAMA), or $z\lesssim 1$ in smaller regions from deeper surveys like the Deep Extragalactic Visible Legacy Survey \citep[DEVILS,][]{2018MNRAS.480..768D} and having some overlap with northern projects including the Dark Energy Spectroscopic Instrument {\citep[DESI,][]{2019AJ....157..168D, 2022MNRAS.512.3662D}} and WEAVE-LOFAR \citep{2016sf2a.conf..271S}. Photometric redshifts will also be available in selected areas, such as from KiDs, covering wider areas from programs such as the Dark Energy Survey \citep [DES,][]{2018ApJS..239...18A}.
Additional spectroscopic redshifts will become available with the advent of the 4MOST surveys, including ORCHIDSS \citep{2023Msngr.190...25D} and WAVES \citep{2019Msngr.175...46D}.

Here we combine the optical spectroscopy from GAMA with early science data taken for EMU in GAMA's G23 region \citep{2022MNRAS.512.6104G}. We use the most recent radio observations of G23 \citep{2022MNRAS.512.6104G}, which supersede earlier data taken with ASKAP \citep{2019PASA...36...24L}. 
These EMU early science observations of G23 cover an area of 82.7 deg$^2$ centered around $\alpha = 23$\,h and $\delta = -32^{\circ}$ with $\sigma \approx 0.038\,$mJy beam$^{-1}$, at $887.5$\,MHz. There are 55\,247 radio sources catalogued of which 39\,812 have ${\rm S/N}\ge5$ detections \citep{2022MNRAS.512.6104G}.

Our radio detected subset (GKV - Radio) consists of radio sources from that catalogue cross-matched against the optical parent sample using a matching radius of $5$\,arcsec. Figure~\ref{Figure 1} shows the differential number of cross-matched galaxies as a function of cross-matching radii. We record the number of cross-matches found at each radius in order to identify the appropriate cross-matching radius to adopt. The flattening at a separation of $5$\,arcsec, our adopted matching radius, is associated with the point at which no new genuine counterparts are being added, only spurious cross-matches are contributing. Where multiple potential counterparts are identified within this matching radius, we adopt the closest as the preferred counterpart, which has been shown in earlier EMU analyses to be the most reliable approach \citep[e.g.,][]{2021PASA...38...46N}. We estimate our spurious cross-match contamination level following \cite{2020MNRAS.497.2730G}, suggesting that our total contamination rate is $\sim 2.5\%$. This results in a catalogue of 8303 GKV - Radio galaxies, of which 7999 galaxies have $nQ$ $\geq$ $3$. The sample of 1207 galaxies defined as GKV - Radio (gold) are those meeting the more stringent spectroscopic quality requirements described in \S\,\ref{subsec:gama}. These samples are summarised in Table \ref{tab:Table 1}.

\begin{figure}[h!]
\centering
\includegraphics[width=\textwidth]{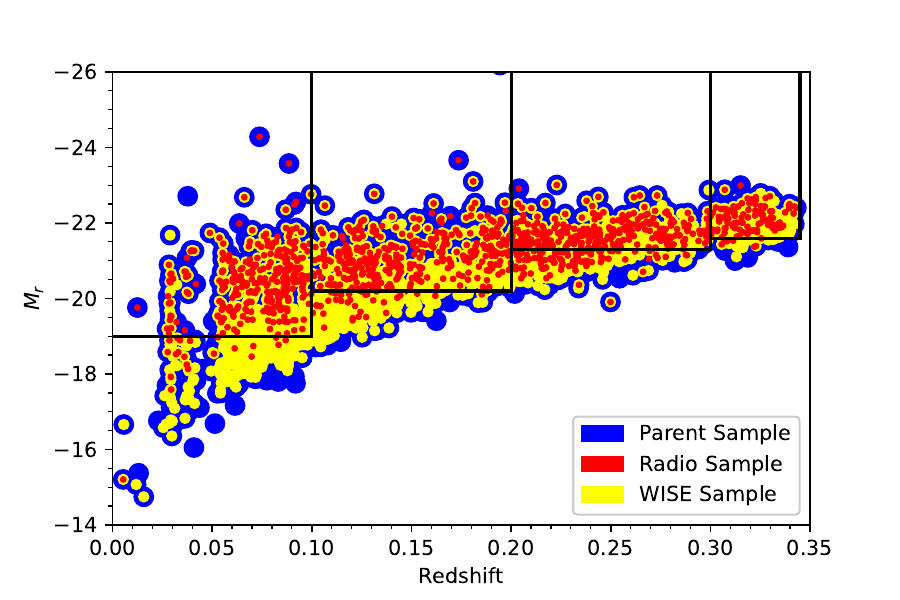}
\caption{Distribution of $M_r$ with redshift, illustrating the four volume limited samples with all the optical parent GKV galaxies (blue), GKV - Radio detections (red), and those with GKV - WISE detections (yellow).}
\label{Figure 2}

\end{figure}

\begin{figure}[hbt!]
\centering
\includegraphics[width=\textwidth]{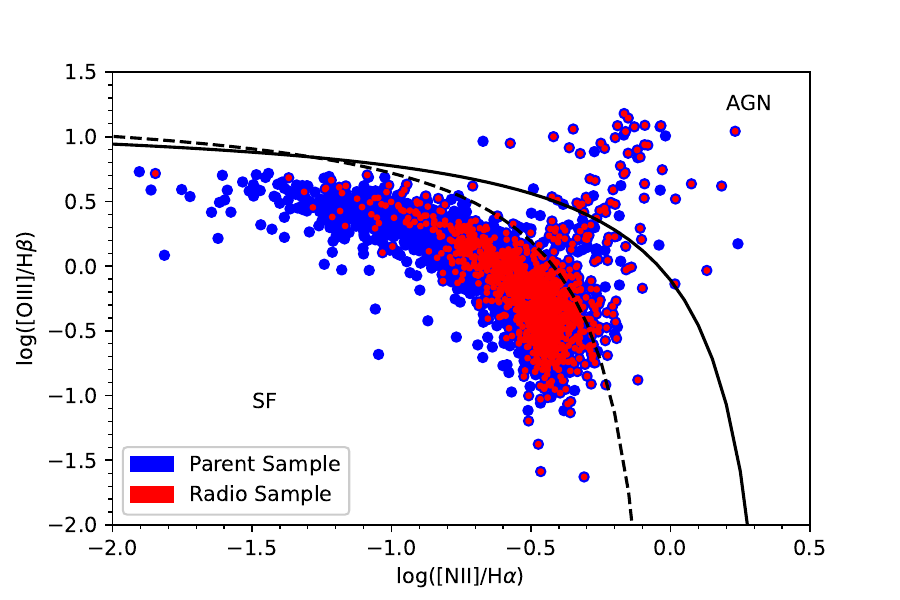}
\caption{Spectral diagnostic diagram illustrating the selection of star forming galaxies (SFGs) using the criteria given by \citet{2001ApJ...556..121K} (solid) and \citet{2003MNRAS.346.1055K} (dashed) which presents the GKV sample, showing the parent sample (blue) and the GKV - Radio sample (red). 
}
\label{Figure 3}

\end{figure}

\subsection{Infrared Data: WISE}
\label{subsec:wise}

We adopt the IR counterparts to the GKV parent sample coming from the ALLWISE catalogue \citep{2012yCat.2311....0C} as detailed by \citet{2020ApJ...903...91Y}. WISE mapped the whole sky achieving 5$\sigma$ point source sensitivities of $0.08$, $0.11$, $1$, and $6$\,mJy in four infrared bands W1, W2, W3, and W4 centered at $3.4$\,$\mu$m, $4.6$\,$\mu$m, $12$\,$\mu$m, and $22$\,$\mu$m, respectively \citep{2010AJ....140.1868W, 2011ApJ...735..112J, 2017ApJ...836..182J}.
Positional cross-matching was used to identify WISE counterparts to the parent sample, using a $3$\,arcsec matching radius \citep{2014ApJ...782...90C,2020ApJ...898...20C, 2017ApJ...836..182J, 2019ApJS..245...25J}. 


Our GKV-WISE sample has 40\,774 galaxies, of which 38\,581 galaxies have $nQ$ $\geq$ $3$ (Table~\ref{tab:Table 1}).
This sample is also matched to the GKV-Radio sample through the common optical counterparts producing a catalogue of 7336 GKV-Radio-WISE galaxies, of which 7059 galaxies have $nQ$ $\geq$ $3$. A separate catalogue of 7583 Radio-WISE galaxies is produced by cross-matching the WISE galaxies directly against the radio catalogue independent of the GKV parent sample, again adopting a 5 arcsec matching radius (Table \ref{tab:Table 1}).






\begin{table}
	\centering
	\caption{The number of galaxies in the optical parent sample brighter than $M_r$ in four different redshift bins.}
	\label{tab:Table 2}
	\begin{tabular}{lcccccr} 
		\hline
		\hline
		$z$ range & $M_r$ & $N$ & $N\le M_r$ & $N\le M_r$ \\
		& & & & (gold)\\
		\hline
		$0$ $<$ $z$ $\leq$ $0.1$ & $-19$ & $6481$ & $4541$ & $529$\\
		$0.1$ $<$ $z$ $\leq$ $0.2$ & $-20.2$ & $13\,446$ & $9947$ & $615$\\
		$0.2$ $<$ $z$ $\leq$ $0.3$  & $-21.3$ & $15\,627$ & $9410$ & $492$\\
		$0.3$ $<$ $z$ $\leq$ $0.345$ & $-21.6$ & $4549$ & $3746$ & $186$\\
		\hline
	\end{tabular}
\end{table}

\begin{table}
	\centering
	\caption{The number of SF galaxies and AGNs in the gold sample for the four different redshift bins.}
	\label{tab:Table 3}
	\begin{tabular}{lcccr} 
		\hline
		\hline
		$z$ range & SFG & AGN \\
		\hline
		$0$ $<$ $z$ $\leq$ $0.1$ & $522$ & $7$ \\
		$0.1$ $<$ $z$ $\leq$ $0.2$ & $605$ & $10$ \\
		$0.2$ $<$ $z$ $\leq$ $0.3$  & $466$ & $26$ \\
		$0.3$ $<$ $z$ $\leq$ $0.345$ & $172$ & $14$ \\
		\hline
	\end{tabular}
\end{table}

\subsection{Sample Selection}
\label{subsec:sample}

Figure~\ref{Figure 2} shows the distribution of $r$-band absolute magnitude, $M_r$, with redshift for the GKV catalogue (blue), radio counterparts (red), and WISE counterparts (yellow) which form the basis of our subsequent analysis.
The upper redshift limit for the GAMA sample corresponds to the redshift at which H$\alpha$ is shifted to wavelengths beyond the spectral limit of the AAOmega camera used in the GAMA observations \citep{2011MNRAS.413..971D}.
We construct four independent volume limited samples in four redshift bins, illustrated in Figure~\ref{Figure 2} as the black boundary lines. The redshift and magnitude limits are given in  Table \ref{tab:Table 2} along with, for each redshift bin, the numbers of galaxies after each of a sequence of cuts. This excludes the 4917 galaxies (having $n$Q$\ge3$) with redshifts $z>0.345$. We show the total number of galaxies, $N$, passing our redshift quality requirement ($n$Q$\ge3$), the number of those galaxies brighter than the $M_r$ threshold, and the number brighter than the $M_r$ threshold after imposing the spectroscopic quality requirements, including S/N (H$\alpha$) $\geq$ $5$ and S/N (H$\beta$) $\geq$ $5$, detailed in \S\,\ref{subsec:gama}.

Our galaxy sample will include both SF and active galactic nuclei (AGN) systems, and radio detected galaxies will certainly include an AGN population.
As we are interested in measuring the obscuration properties in SFGs here, we need to distinguish between these, and exclude the AGN dominated systems, which we do using the standard diagnostic diagram of Baldwin, Phillips and Terlevich \citep{1981PASP...93....5B, 1987ApJS...63..295V,2001ApJ...556..121K}, shown in Figure~\ref{Figure 3} and referred to hereafter as the BPT diagram. The separation lines shown are from \citet{2001ApJ...556..121K} (solid), and \citet{2003MNRAS.346.1055K} (dashed). This figure distinguishes between star-forming galaxies (below the dashed line) from those where the ionisation arises from an AGN (above the solid line). Galaxies between these separators are commonly interpreted as composite systems, with some contribution from both SF and AGN. We include these in our analysis, but they are small in number and excluding them does not qualitatively change our results. 

Figure~\ref{Figure 3} shows only the G23 sample, with the parent sample in blue and the radio sample in red, illustrating an indication of some differences in the radio-detected subsample, which we explore further below (see \S\,\ref{BD_BPT}).
Table \ref{tab:Table 3} shows the number of galaxies in each volume limited sample after classification as SF or AGN. We exclude the AGN from further analysis in this work in the interests of simplicity, although they will be an important element of future work with EMU.
It is worth noting that there may still be low luminosity radio AGN present in some small fraction of our sample, which appear in the SF region of the BPT diagram. We do not attempt to exclude these, as they will not bias our analysis. Since the Balmer lines used to measure BD are those leading to an SF classification in the BPT diagram, they must originate from regions of the galaxy dominated by SF, and will provide robust BD estimates for our analysis. Radio luminosities for such systems may not be accurate estimators of SFR, but this is not the focus of the current analysis anyway.

The samples we use for the bulk of our analysis are these four volume limited samples, selected from those labelled ``gold'', and defined by meeting the following criteria:
\begin{enumerate}
 \item Reliable redshift estimates, defined by $nQ\ge3$, well-fit emission lines defined by ${\rm NPEG}=0$ for H$\alpha$ and H$\beta$, signal-to-noise ratio ${\rm S/N} \geq 5$ for H$\alpha$ and H$\beta$,  as well as detection of the [N{\sc ii}] and [O{\sc iii}] lines to accurately classify the galaxies as star-forming or AGN.
 \item Redshifts between $0 < z \leq 0.345$. The upper redshift limit is due to the maximum detectable redshift of H$\alpha$.
 \item Having absolute $r$-band magnitude brighter than the $M_r$ threshold as shown in Table \ref{tab:Table 2}.
 \item Excluding AGN based on the BPT diagram (Figure~\ref{Figure 3}).
\end{enumerate}

\begin{figure*}[hbt!]
\centering
\includegraphics[width=0.45\textwidth]{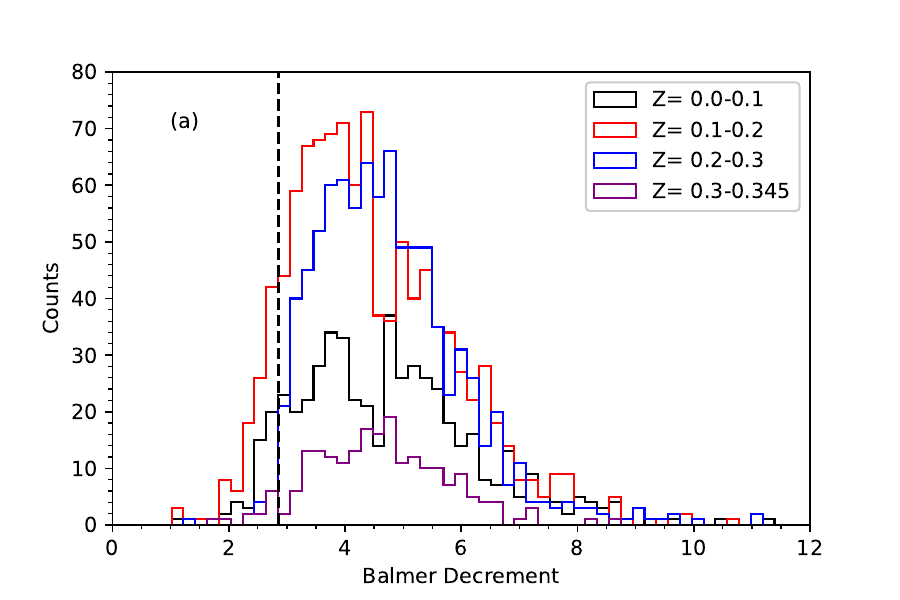}
\includegraphics[width=0.45\textwidth]{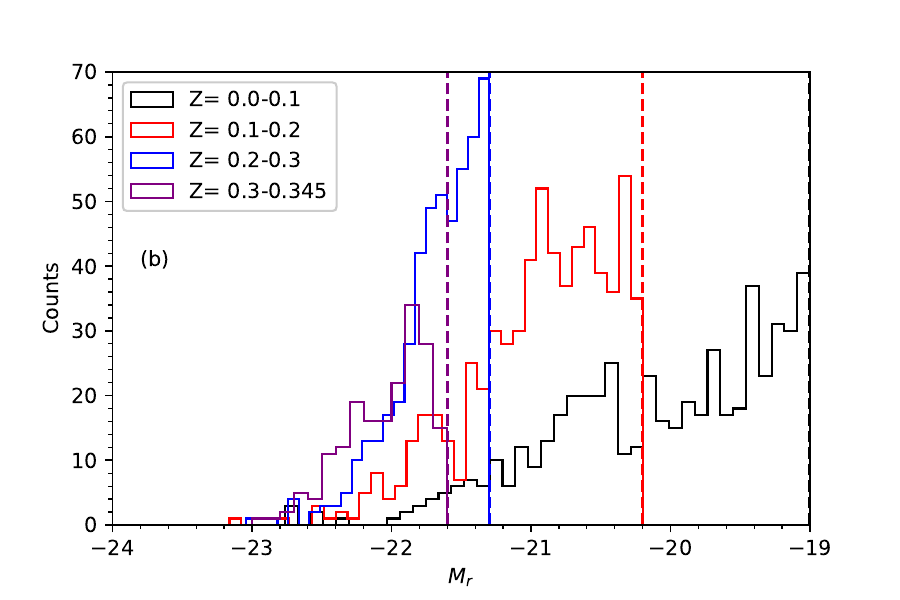}
    \caption{(a) Distribution of Balmer decrement in four redshift bins for the full G23 sample. The vertical dotted line represents the nominal Case B value of \rm{BD}=2.86 \citep{1971MNRAS.153..471B}. (b) Distribution of $M_r$ for the four independent volume limited samples. The vertical dashed lines represent our $M_r$ limits for each redshift bin.}
\label{Figure 4}

\end{figure*}

\begin{figure*}[hbt!]
\centering
\includegraphics[width=0.45\textwidth]{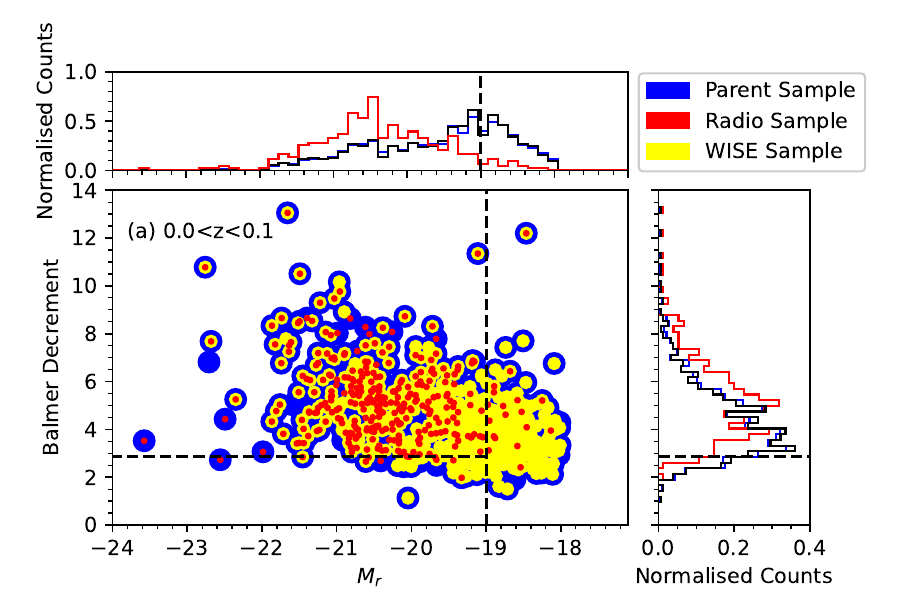}
\includegraphics[width=0.45\textwidth]{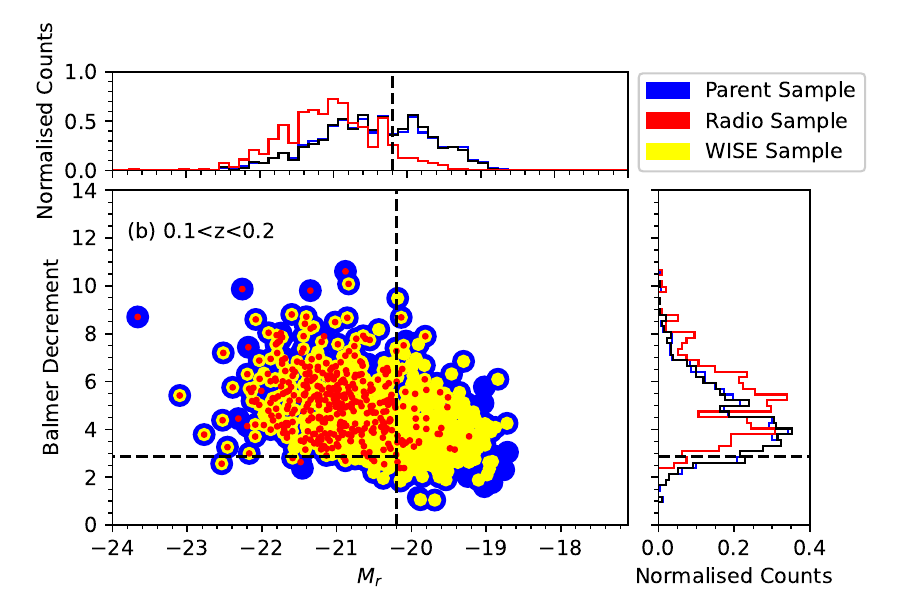}
\includegraphics[width=0.45\textwidth]{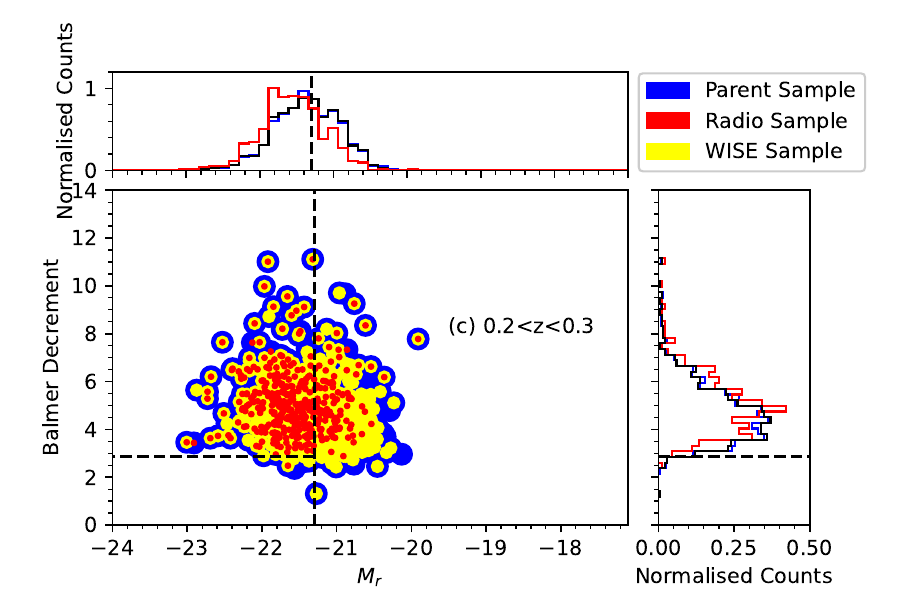}
\includegraphics[width=0.45\textwidth]{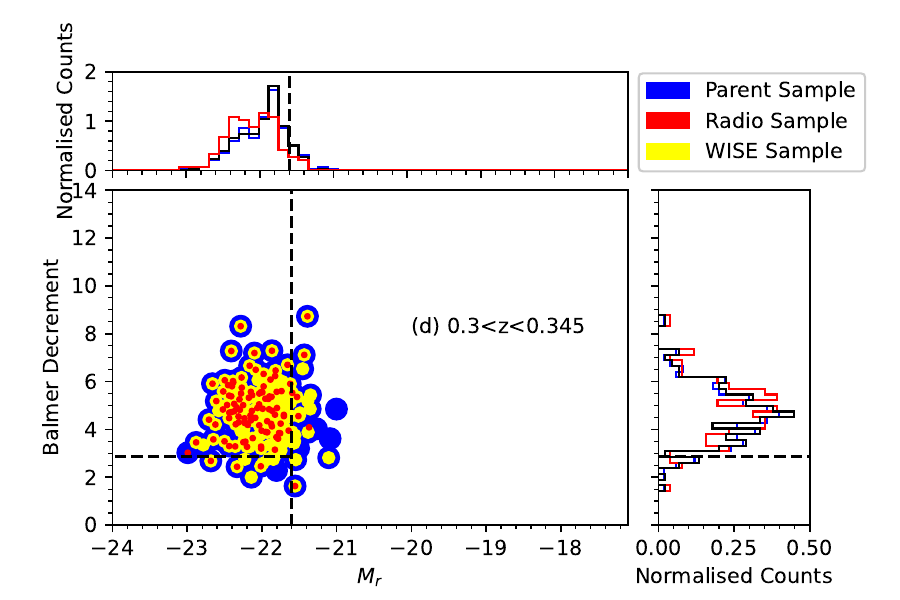}
\caption{Comparison of $M_r$-BD distribution between the gold samples of GKV (blue), GKV - WISE (yellow), and GKV - Radio (red). The samples shown here extend below the volume-limited sample magnitude limits for illustrative purposes, and to demonstrate the impact of our selection limits. The vertical dotted lines represent the $M_r$ limits for each redshift bin, and the horizontal dotted lines represent the nominal Case B value of \rm{BD}=2.86. The histograms are shown as normalised counts to aid visual comparison of the shapes of the distributions for each subsample, with the same colour-coding, except that GKV - WISE, is presented in black to improve visibility.}
\label{Figure 5}

\end{figure*}

\begin{figure*}[hbt!]
\centering
\includegraphics[width=0.45\textwidth]{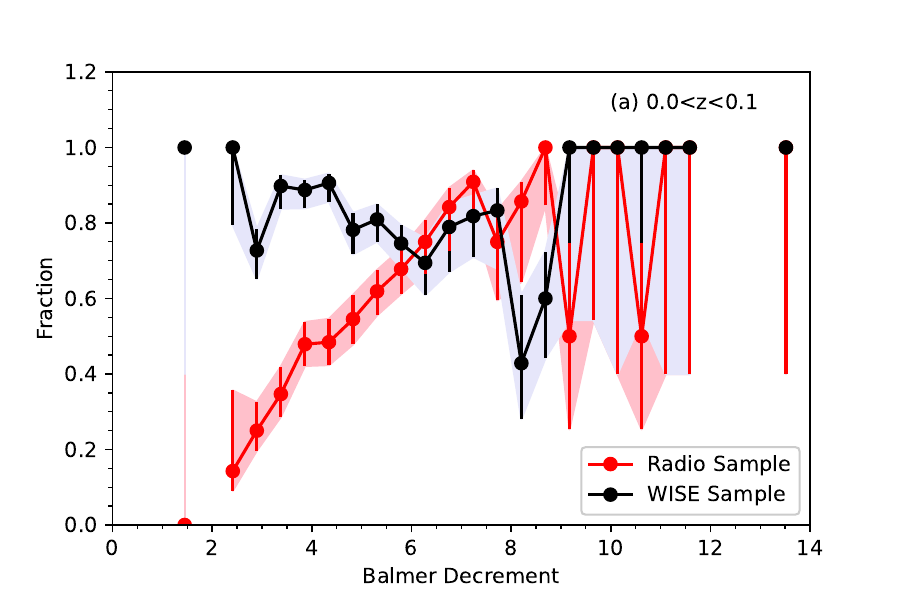}
\includegraphics[width=0.45\textwidth]{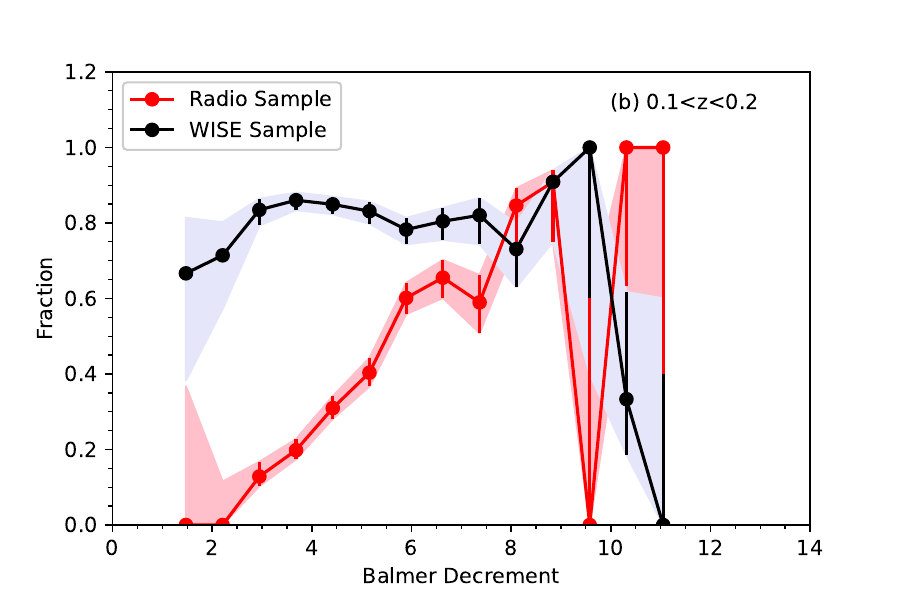}
\includegraphics[width=0.45\textwidth]{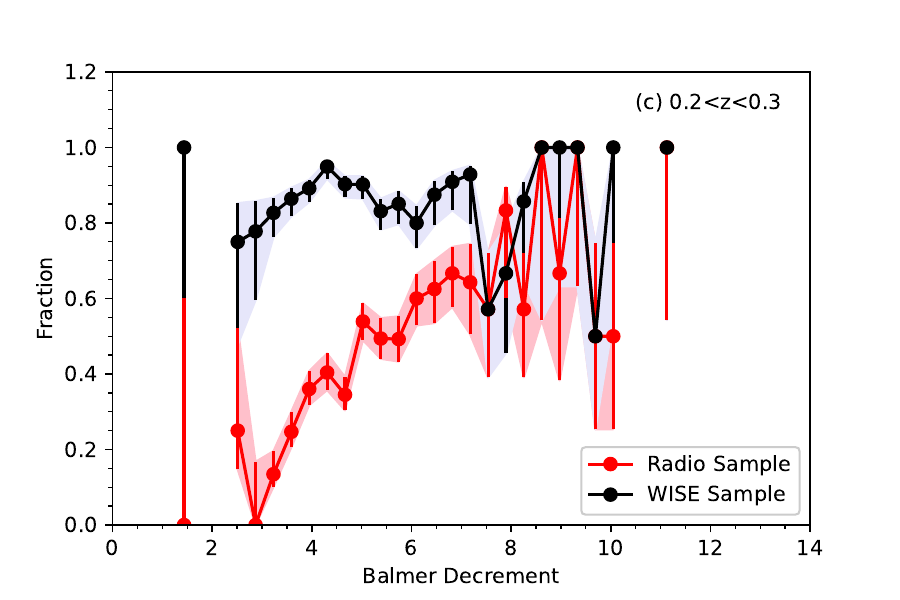}
\includegraphics[width=0.45\textwidth]{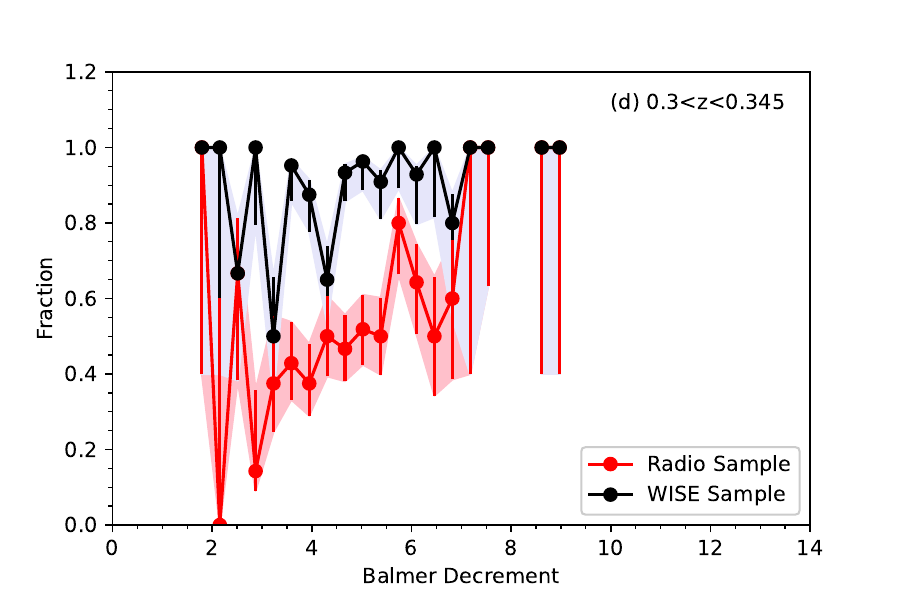}
\caption{Fractions of the (gold) radio (GKV - radio, in red), and WISE (GKV - WISE, in black) subsets over the parent sample (GKV) as a function of BD. This shows that the radio detected subset is lacking the lowest BD galaxies, compared to the parent sample. The error bars are estimated using the method of \citet{2011PASA...28..128C}, which correspond to 1$\sigma$ binomial uncertainties. The shaded regions indicate these uncertainties.}
\label{Figure 6}

\end{figure*}

\begin{figure*}[hbt!]
\centering
\includegraphics[width=0.45\textwidth]{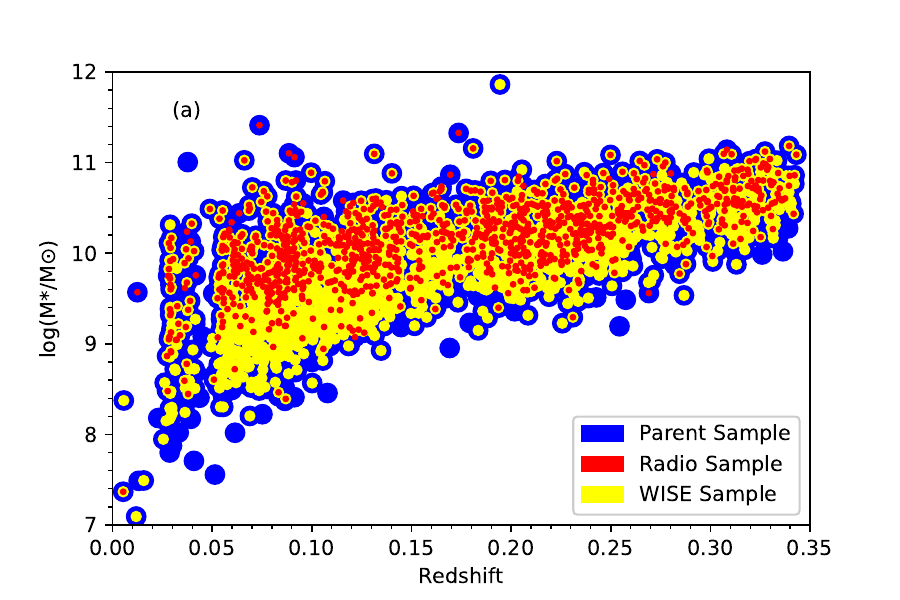}
\includegraphics[width=0.45\textwidth]{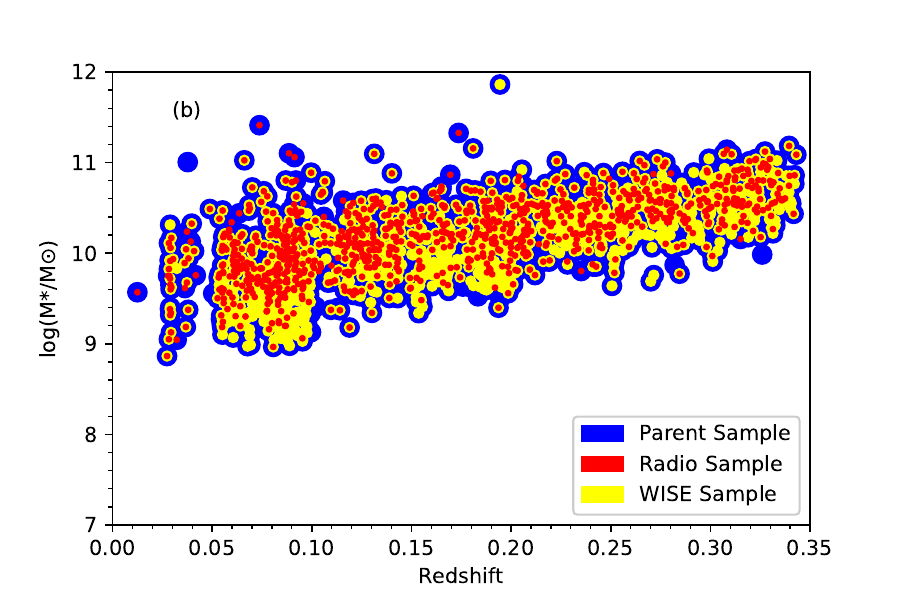}
\caption{(a) Stellar mass as a function of redshift for the parent sample (GKV, in blue), the radio detected subset (GKV - radio, in red), and the WISE subset (GKV - WISE, in yellow). (b) The same measurements restricted to the data in our volume limited samples.}
\label{Figure 7}
\end{figure*}

\begin{figure}[hbt!]
\centering
	\includegraphics[width=\textwidth]{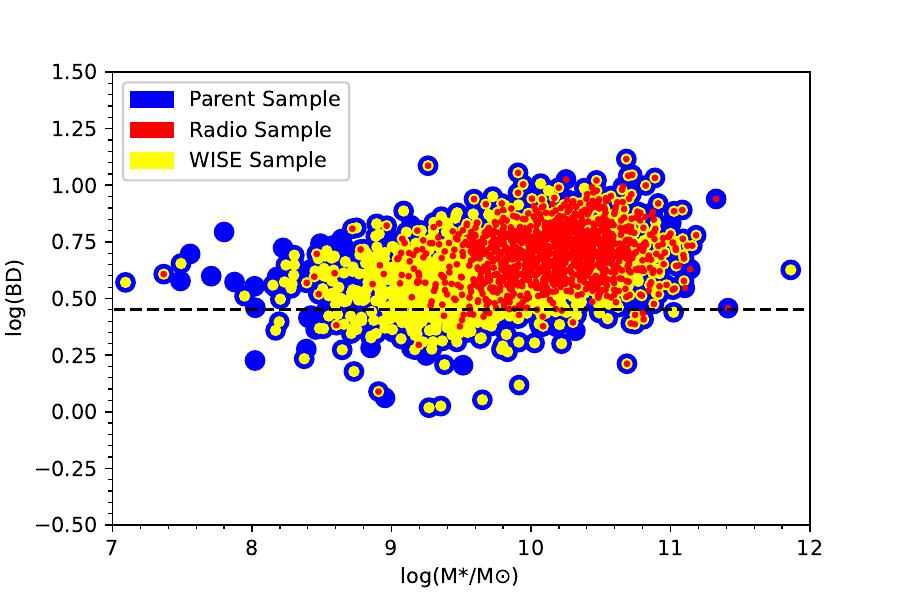}
    \caption{Balmer decrement as a function of stellar mass for the GKV parent sample (blue), radio detected subset (GKV - radio, in red), and WISE subset (GKV - WISE, in yellow). The horizontal dashed line represents the nominal value of {\rm BD}=2.86.}
    \label{Figure 8}
\end{figure}

\begin{figure*}[hbt!]
\centering
\includegraphics[width=0.45\textwidth]{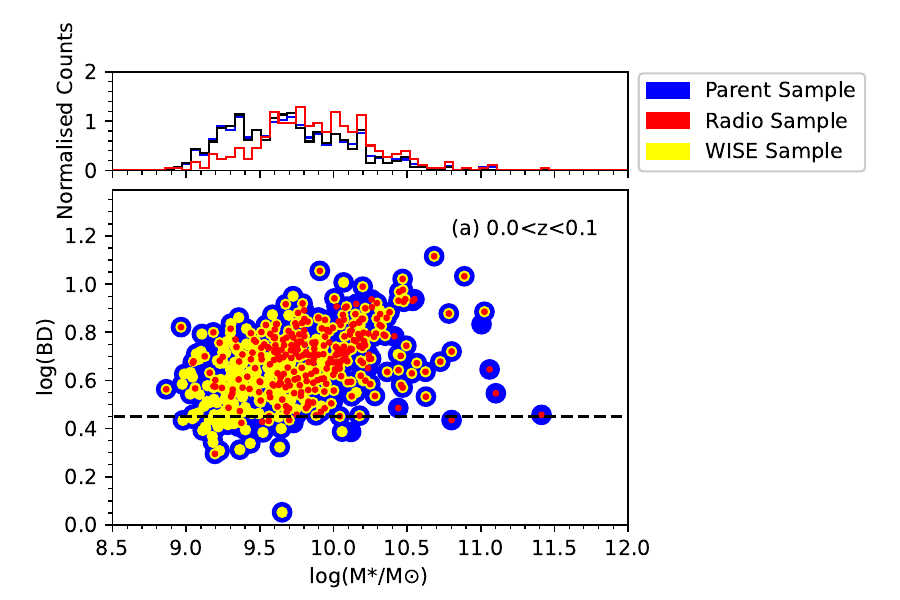}
\includegraphics[width=0.45\textwidth]{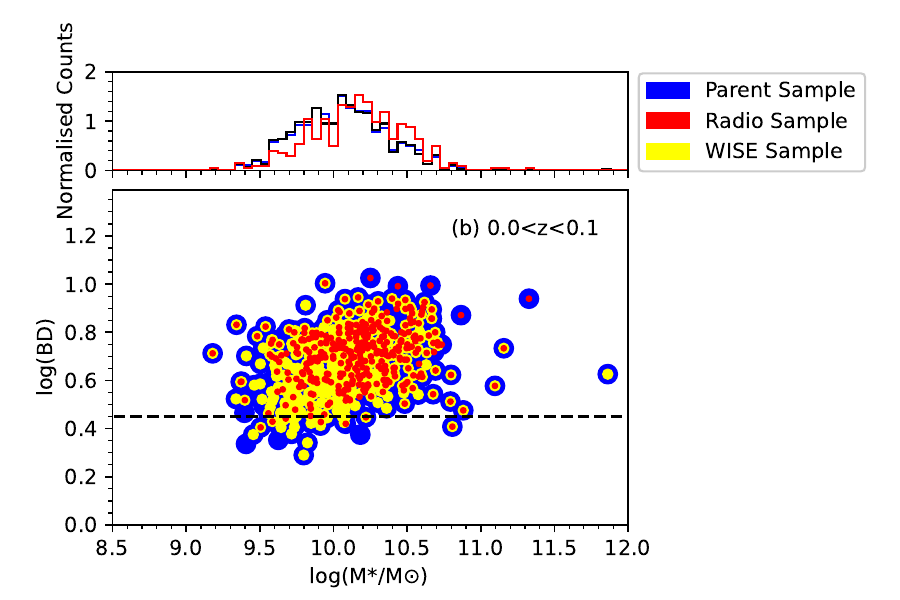}
\includegraphics[width=0.45\textwidth]{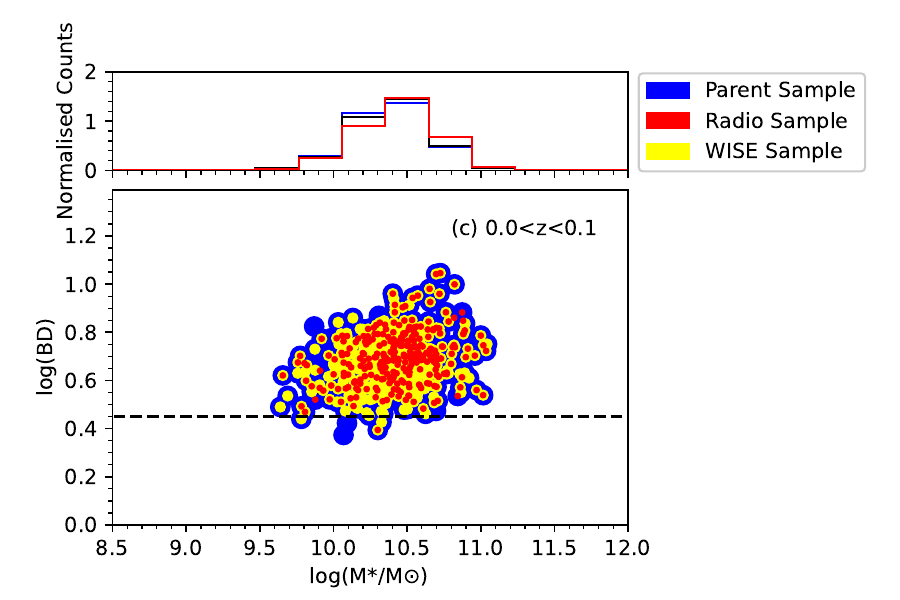}
\includegraphics[width=0.45\textwidth]{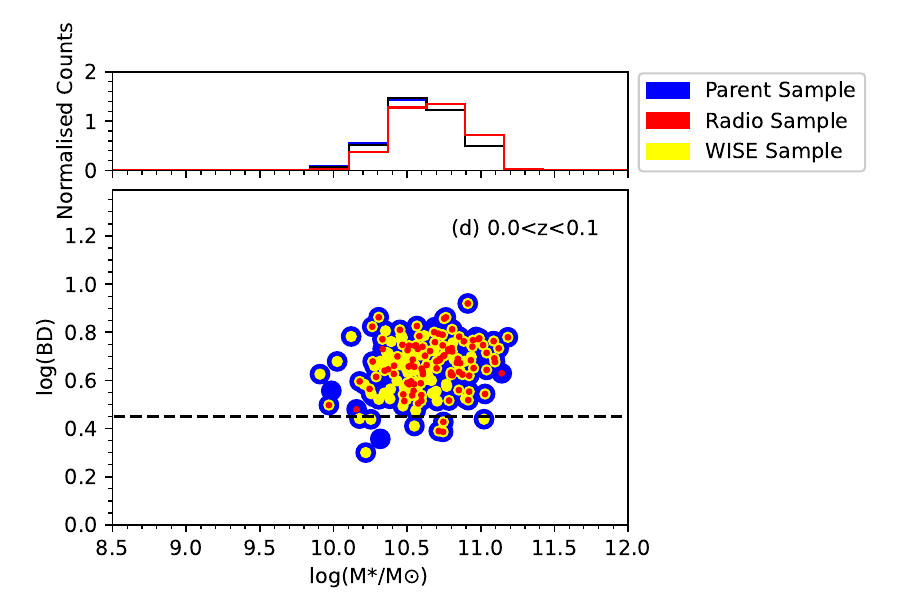}
    \caption{BD as a function of stellar mass in four different redshift bins for the parent sample (blue), radio detected subset (red), and WISE subset (yellow). The horizontal dashed line represents the nominal value of {\rm BD}=2.86. The histograms are shown as normalised counts with the same colour coding, except that GKV - WISE, is presented in black to improve visibility. Again the WISE subset closely follows the parent sample, while the radio detected subset is restricted to the higher mass systems.}
    \label{Figure 9}
\end{figure*}


\begin{figure*}[hbt!]
\centering
	\includegraphics[width=0.45\textwidth]{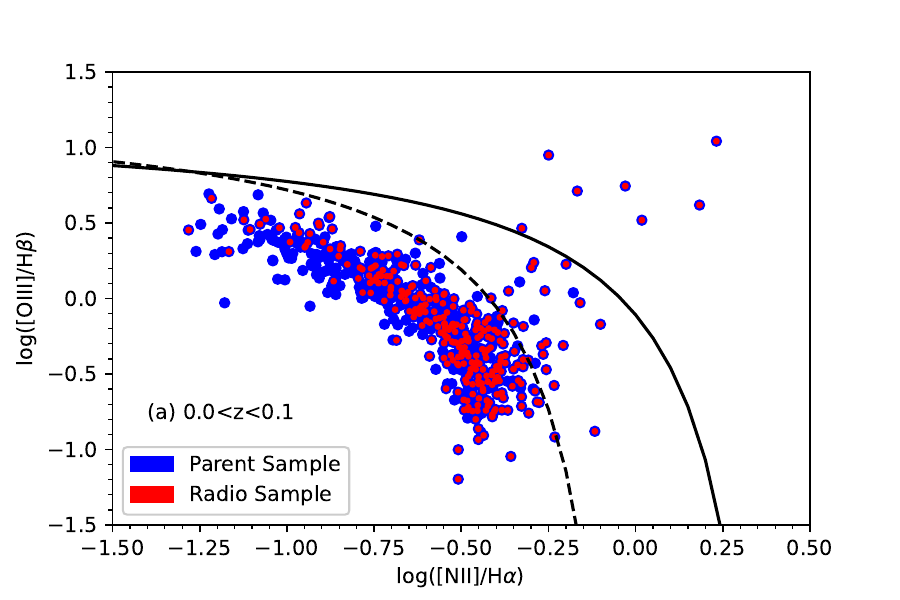}
        \includegraphics[width=0.45\textwidth]{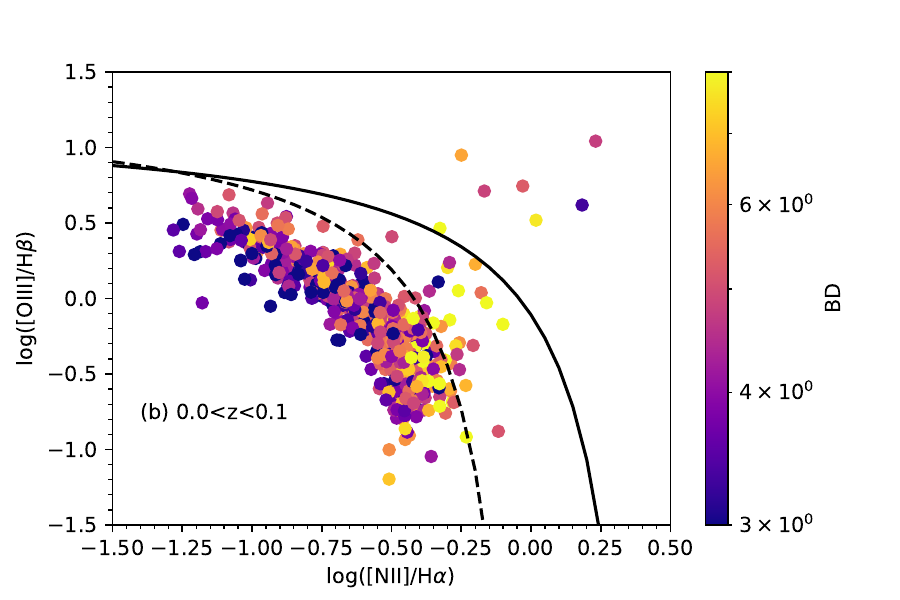}
        \includegraphics[width=0.45\textwidth]{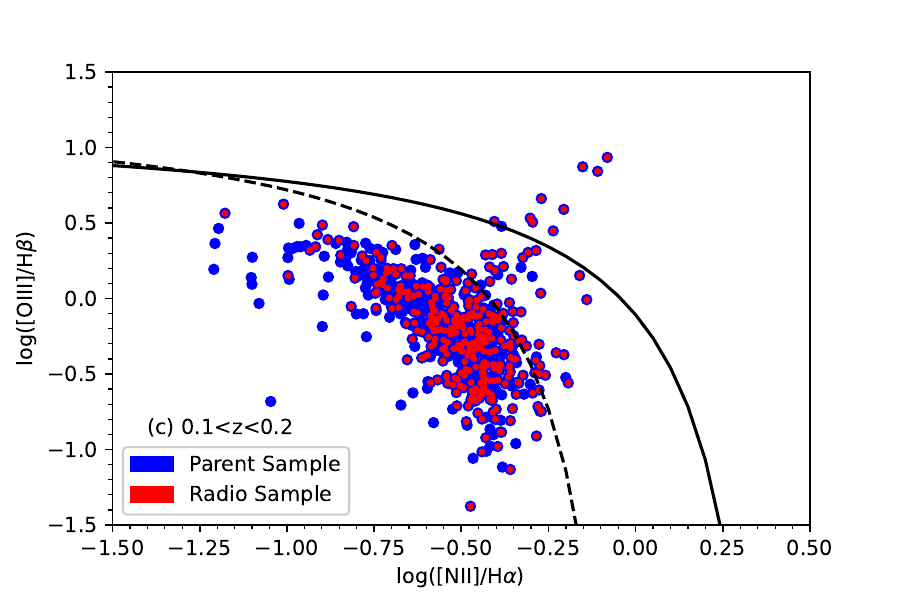}
	\includegraphics[width=0.45\textwidth]{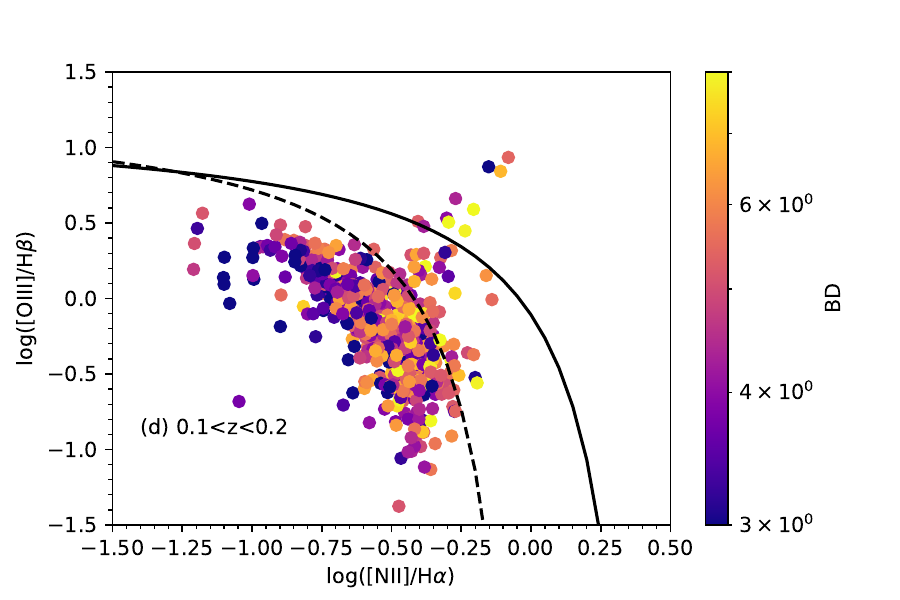}
        \includegraphics[width=0.45\textwidth]{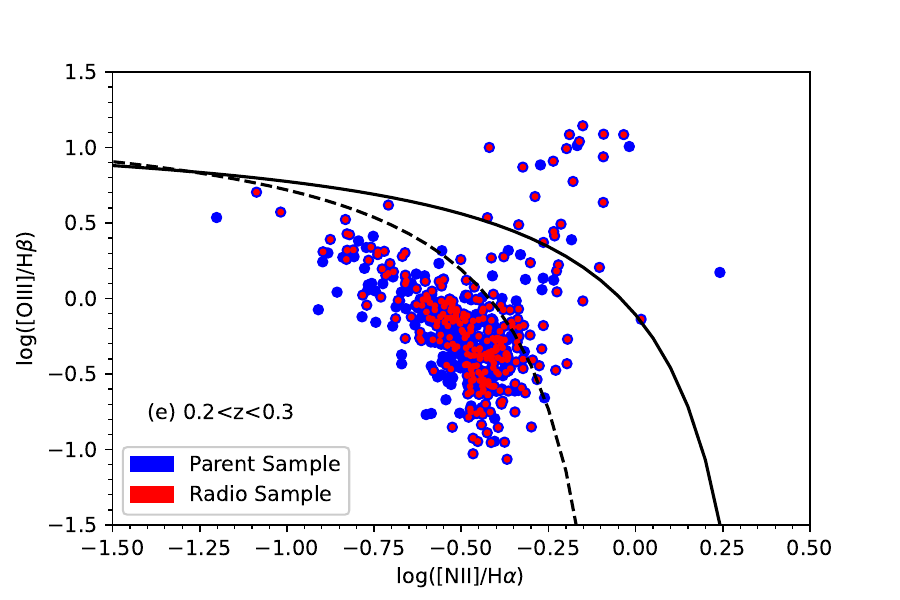}
	\includegraphics[width=0.45\textwidth]{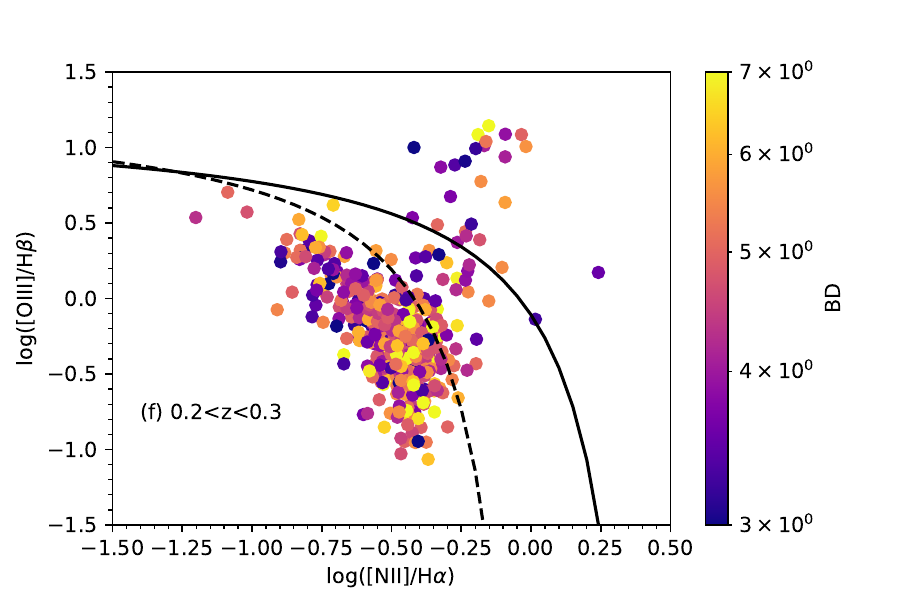}
        \includegraphics[width=0.45\textwidth]{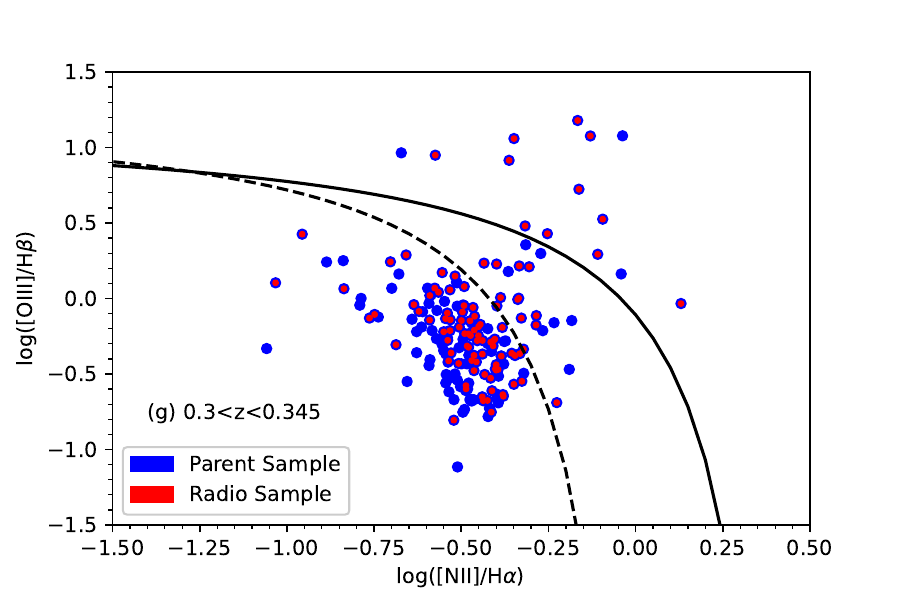}
	\includegraphics[width=0.45\textwidth]{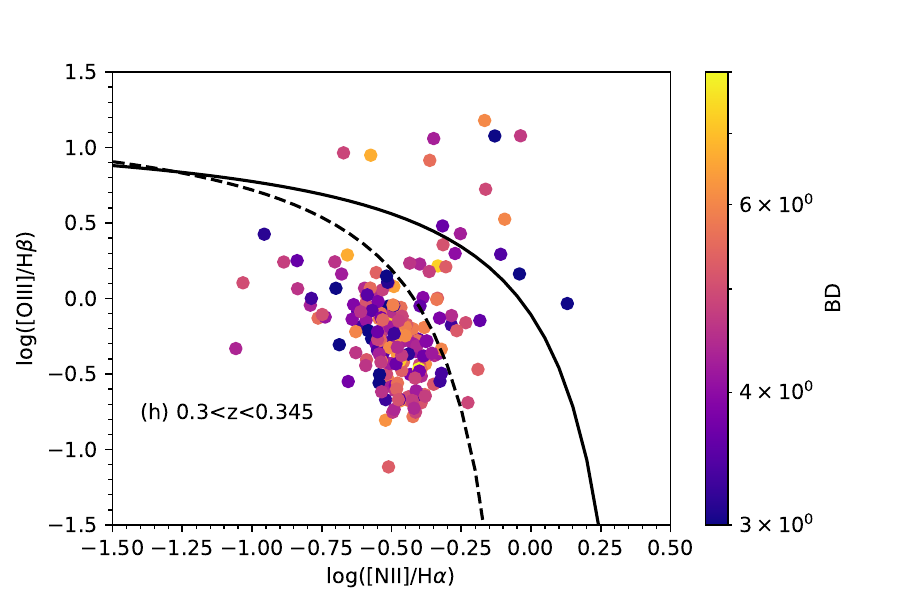}
    \caption{The BPT diagram presented in each redshift bin. The left panels compare the parent sample (GKV gold) in blue with the radio subset (GKV - radio gold) in red. The radio detected systems preferentially populate the upper locus and lower right, corresponding to systems with higher BD, metallicity, and mass. The right panel presents the same data for the  parent (GKV gold) sample only, here colour coded by BD value. The BD can be seen to generally increase from the top left to the bottom right of the diagram.}
    \label{Figure 10}
\end{figure*}



\begin{figure*}[hbt!]
\centering
\includegraphics[width=0.45\textwidth]{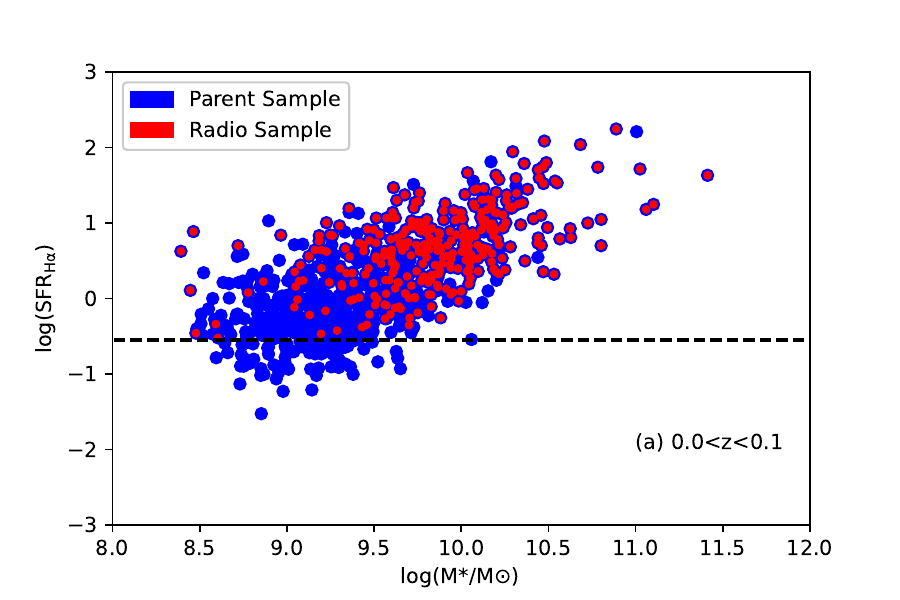}
\includegraphics[width=0.45\textwidth]{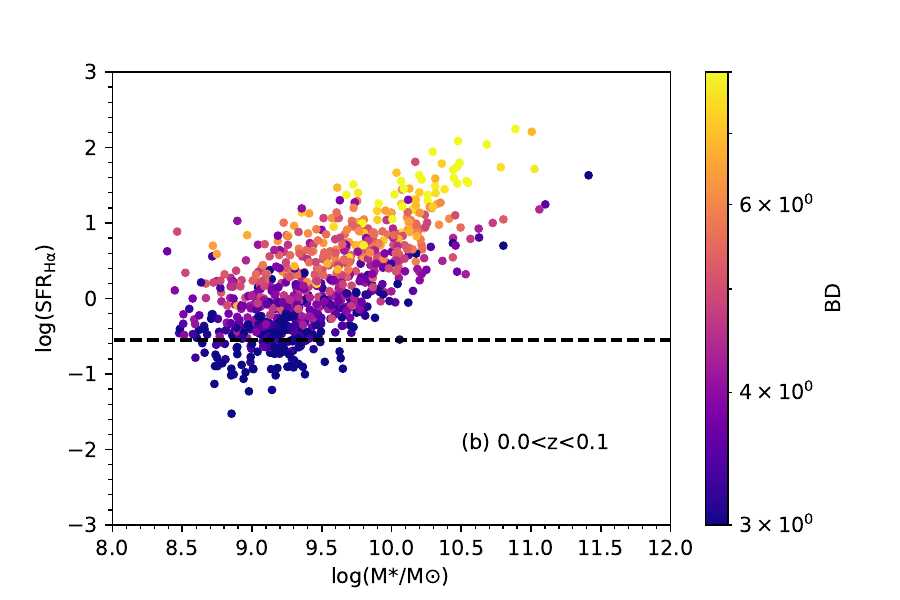}
\includegraphics[width=0.45\textwidth]{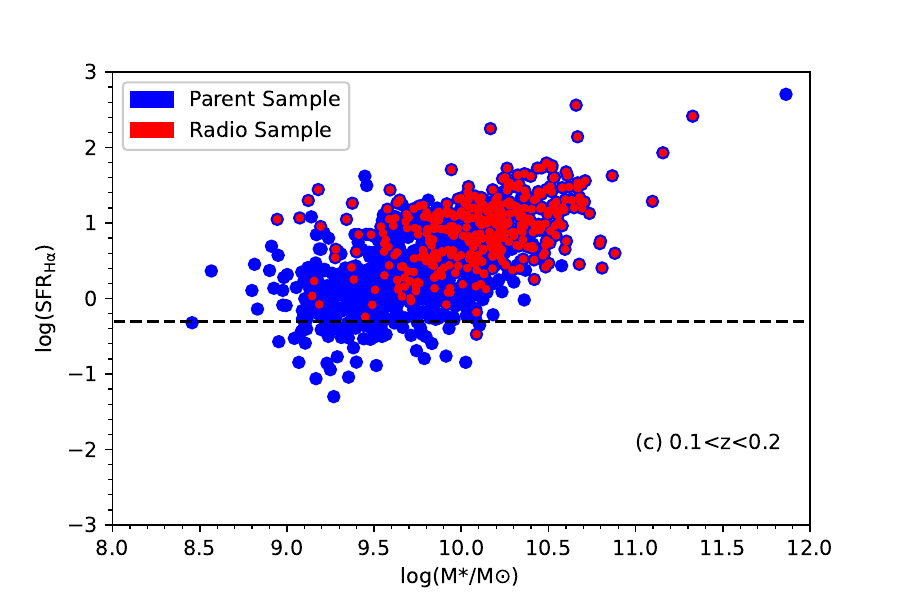}
\includegraphics[width=0.45\textwidth]{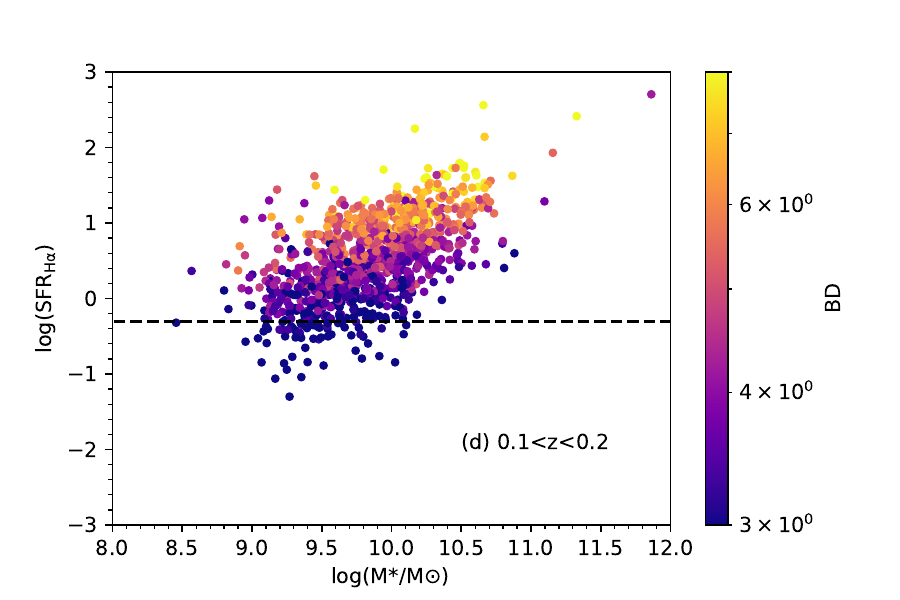}
\includegraphics[width=0.45\textwidth]{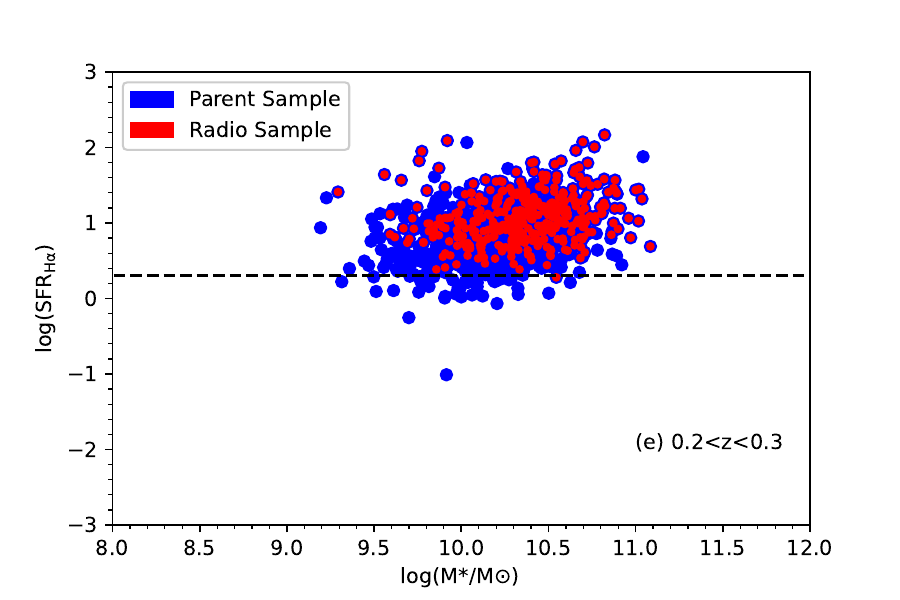}
\includegraphics[width=0.45\textwidth]{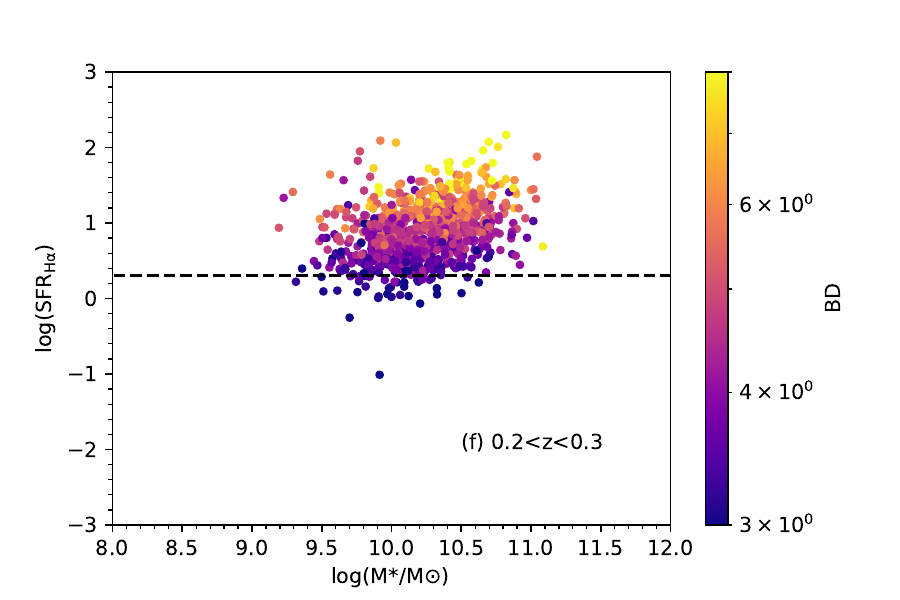}
\includegraphics[width=0.45\textwidth]{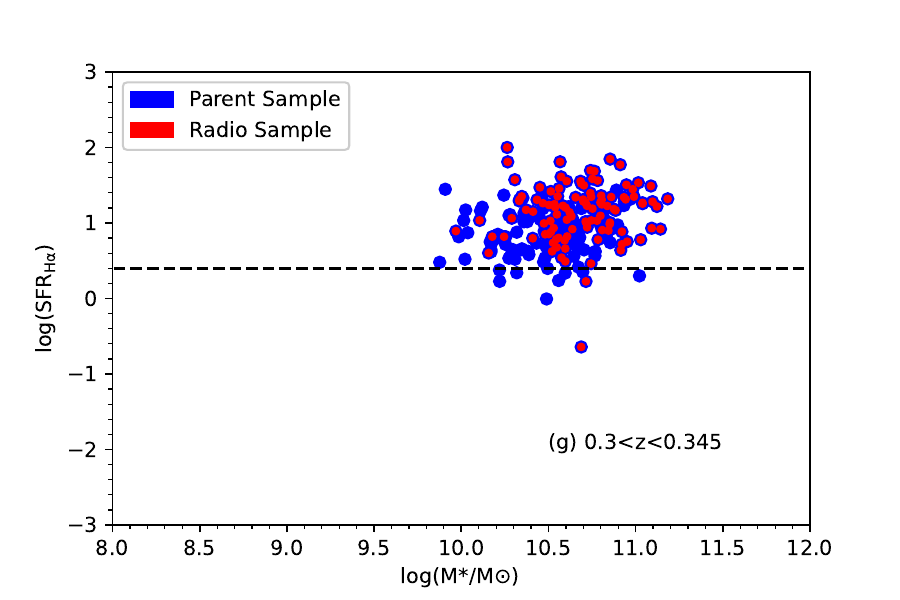}
\includegraphics[width=0.45\textwidth]{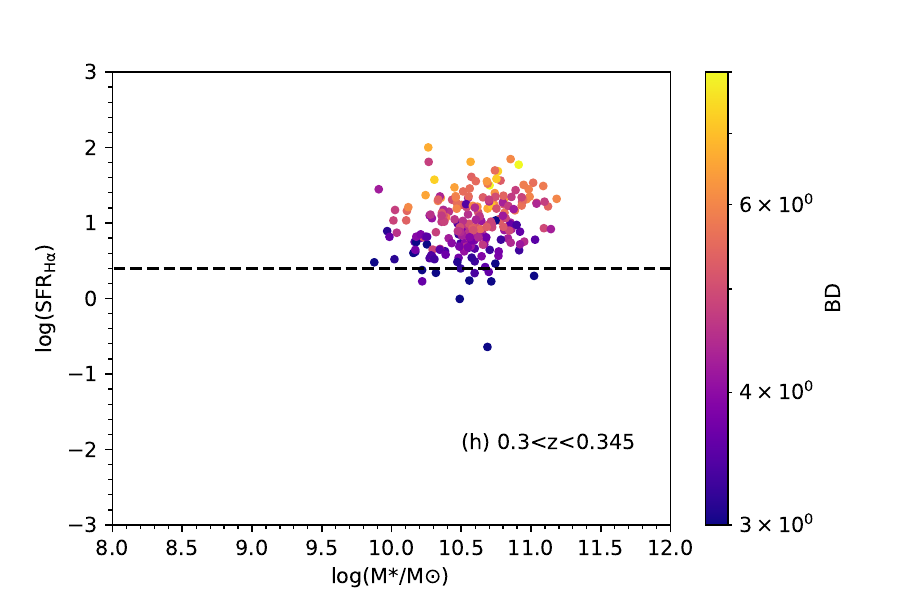}
    \caption{The main sequence of H$\alpha$ SFR as a function of $M_*$ in galaxies (left panels), showing all the optical parent galaxies (GKV gold) in blue with the radio subset (GKV - radio gold) in red. In the right hand panels, the data, for the parent sample only, are reproduced, and here colour coded by BD. The empirical thresholds in H$\alpha$ SFR are marked by the horizontal dashed lines in each redshift bin.}
    \label{Figure 11}
\end{figure*}

\begin{figure}[h!]
\centering
\includegraphics[width=\textwidth]{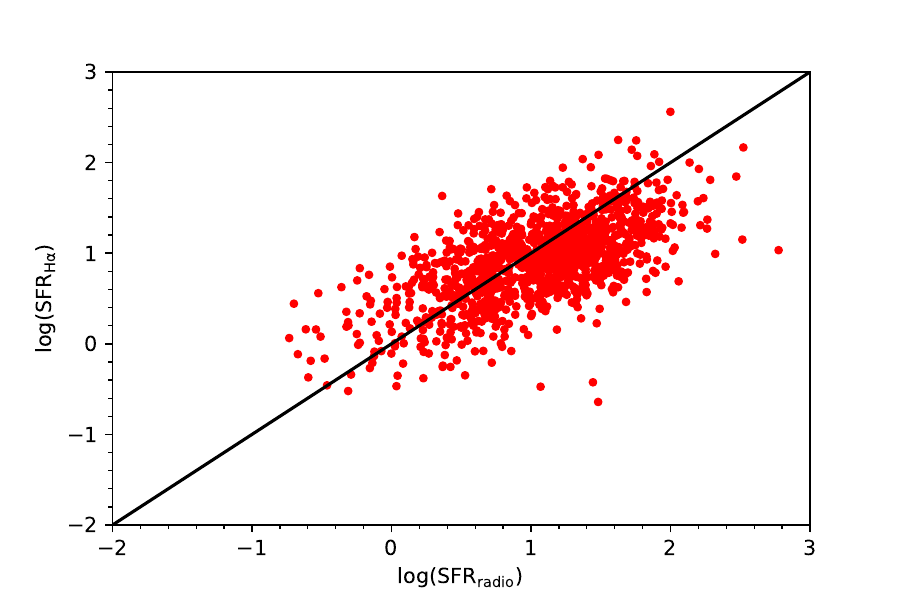}
\caption{Comparison of H$\alpha$ SFR and radio-SFR for the full optical parent sample (GKV gold). As is well-known \citep[e.g.,][]{2003AJ....125..465H,2016MNRAS.461..458D}, there is only a general trend, rather than a tight correlation.}
\label{Figure 12}
\end{figure}

\begin{figure*}[hbt!]
\centering
\includegraphics[width=0.45\textwidth]{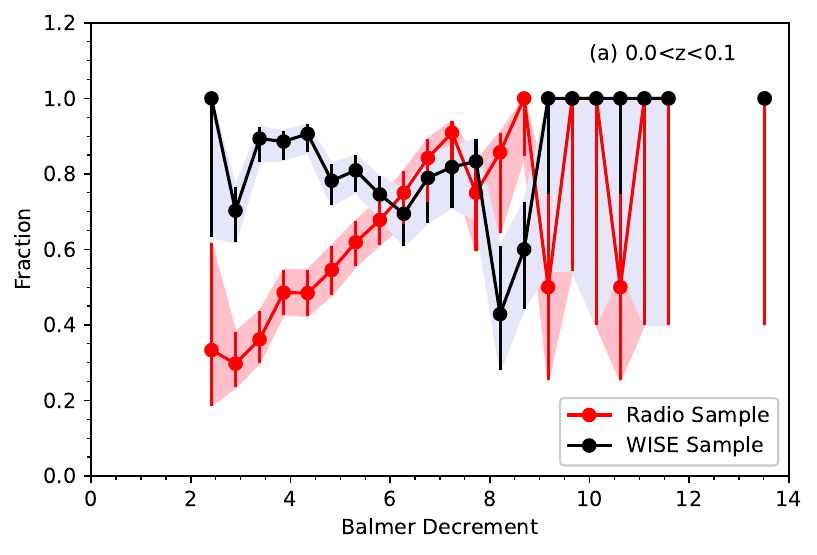}
\includegraphics[width=0.45\textwidth]{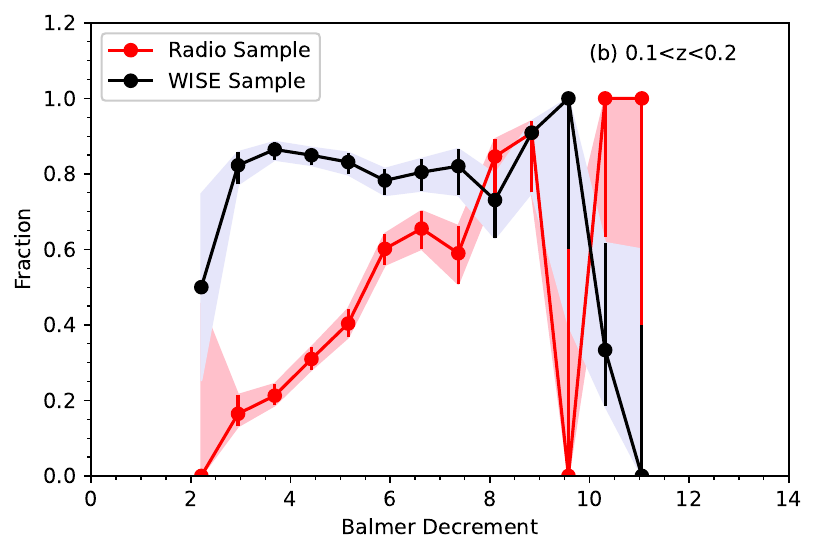}
\includegraphics[width=0.45\textwidth]{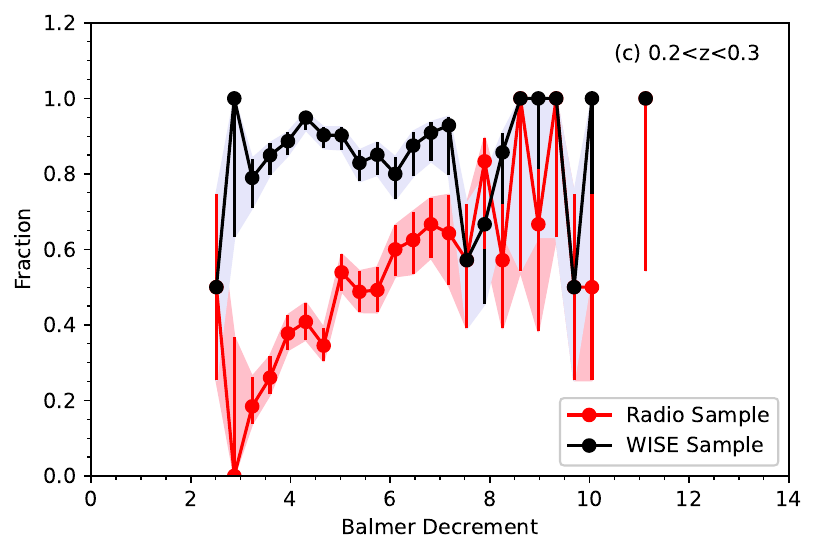}
\includegraphics[width=0.45\textwidth]{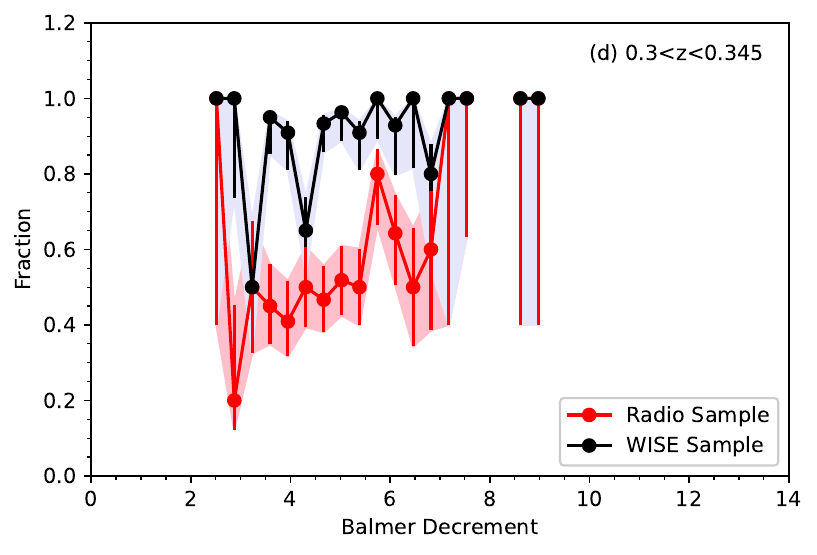}
\caption{{An updated version of Figure~\ref{Figure 6}, again showing fractions of the (gold) radio (GKV - radio, in red), and WISE (GKV - WISE, in black) subsets over the parent sample (GKV), but now excluding the galaxies below the H$\alpha$ SFR threshold defined in Figure 11. Excluding those galaxies that may fall below a nominal radio sensitivity limit does not change the key result, that the radio detected subset is lacking the lowest BD galaxies. The error bars and shaded error regions are estimated as in Figure~\ref{Figure 6}, corresponding to 1$\sigma$ binomial uncertainties, following \citet{2011PASA...28..128C}.}}
\label{Figure 13}

\end{figure*}

\begin{figure*}[hbt!]
\centering
\includegraphics[width=0.45\textwidth]{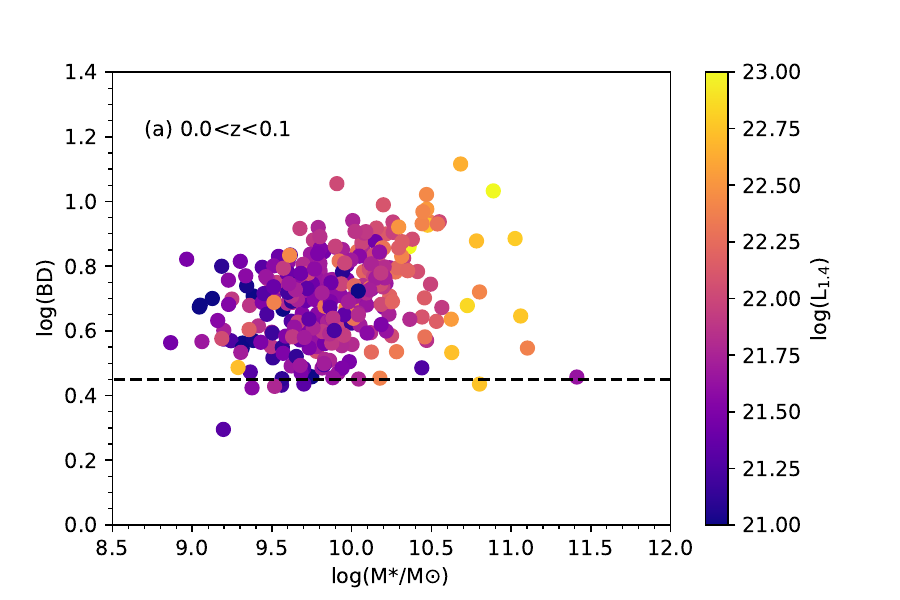}
\includegraphics[width=0.45\textwidth]{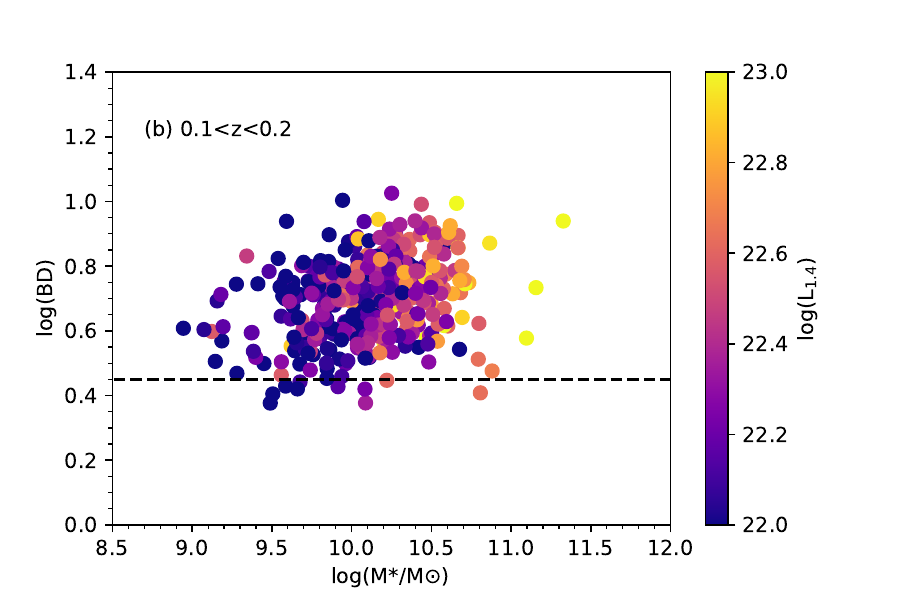}
    \caption{BD as a function of stellar mass colour coded by radio luminosity at $1.4\,$GHz (W\,Hz$^{-1}$) calculated from the measured $888\,$MHz luminosities assuming a spectral index $\alpha=-0.7$ as detailed in \S\,\ref{MS} for the two lowest redshift bins. Showing radio luminosity at $888\,$MHz gives a largely identical result. Note the colour scales are different between panels, to better emphasise the measured range. The horizontal dotted line represents the nominal value of {\rm BD}=2.86. The strong relationship between radio luminosity and stellar mass is evident, but no clear trend between radio luminosity and BD is present.}
    \label{Figure 14}
\end{figure*}

\section{RESULTS AND ANALYSIS}
\label{sec:analysis}

We are aiming to understand the obscuration properties of radio-detected SFGs, and have constructed samples defined by selection from an optical parent sample. This is somewhat different from earlier work \citep[e.g.,][]{2003ApJ...597..269A}, which demonstrates that radio selection of galaxy samples is effective at detecting objects that are too obscured to enter into optical samples. Here, by contrast, we are looking to compare the obscuration properties {\em from the same optically selected parent sample\/} between those galaxies that are radio detected and those that are not. This approach removes the difference in sample selection, allowing us to explore whether there is any intrinsic difference, for a given population, in the obscuration of those galaxies that are radio-detected.

For the purpose of this investigation we use the BD as the obscuration-sensitive parameter. This is not the same as a parameter (such as infrared luminosity or spectral shape) that is representative of the total dust mass. We are interested, however, in understanding the impact of dust on sample selection, and as the BD is a direct tracer of the effect of dust on that fraction of emission that escapes a galaxy, this is the appropriate parameter to use. We note that the nominal Case B value of ${\rm BD} = 2.86$ \citep{1971MNRAS.153..471B} indicates no obscuration. Many measurements, though, give values lower than this due to measurement uncertainties or physical properties (including temperature and metallicity) that differ from the assumed Case B values \citep[e.g.,][]{2015MNRAS.450.3381L}. We do not exclude any galaxies on the basis of their BD value, although we show the nominal ${\rm BD} = 2.86$ for reference in our figures as appropriate.

We explore the relationship between BD and $M_r$ in \S~\ref{BD_Mr}, and the relation between BD and stellar mass in \S~\ref{BD_SM} comparing the radio detected subset against the parent optical and the WISE-detected samples. We investigate how radio-detected galaxies populate the spectral diagnostics in \S~\ref{BD_BPT}.
In \S~\ref{MS} we introduce the galaxy ``main sequence'' to assess the potential impact of our radio sensitivity limit.
In each of the analyses here we restrict ourselves to the GKV (gold), GKV - Radio (gold), and GKV - Radio - WISE (gold) samples, unless noted otherwise.

\subsection{BD - $M_r$ relation}
\label{BD_Mr}

In order to measure the BD for our galaxies, we not only require the H$\alpha$ and H$\beta$ emission lines, but also need to correct their flux measurements for the effect of stellar absorption \citep[e.g.,][]{2003AJ....125..465H,2004MNRAS.351.1151B,2009A&A...508..615L,2012MNRAS.419.1402G}. As with many analyses using GAMA data, we adopt a simple correction to the Balmer emission line equivalent widths of $EW_c =2.5\,$\AA\ \citep[following][]{2013MNRAS.433.2764G,2013MNRAS.430.2047H} to correct the H$\alpha$ and H$\beta$ fluxes, using Equation~4 from \citet{2003AJ....125..465H}. 

The distributions of BD and $M_r$ are shown in Figure~\ref{Figure 4} directly comparing our four redshift bins. The distributions of BD (Figure~\ref{Figure 4}a) are largely similar between all four redshift bins.
Figure~\ref{Figure 4}b illustrates the distribution of optical luminosity range spanned for four volume limited samples. The vertical dashed line represents our magnitude limits for each redshift bin. The lowest redshift bin limit of $M_r=-19$ corresponds to the right edge of the figure. As is typical for flux-limited datasets, the sample in the lowest redshift bin spans the broadest absolute magnitude distribution, with narrower and brighter magnitude ranges accessible in each higher redshift bin, and with the highest luminosities only probed in the higher redshift bins. 

The distributions in BD and $M_r$ comparing the radio (GKV - radio gold) and WISE (GKV - WISE gold) subsets with the parent sample (GKV gold), for the four volume-limited samples, are presented in Figure~\ref{Figure 5}. 

Figure~\ref{Figure 5} shows the distribution in $M_r$. In what will be evident as a recurring theme throughout this analysis, the WISE subset closely follows the distribution of the parent optical sample, but the radio detected population shows some differences. Here the differences are visible as the radio subset favouring the brighter optical galaxies, with proportionally fewer of the fainter optical systems being radio-detected.

Figure~\ref{Figure 5} also shows the distribution in both obscuration and magnitude. Again, the WISE subset closely follows the parent sample, while the radio subset shows some differences. First it is clear that radio counterparts are detected spanning all absolute optical magnitudes in each bin. While fewer radio sources are detected at the faintest magnitudes (seen most easily in panels a and b), reflecting the result in the distribution in magnitude, they are still detected down to and below the magnitude limits of each volume-limited sample. Where radio sources appear to be missing is at the low BD end of the faint population. This is, again, most apparent in panels a and b, the two lowest redshift bins, probing the fainter galaxies. So the radio detected subset appears to be lacking in low obscuration galaxies. Radio counterparts do exist for objects at our faint magnitude limits, so it is not just a magnitude effect that causes them to be missing. We explore this result further below. 

Histograms of the BD are presented in Figure~\ref{Figure 5}, demonstrating explicitly the lack of the lowest BD systems in the radio subset inferred from the distribution of BD and magnitude. In our analysis throughout, the highest redshift bin is typically less significant due to the limited number of targets but retained for completeness. While the WISE subset again reliably traces the parent population, the radio detected systems tend to have proportionally fewer of the lower value BD galaxies. Said another way, the least obscured systems are preferentially not radio detected. The radio detected population seems to trace the high BD end of the parent population, while under-sampling the least obscured systems.

To summarise briefly, then, the radio detected subset traces the brighter galaxies but begins to miss galaxies at progressively fainter magnitudes. Of the fainter galaxies that are missed, it is preferentially the low obscuration systems that are excluded, leading to an overall lack of low BD systems in the radio detected population. 
Together these results demonstrate that the radio-detected subset is missing the lowest obscuration systems compared to the parent population. This is true in each of our four independent volume limited samples, although most visually apparent in the two or three lowest redshift bins.

We quantify the difference in the BD distribution of the parent (GKV gold) and radio detected subset (GKV - radio gold), in each of the four volume-limited samples, using a Kolmogorov-Smirnov (KS) two sample test. For the three lowest redshift volume limited samples, the $p$-value is essentially zero, statistically rejecting the null hypothesis that both samples are drawn from the same population. The highest redshift bin has $p=0.3$, implying no statistical difference between the distributions there, although this bin is also limited to the highest luminosity systems only, and with the fewest measurements. By contrast, the KS test comparison between the WISE subset and the parent sample shows $p$-values close to unity for each redshift bin, implying the strong probability that this subset is drawn from the same underlying population as the parent sample. This confirms that the radio sample is statistically different from the optical parent population in terms of its BD distribution. We explore in more detail below the origin of this discrepancy.

First, to quantify this effect in a more visual fashion, we show in Figure~\ref{Figure 6} the BD distribution in terms of the fraction of radio or WISE detected sources compared to the parent sample. As before, the WISE population closely traces the parent sample, reflected in the largely constant fraction with BD. By contrast, the fraction of galaxies with radio counterparts increases with increasing BD in each redshift bin, including the highest, although again that bin has the largest variations and degree of uncertainty. The uncertainties in Figure~\ref{Figure 6} are obtained by computing the confidence interval at $c = 0.68$ for a beta distribution using the method of \citet{2011PASA...28..128C}, which corresponds to 1$\sigma$ binomial uncertainties. The radio detected galaxies are predominantly associated with the parent galaxies having high values of BD and have progressively lower fractions of counterparts at the low BD end. This again emphasises that the radio sources preferentially favour the high BD end compared to the parent sample. The clear consequence of this result is that the average or median BD for the radio detected subset is higher than that of the parent sample, leading to our claim that radio detected galaxies are more obscured than optically selected galaxies.

\subsection{Stellar Mass - BD dependences}
\label{BD_SM}

The stellar masses used here are taken from the analysis of \citet{2011MNRAS.418.1587T}, derived using a Bayesian analysis drawing on the population synthesis models of \citet{2003MNRAS.344.1000B}, assuming a Chabrier stellar initial mass function  \citep{2003ApJ...586L.133C}, exponentially-declining star formation histories, and the dust obscuration law from \citet{2000ApJ...533..682C}.
Figure~\ref{Figure 7} shows the stellar mass of our samples as a function of redshift.
Figure~\ref{Figure 7}a presents the stellar mass distribution as a function of redshift of the full GKV sample (blue), radio counterparts (GKV - Radio, in red), and WISE counterparts (GKV - WISE, in yellow). The mass range spanned by the radio detected subset is much narrower than the parent and WISE samples, reflecting the same trend in $M_r$. Figure~\ref{Figure 7}b shows these same results restricted to our volume limited samples, reinforcing our magnitude limit choices that correspond to ensuring radio detectability at the lowest luminosity, or mass, limits.
Figure~\ref{Figure 8} shows BD as a function of mass,  demonstrating that the range of obscuration values measured spans the full range of BDs at the highest masses, but at the lowest mass end, the radio detections preferentially favour the higher BD systems. The WISE detected galaxies, again, tend to have similar properties to the parent sample.

Figure~\ref{Figure 9} presents comparisons between BD and stellar mass for the four volume-limited samples. The radio detected population favours the higher mass end compared to the parent and WISE samples. In each of the volume limited samples, the radio detected galaxies are still sensitive to the full mass range, but exclude the low BD systems  at the low mass end, leading to undersampling of the lowest mass galaxies.
This illustration emphasises most strongly the lack of low mass, low BD systems in the radio detected subset, most visible by eye in the lowest two redshift bins. In general terms, with an increase in stellar mass galaxies tend to show higher levels of obscuration, and this is clearly seen in all four redshift bins, although the range of obscuration at any given mass may still be large. Again, the WISE detected subset closely follows the parent sample distribution, but it is clear that the radio detected subset misses more of the lower mass systems with the lowest values of BD. Therefore, the radio detected galaxies are preferentially associated with the higher mass and higher BD parent galaxies. This result is not just a sensitivity effect, since we have carefully defined our volume-limited samples to ensure that the radio detection limits are sensitive to the lowest optical luminosity systems. The implication then is that the low mass and low BD systems without radio counterparts must have radio luminosities, if any, below our detection threshold, and consequently significantly lower than their higher BD counterparts.
 
\subsection{Spectral diagnostics}
\label{BD_BPT}

Many authors have used the BPT diagram to explore the properties of SFGs  \citep[e.g.,][]{2008MNRAS.385..769B, 2013MNRAS.430.2047H, 2015A&A...574A..47S, 2016ApJ...828...18M, 2017ApJ...847...18Z, 2018ApJ...855..132F}, including investigating how galaxies populate this diagnostic as a function of different physical properties, such as mass, metallicity, obscuration, and SFR.


Similarly, we explore our sample properties using this diagnostic (Figure~\ref{Figure 10}).
For clarity we note here that Figure~\ref{Figure 10} shows the AGN population as well as the SF systems, although they are not included in our analysis or discussion, as the Balmer lines in these systems may not be accurate tracers of the obscuration if they are influenced by the AGN ionisation. 
 
The distribution of BD is shown for our parent (GKV gold) sample across the BPT diagram, in our four volume-limited redshift bins, in the right hand panels of Figure~\ref{Figure 10}. It can be seen that there is a tendency for the higher BD systems to populate the upper locus and bottom right of the diagram, with a trend for BD to be higher for galaxies with higher values of [N{\sc ii}]/H$\alpha$ and [O{\sc iii}]/H$\beta$.
This trend is likely to be reflective of higher metallicity in the systems toward the bottom right, since,
as described by other authors \citep[e.g.,][]{2013MNRAS.430.2047H,2015MNRAS.450.3381L}, there is a known metallicity trend in the BPT diagram, with metallicity increasing from the top left towards the bottom right.

We compare the parent (GKV gold) and radio (GKV - radio gold) populations in this diagnostic (left hand panels of Figure~\ref{Figure 10}).
Here it is clear that the radio detected sources preferentially  populate the upper locus and right hand side of the distribution compared to the parent population, which extends much further to the left and further from the Kauffmann line. The radio detected systems are clearly visible among the AGN population, and are well represented in the composite region. The majority of the radio detected SFGs are in that region corresponding to higher metallicity host galaxies, and these will also correspond typically to higher mass systems, because of the mass-metallicity relationship \citep[e.g.,][]{2004ApJ...613..898T,2013MNRAS.434..451L}. The radio detected population is clearly coming preferentially from this higher mass and higher metallicity region, corresponding as well to higher BD.

\subsection{SFR -  Stellar Mass relation}
\label{MS}

Many studies use the scaling relation between star formation rate and galaxy stellar mass as a key diagnostic to explore the role of star formation in understanding galaxy formation \citep{2004MNRAS.351.1151B, 2007ApJS..173..267S, 2007ApJ...670..156D, 2007ApJ...660L..47N, 2007ApJ...660L..43N, 2013MNRAS.434..209B, 2015A&A...575A..74S, 2016MNRAS.461..458D}. This well-known relation, referred to as the ``galaxy main sequence'' represents the link between star formation and the mass growth of galaxies \citep{2011ApJ...739L..40R, 2014ApJS..214...15S}. This relation between SFR and $M_*$ in actively star-forming galaxies changes with increasing redshift and has been extensively explored in approaches aiming to understand models of galaxy evolution \citep[e.g.,][]{2007ApJ...660L..47N}.

We use this SFR-$M_*$ relation for our sample in order to understand whether our radio sensitivity limit is a significant contributor to the results identified above.
To calculate H$\alpha$ SFRs, we follow \citet{2003AJ....125..465H} and \citet{2011MNRAS.415.1647G}, to measure aperture-, obscuration-, and stellar-absorption-corrected H$\alpha$ luminosities: 

\begin{multline} 
    L_{H\alpha}  = (EW_{H\alpha}  +  EW_c)  \times  10^{-0.4(M_r - 34.10)}\\
    \times \frac{3 \times 10^{18}}{[6564.61(1+z)]^2}\times \left(\frac{\frac{F_{H\alpha}}{F_{H\beta}}}{2.86}\right)^{2.36}
\label{eq:eq1}
\end{multline}

\noindent where $EW_{{\rm H}\alpha}$ denotes the H$\alpha$ equivalent width, $EW_c$ is the equivalent width correction for stellar absorption, $M_r$ is the $r$-band absolute magnitude and $F_{H\alpha}$ / $F_{H\beta}$ denotes the Balmer decrement used to correct for dust obscuration. From this luminosity, the H$\alpha$ SFR is determined from the \citet{1998ApJ...498..541K} relation:
\begin{equation}
    SFR_{H\alpha} [M_{\odot} yr^{-1}] =  \frac{L_{H\alpha}}{1.27 \times 10^{34}}.
	\label{eq:eq2}
\end{equation}

We present the ``main sequence'' with these SFRs as a function of $M_*$ in each of the four volume limited redshift bins of our GKV gold sample (Figure~\ref{Figure 11}). In this Figure, the left-hand panels show the parent sample (blue) overlaid with the radio detected sample (red). The right hand panels show the parent sample colour coded by BD. From the right hand panels it is explicitly evident that the low BD galaxies are also those with the lowest mass and SFR. At the high mass end essentially all of the optical parent sample is identified with radio counterparts. At the low mass and low SFR end, though, we start losing radio counterparts, and these are also those galaxies that have the lowest BD.

It is apparent in Figure~\ref{Figure 11} that there is an empirical threshold in terms of SFR$_{H\alpha}$ below which we are not identifying radio counterparts. In principle we could attempt to estimate such a threshold quantitatively, based on the radio flux density detection limit, by converting that to a radio-derived SFR and then to an H$\alpha$ SFR. In practice, though, this is impractical due to the large scatter between these SFR estimators. In other words, the radio flux density threshold does not convert simply to a hard H$\alpha$ SFR threshold. We demonstrate this in Figure~\ref{Figure 12}, which explicitly compares the radio and H$\alpha$ derived SFRs.
The radio SFR calibration is adopted from the \citet{2003ApJ...586..794B} relation:
\begin{equation}
    SFR_{1.4GHz} [M_{\odot} yr^{-1}] =  \frac{f L_{1.4GHz}}{1.81 \times 10^{21} {{\rm WHz}^{-1}}}.
	\label{eq:eq3}
\end{equation}
where
\begin{equation}
f = 
\left\{
\begin{array}{ll}
 1 & L_{\rm 1.4GHz} > L_c \\
 (0.1 + 0.9 (L_{\rm 1.4GHz}/L_c)^{0.3})^{-1} & L_{\rm 1.4GHz} \le L_c,
\end{array}
\right.
\end{equation}
and we assume a radio spectral index\footnote{We adopt the convention that spectral index, $\alpha$, links flux density, $S$, and frequency, $\nu$, through $S\propto\nu^{\alpha}$.} of $\alpha=-0.7$ to convert our $888\,$MHz luminosities to equivalent $1.4\,$GHz luminosities.
The order-of-magnitude scatter between these quantities prevents a simplistic linear-calibration conversion to scale from a radio flux density limit to an equivalent SFR limit in Figure~\ref{Figure 11}.
Despite this, we can use the apparent empirical limit, which we mark with horizontal dashed lines, to explore the effect of our radio sensitivity limit. By excluding galaxies with SFRs below this threshold, we can establish whether the result shown in Figure~\ref{Figure 6} arises primarily from the radio sensitivity limit or not. After excluding galaxies below this empirical threshold, the revised histogram fractions, shown in Figure~\ref{Figure 13}, demonstrate that the result is essentially unchanged, and the lack of low BD galaxies with radio counterparts is not a consequence of the radio sensitivity limit. Differences between Figure~\ref{Figure 6} and Figure~\ref{Figure 13} are primarily visible in the lowest two redshift bins, where more sources have been excluded, compared to the higher two redshift bins. The uncertainty measurements are also slightly larger in Figure~\ref{Figure 13} due to the smaller numbers of objects in each bin. Otherwise, there are very few differences.

Since this result is not a simple consequence of our radio detection limits, those galaxies that are missing radio detections must have radio luminosities lower than expected compared to their radio detected counterparts at a given mass, SFR, or optical luminosity. The systems with the lowest BDs are also those with the lowest SFRs \citep[e.g.,][]{2011MNRAS.415.1647G,2023MNRAS.518.5475P}, and hence the lowest radio luminosities \citep{1992ARA&A..30..575C}. To explore this further, we look at the distribution of radio luminosity with BD and stellar mass in Figure~\ref{Figure 14}. For those galaxies that are radio detected it is evident that, while there is a strong relationship between radio luminosity and stellar mass, there is no clear or strong trend between radio luminosity and the BD. This implies that for a given stellar mass we might expect a more or less constant radio luminosity, independent of the BD. This expectation clearly breaks down at the lowest mass end.

\section{DISCUSSION}
\label{sec:discussion}
\citet{2001AJ....122..288H} proposed an optical luminosity dependent approach to correcting for obscuration in galaxies in the absence of more direct measurements of the obscuration itself, building on earlier work  \citep{1996ApJ...457..645W}, and later updated by \citet{2010MNRAS.402.2017G} for galaxies at higher redshift. This approach was refined by \citet{2003ApJ...597..269A}, using a radio selected galaxy sample and \citet{2003ApJ...599..971H} using an optically selected sample. \citet{2003ApJ...599..971H} suggested that the differences found between the luminosity dependence on the obscuration was a consequence of radio selection preferentially being more sensitive to heavily obscured populations missing from the optically-selected samples.

Our approach here has been to begin with an optically selected sample (GKV) and explore the obscuration properties of the radio detected population, to see whether, and if so how, they differ. We do find a statistically significant difference in the obscuration of the radio detected population (\S,\ref{BD_Mr}), with these galaxies being on average more obscured than the parent sample (Figures~\ref{Figure 5} and \ref{Figure 6}). We identify that the difference arises from the lack of the lowest mass and least obscured (lowest BD) galaxies in the radio detected subset (Figure~\ref{Figure 14}).

To summarise our main results:
\begin{enumerate}
 \item We analysed four volume-limited samples, defined to ensure that galaxies should lie above our radio detection limits (Figure~\ref{Figure 2}).
 \item The radio detected systems recover almost all of the galaxies at high optical luminosities, but lack the low optical luminosity and low BD galaxies (Figure~\ref{Figure 5}). 
 \item The radio detected subsets are lacking the least obscured (lowest BD) galaxies at the lower mass end of the samples (Figure~\ref{Figure 9}).
 \item The distribution of SFGs in the BPT diagrams primarily trace a sequence in metallicity. It is evident that the radio detected objects lack those toward the upper left in Figure~\ref{Figure 10}, which correspond to low BD and low metallicity systems.
 \item Using the SFR$_{{\rm H}\alpha}$ - $M_*$ ``main sequence'' relation we defined an empirical SFR limit for our radio detections, and used this to demonstrate that the lack of low SFR, low BD galaxies with radio counterparts is not a consequence of this sensitivity limit (Figure~\ref{Figure 13}).
 
\end{enumerate}
Together these results indicate that the radio detected subset is missing those galaxies with the lowest mass, SFR, and obscuration. Here we explore further why this might be the case.

The ``staged galaxy formation'' model of \citet{2007ApJ...660L..47N, 2007ApJ...660L..43N} describes the evolution in galaxy star formation through a gradual transition from high SFR in high mass galaxies at early times that rapidly declines, while low mass galaxies start forming the bulk of their stars later on average and with a longer timescale. This broad scenario is also often referred to as ``galaxy downsizing'' \citep{1996AJ....112..839C}. The challenge of understanding the link between the gas and the star formation that it fuels is explored by \citet{2008ApJ...682L..13H}, establishing a need for gas replenishment in galaxies to sustain ongoing star formation over cosmic time. This broad picture establishes a difference in the formation histories of the lowest mass galaxies compared to the higher mass population, although it does not directly address why their radio properties may differ in the way we see.

It is well known that for low luminosity (or low mass, or low metallicity) galaxies the relationship between radio luminosity and SFR changes \citep {2003ApJ...586..794B, 2017ApJ...847..136B}, and probably arises through the greater tendency for cosmic ray electrons to escape such low mass systems, as shown in samples of nearby dwarf galaxies \citep[e.g.,][]{2018ApJS..234...29H,2022A&A...664A..83H}. This may be related to the effect we see in Figure~\ref{Figure 9} but because this ``leaky box'' description is explicitly mass dependent, by itself it does not explain why we see a difference with BD at a common mass.
\citet{2010ApJ...714.1305S}, using the ratio between radio and ultraviolet luminosities as a proxy for obscuration, find that there is a galaxy type dependence on the obscuration. They show that quiescent galaxies seem to have higher levels of obscuration for a given radio luminosity than SFGs. We may be seeing this effect here at the lowest mass end of our population, and this would be consistent with the implication that the systems we are missing are similar to the ``slowest forming galaxies'' of \citet{2011MNRAS.413.1236B}.
With more sensitive radio data we would be more likely to detect these otherwise missing systems, and could quantitatively explore such relations between radio luminosity and BD for low mass galaxies.

It is also true that there is a broad range of SFR for any chosen galaxy mass, given the finite width in the SFR$_{{\rm H}\alpha}$ - $M_*$ ``main sequence'' relation \citep[e.g.,][]{2018MNRAS.475.3010G}. In particular, the low mass, low SF systems with shallow potential wells experience SF in a very bursty mode \citep{2007ApJ...671L.113L, 2009ApJ...706..599L, 2011MNRAS.413.1236B, 2013MNRAS.434..209B}. Such  stochastic SF, for any value of SFR, may in principle lead to a time-delay between the radio emission (arising from supernova remnants) compared to the H$\alpha$ emission from an active burst of star formation. This in turn may result in some low mass galaxies with measurable H$\alpha$ having no radio detection, because they are serendipitously caught at a point before significant radio emission has been generated. This effect may also contribute to the scatter seen in the SFR comparison (Figure~\ref{Figure 12}) at the low SFR end. It is worth noting, though, that the low mass galaxies considered here have masses in the range of $8.5 \lesssim \log(M_*) \lesssim 9.5$, which is larger by an order or magnitude or more than those in the studies referenced above.

With that in mind, one tentative explanation for an increase in the scatter seen in the lowest mass galaxies \citep[e.g.,][]{2007ApJ...671L.113L, 2009ApJ...706..599L,2009ApJ...695..765M} relates to possible variations in the stellar initial mass function (IMF). We can speculatively draw on those results to consider whether differences in the IMF may also contribute to our result.
Recall that the radio luminosity from star formation is associated with high mass stellar populations \citep{1992ARA&A..30..575C}. It is the high mass stars that produce the supernovae, whose ejecta accelerate the cosmic ray electron population to the relativistic speeds needed to produce synchrotron emission at radio wavelengths. The highest mass stars are also associated with increased levels of obscuration \citep {1998ApJ...501..643D}. Now, if there are fewer high mass stars present than expected, the radio luminosity will be lower than expected. So our result could be a consequence of low mass, low obscuration galaxies having IMFs that are poorer in high mass stars compared to their higher BD counterparts. In other words, they would have a steeper high mass IMF slope, as also suggested by \citet{2011MNRAS.413.1236B}.

This speculation is qualitatively consistent with other analyses looking at IMF variations in SFGs. For example, \citet{2011MNRAS.415.1647G} find a tendency for low SFR or low specific SFR galaxies (which are also low mass systems) to have IMFs preferentially lacking in high mass stars (steeper high mass IMF slopes) compared to their high SFR counterparts. Similar results, although quantitatively different, are found by \citet{2017MNRAS.468.3071N}, \citet{2009ApJ...695..765M}, \citet{2009ApJ...706..599L}, and \citet{2009ApJ...706.1527B} \citep[and see][for a detailed review]{2018PASA...35...39H}.



Two directions for further exploration make themselves apparent here. The first is related to understanding the impact of obscuration on radio or optical selected samples. This requires us to invert the approach, and begin with a complete radio selected sample, with well-defined obscuration measures for all radio sources, to explore how the obscuration properties differ as a function of optical magnitude or survey sensitivity.
The second is a deeper exploration into whether there may be any measurable IMF effect, associated with the underlying star formation, and potentially its physical distribution throughout a galaxy.

A wider range of radio measurements from the main survey data of the EMU \citep{2011PASA...28..215N, 2021PASA...38...46N} will support the expansion of this type of analysis to a much wider set of galaxies with available optical spectra, including spatially resolved spectroscopy. The technique of \citet{2011MNRAS.415.1647G} can be directly applied to large samples of radio galaxies from the EMU main survey in order to directly explore IMF variations.
Such an analysis may be expanded by explicitly looking at any dependence on the spatial distribution of SF throughout a galaxy, through the current and newer generations of integral field spectroscopy survey projects, such as SAMI \citep{2012MNRAS.421..872C}, CALIFA \citep{2012A&A...538A...8S}, MANGA \citep{2015ApJ...798....7B}, and MAGPI \citep{2021PASA...38...31F}. This approach, requiring H$\alpha$ line measurements, is of necessity limited with optical spectroscopic datasets to the low redshift Universe. Alternative IMF-sensitive metrics, especially those accessible at high redshifts, would be valuable to establish in order to further extend this work.

\section{CONCLUSION}
\label{sec:conclusion}

The motivation for this analysis originated in part from the early work of \citet{2003ApJ...599..971H} and \citet{ 2003ApJ...597..269A} showing that one consequence of increased dust obscuration in higher SFR systems is that radio selection is sensitive to more heavily obscured galaxies that may potentially be excluded from optically selected samples. To extend these early analyses, we explore the obscuration properties of radio detected galaxies in the local universe, using a sample of SFGs from the GAMA GKV catalogue combined with EMU early science data. Our aim is to explore the properties of obscuration in galaxies using the BD, and examine how they differ between a radio detected sample and its parent optical sample.

We find that while the majority of high mass galaxies have radio counterparts, in the low mass galaxy population only the highest obscuration systems appear to. This leads to a statistically significant difference in the obscuration properties for the radio detected subset. The low obscuration, low mass galaxies do not show radio emission, although it would be detectable if present at the same proportional level as their higher obscuration counterparts. 
We have demonstrated that this result is robust, and not a consequence of the sensitivity limit of the radio data.

We discuss possible causes of the lower than expected radio luminosity in the low mass, low BD population. This may be due in part to the suggestion of a galaxy type dependence with BD \citep{2010ApJ...714.1305S}, as well the possibility that the radio detections are impacted by the bursty nature of star formation in low mass galaxies \citep[e.g.,][]{2011MNRAS.413.1236B,2013MNRAS.434..209B}.
We also speculate that these systems may have steeper high mass IMF slopes, leading to relatively fewer high mass stars, and consequently to lower radio luminosities than their higher BD counterparts at the same stellar mass.

\begin{acknowledgement}
We thank Ian Smail for helpful feedback and input as this work was being prepared. Y.A.G. is supported by US National Science Foundation (NSF) grants AST 20-09441 and AST 22-06053. M.B. is supported by the Polish National Science Center through grants no.\ 2020/38/E/ST9/00395, 2018/30/E/ST9/00698, 2018/31/G/ST9/03388 
and 2020/39/B/\-ST9/03494, and by the Polish Ministry of Science and Higher Education through grant DIR/WK/2018/12. L.M. acknowledges financial support from the Italian Ministry of Foreign Affairs and International Cooperation (MAECI Grant Number ZA18GR02) and the South African NRF (Grant Number 113121) as part of the ISARP RADIOSKY2020 Joint Research Scheme. I.P. acknowledges support from CSIRO under its Distinguished Research Visitor Programme, to work on the GAMA23 EMU Early Science project. E.V. acknowledges support by the Carl Zeiss Stiftung with the project code KODAR.

GAMA is a joint European-Australasian project based around a spectroscopic campaign using the Anglo-Australian Telescope. The GAMA input catalogue is based on data taken from the Sloan Digital Sky Survey and the UKIRT Infrared Deep Sky Survey. Complementary imaging of the GAMA regions is being obtained by a number of independent survey programmes including {\em GALEX\/} MIS, VST KiDS, VISTA VIKING, {\em WISE\/}, {\em Herschel}-ATLAS, GMRT and ASKAP providing UV to radio coverage. GAMA is funded by the STFC (UK), the ARC (Australia), the AAO, and the participating institutions. The GAMA website is \url{http://www.gama-survey.org/}.


This scientific work uses data obtained from Inyarrimanha Ilgari Bundara / the Murchison Radio-astronomy Observatory. We acknowledge the Wajarri Yamaji People as the Traditional Owners and native title holders of the Observatory site. The Australian SKA Pathfinder is part of the Australia Telescope National Facility (https://ror.org/05qajvd42) which is managed by CSIRO. Operation of ASKAP is funded by the Australian Government with support from the National Collaborative Research Infrastructure Strategy. ASKAP uses the resources of the Pawsey Supercomputing Centre. Establishment of ASKAP, the Murchison Radio-astronomy Observatory and the Pawsey Supercomputing Centre are initiatives of the Australian Government, with support from the Government of Western Australia and the Science and Industry Endowment Fund.

This paper includes archived data obtained through the CSIRO ASKAP Science Data Archive, CASDA (\url{http://data.csiro.au}).

\end{acknowledgement}


\bibliography{ms}

\begin{thebibliography}{}
\expandafter\ifx\csname natexlab\endcsname\relax\def\natexlab#1{#1}\fi

\bibitem[{{Abbott} {et~al.}(2018){Abbott}, {Abdalla}, {Allam}, {Amara},
  {Annis}, {Asorey}, {Avila}, {Ballester}, {Banerji}, {Barkhouse}, {Baruah},
  {Baumer}, {Bechtol}, {Becker}, {Benoit-L{\'e}vy}, {Bernstein}, {Bertin},
  {Blazek}, {Bocquet}, {Brooks}, {Brout}, {Buckley-Geer}, {Burke}, {Busti},
  {Campisano}, {Cardiel-Sas}, {Carnero Rosell}, {Carrasco Kind}, {Carretero},
  {Castander}, {Cawthon}, {Chang}, {Chen}, {Conselice}, {Costa}, {Crocce},
  {Cunha}, {D'Andrea}, {da Costa}, {Das}, {Daues}, {Davis}, {Davis}, {De
  Vicente}, {DePoy}, {DeRose}, {Desai}, {Diehl}, {Dietrich}, {Dodelson},
  {Doel}, {Drlica-Wagner}, {Eifler}, {Elliott}, {Evrard}, {Farahi}, {Fausti
  Neto}, {Fernandez}, {Finley}, {Flaugher}, {Foley}, {Fosalba}, {Friedel},
  {Frieman}, {Garc{\'\i}a-Bellido}, {Gaztanaga}, {Gerdes}, {Giannantonio},
  {Gill}, {Glazebrook}, {Goldstein}, {Gower}, {Gruen}, {Gruendl}, {Gschwend},
  {Gupta}, {Gutierrez}, {Hamilton}, {Hartley}, {Hinton}, {Hislop}, {Hollowood},
  {Honscheid}, {Hoyle}, {Huterer}, {Jain}, {James}, {Jeltema}, {Johnson},
  {Johnson}, {Kacprzak}, {Kent}, {Khullar}, {Klein}, {Kovacs}, {Koziol},
  {Krause}, {Kremin}, {Kron}, {Kuehn}, {Kuhlmann}, {Kuropatkin}, {Lahav},
  {Lasker}, {Li}, {Li}, {Liddle}, {Lima}, {Lin}, {L{\'o}pez-Reyes}, {MacCrann},
  {Maia}, {Maloney}, {Manera}, {March}, {Marriner}, {Marshall}, {Martini},
  {McClintock}, {McKay}, {McMahon}, {Melchior}, {Menanteau}, {Miller},
  {Miquel}, {Mohr}, {Morganson}, {Mould}, {Neilsen}, {Nichol}, {Nogueira},
  {Nord}, {Nugent}, {Nunes}, {Ogando}, {Old}, {Pace}, {Palmese},
  {Paz-Chinch{\'o}n}, {Peiris}, {Percival}, {Petravick}, {Plazas}, {Poh},
  {Pond}, {Porredon}, {Pujol}, {Refregier}, {Reil}, {Ricker}, {Rollins},
  {Romer}, {Roodman}, {Rooney}, {Ross}, {Rykoff}, {Sako}, {Sanchez}, {Sanchez},
  {Santiago}, {Saro}, {Scarpine}, {Scolnic}, {Serrano}, {Sevilla-Noarbe},
  {Sheldon}, {Shipp}, {Silveira}, {Smith}, {Smith}, {Smith}, {Soares-Santos},
  {Sobreira}, {Song}, {Stebbins}, {Suchyta}, {Sullivan}, {Swanson}, {Tarle},
  {Thaler}, {Thomas}, {Thomas}, {Troxel}, {Tucker}, {Vikram}, {Vivas},
  {Walker}, {Wechsler}, {Weller}, {Wester}, {Wolf}, {Wu}, {Yanny}, {Zenteno},
  {Zhang}, {Zuntz}, {DES Collaboration}, {Juneau}, {Fitzpatrick}, {Nikutta},
  {Nidever}, {Olsen}, {Scott}, \& {NOAO Data Lab}}]{2018ApJS..239...18A}
{Abbott}, T.~M.~C., {Abdalla}, F.~B., {Allam}, S., {et~al.} 2018, \apjs, 239,
  18

\bibitem[{{Afonso} {et~al.}(2003){Afonso}, {Hopkins}, {Mobasher}, \&
  {Almeida}}]{2003ApJ...597..269A}
{Afonso}, J., {Hopkins}, A., {Mobasher}, B., \& {Almeida}, C. 2003, \apj, 597,
  269

\bibitem[{{Baldry} {et~al.}(2010){Baldry}, {Robotham}, {Hill}, {Driver},
  {Liske}, {Norberg}, {Bamford}, {Hopkins}, {Loveday}, {Peacock}, {Cameron},
  {Croom}, {Cross}, {Doyle}, {Dye}, {Frenk}, {Jones}, {van Kampen}, {Kelvin},
  {Nichol}, {Parkinson}, {Popescu}, {Prescott}, {Sharp}, {Sutherland},
  {Thomas}, \& {Tuffs}}]{2010MNRAS.404...86B}
{Baldry}, I.~K., {Robotham}, A.~S.~G., {Hill}, D.~T., {et~al.} 2010, \mnras,
  404, 86

\bibitem[{{Baldry} {et~al.}(2018){Baldry}, {Liske}, {Brown}, {Robotham},
  {Driver}, {Dunne}, {Alpaslan}, {Brough}, {Cluver}, {Eardley}, {Farrow},
  {Heymans}, {Hildebrandt}, {Hopkins}, {Kelvin}, {Loveday}, {Moffett},
  {Norberg}, {Owers}, {Taylor}, {Wright}, {Bamford}, {Bland-Hawthorn},
  {Bourne}, {Bremer}, {Colless}, {Conselice}, {Croom}, {Davies}, {Foster},
  {Grootes}, {Holwerda}, {Jones}, {Kafle}, {Kuijken}, {Lara-Lopez},
  {L{\'o}pez-S{\'a}nchez}, {Meyer}, {Phillipps}, {Sutherland}, {van Kampen}, \&
  {Wilkins}}]{2018MNRAS.474.3875B}
{Baldry}, I.~K., {Liske}, J., {Brown}, M.~J.~I., {et~al.} 2018, \mnras, 474,
  3875

\bibitem[{{Baldwin} {et~al.}(1981){Baldwin}, {Phillips}, \&
  {Terlevich}}]{1981PASP...93....5B}
{Baldwin}, J.~A., {Phillips}, M.~M., \& {Terlevich}, R. 1981, \pasp, 93, 5

\bibitem[{{Barger} {et~al.}(2007){Barger}, {Cowie}, \&
  {Wang}}]{2007ApJ...654..764B}
{Barger}, A.~J., {Cowie}, L.~L., \& {Wang}, W.~H. 2007, \apj, 654, 764

\bibitem[{{Battisti} {et~al.}(2016){Battisti}, {Calzetti}, \&
  {Chary}}]{2016ApJ...818...13B}
{Battisti}, A.~J., {Calzetti}, D., \& {Chary}, R.~R. 2016, \apj, 818, 13

\bibitem[{{Bauer} {et~al.}(2013){Bauer}, {Hopkins}, {Gunawardhana}, {Taylor},
  {Baldry}, {Bamford}, {Bland-Hawthorn}, {Brough}, {Brown}, {Cluver},
  {Colless}, {Conselice}, {Croom}, {Driver}, {Foster}, {Jones}, {Lara-Lopez},
  {Liske}, {L{\'o}pez-S{\'a}nchez}, {Loveday}, {Norberg}, {Owers}, {Pimbblet},
  {Robotham}, {Sansom}, \& {Sharp}}]{2013MNRAS.434..209B}
{Bauer}, A.~E., {Hopkins}, A.~M., {Gunawardhana}, M., {et~al.} 2013, \mnras,
  434, 209

\bibitem[{{Bell}(2003)}]{2003ApJ...586..794B}
{Bell}, E.~F. 2003, \apj, 586, 794

\bibitem[{{Bellstedt} {et~al.}(2020){Bellstedt}, {Driver}, {Robotham},
  {Davies}, {Bogue}, {Cook}, {Hashemizadeh}, {Koushan}, {Taylor}, {Thorne},
  {Turner}, \& {Wright}}]{2020MNRAS.496.3235B}
{Bellstedt}, S., {Driver}, S.~P., {Robotham}, A. S.~G., {et~al.} 2020, \mnras,
  496, 3235

\bibitem[{{Boselli} {et~al.}(2009){Boselli}, {Boissier}, {Cortese}, {Buat},
  {Hughes}, \& {Gavazzi}}]{2009ApJ...706.1527B}
{Boselli}, A., {Boissier}, S., {Cortese}, L., {et~al.} 2009, \apj, 706, 1527

\bibitem[{{Brinchmann} {et~al.}(2004){Brinchmann}, {Charlot}, {White},
  {Tremonti}, {Kauffmann}, {Heckman}, \& {Brinkmann}}]{2004MNRAS.351.1151B}
{Brinchmann}, J., {Charlot}, S., {White}, S.~D.~M., {et~al.} 2004, \mnras, 351,
  1151

\bibitem[{{Brinchmann} {et~al.}(2008){Brinchmann}, {Pettini}, \&
  {Charlot}}]{2008MNRAS.385..769B}
{Brinchmann}, J., {Pettini}, M., \& {Charlot}, S. 2008, \mnras, 385, 769

\bibitem[{{Brocklehurst}(1971)}]{1971MNRAS.153..471B}
{Brocklehurst}, M. 1971, \mnras, 153, 471

\bibitem[{{Brough} {et~al.}(2011){Brough}, {Hopkins}, {Sharp}, {Gunawardhana},
  {Wijesinghe}, {Robotham}, {Driver}, {Baldry}, {Bamford}, {Liske}, {Loveday},
  {Norberg}, {Peacock}, {Bland-Hawthorn}, {Brown}, {Cameron}, {Croom}, {Frenk},
  {Foster}, {Hill}, {Jones}, {Kelvin}, {Kuijken}, {Nichol}, {Parkinson},
  {Pimbblet}, {Popescu}, {Prescott}, {Sutherland}, {Taylor}, {Thomas}, {Tuffs},
  \& {van Kampen}}]{2011MNRAS.413.1236B}
{Brough}, S., {Hopkins}, A.~M., {Sharp}, R.~G., {et~al.} 2011, \mnras, 413,
  1236

\bibitem[{{Brown} {et~al.}(2017){Brown}, {Moustakas}, {Kennicutt}, {Bonne},
  {Intema}, {de Gasperin}, {Boquien}, {Jarrett}, {Cluver}, {Smith}, {da Cunha},
  {Imanishi}, {Armus}, {Brandl}, \& {Peek}}]{2017ApJ...847..136B}
{Brown}, M. J.~I., {Moustakas}, J., {Kennicutt}, R.~C., {et~al.} 2017, \apj,
  847, 136

\bibitem[{{Bruzual} \& {Charlot}(2003)}]{2003MNRAS.344.1000B}
{Bruzual}, G., \& {Charlot}, S. 2003, \mnras, 344, 1000

\bibitem[{{Bundy} {et~al.}(2015){Bundy}, {Bershady}, {Law}, {Yan}, {Drory},
  {MacDonald}, {Wake}, {Cherinka}, {S{\'a}nchez-Gallego}, {Weijmans}, {Thomas},
  {Tremonti}, {Masters}, {Coccato}, {Diamond-Stanic}, {Arag{\'o}n-Salamanca},
  {Avila-Reese}, {Badenes}, {Falc{\'o}n-Barroso}, {Belfiore}, {Bizyaev},
  {Blanc}, {Bland-Hawthorn}, {Blanton}, {Brownstein}, {Byler}, {Cappellari},
  {Conroy}, {Dutton}, {Emsellem}, {Etherington}, {Frinchaboy}, {Fu}, {Gunn},
  {Harding}, {Johnston}, {Kauffmann}, {Kinemuchi}, {Klaene}, {Knapen},
  {Leauthaud}, {Li}, {Lin}, {Maiolino}, {Malanushenko}, {Malanushenko}, {Mao},
  {Maraston}, {McDermid}, {Merrifield}, {Nichol}, {Oravetz}, {Pan}, {Parejko},
  {Sanchez}, {Schlegel}, {Simmons}, {Steele}, {Steinmetz}, {Thanjavur},
  {Thompson}, {Tinker}, {van den Bosch}, {Westfall}, {Wilkinson}, {Wright},
  {Xiao}, \& {Zhang}}]{2015ApJ...798....7B}
{Bundy}, K., {Bershady}, M.~A., {Law}, D.~R., {et~al.} 2015, \apj, 798, 7

\bibitem[{{Calzetti}(2001)}]{2001PASP..113.1449C}
{Calzetti}, D. 2001, \pasp, 113, 1449

\bibitem[{{Calzetti} {et~al.}(2000){Calzetti}, {Armus}, {Bohlin}, {Kinney},
  {Koornneef}, \& {Storchi-Bergmann}}]{2000ApJ...533..682C}
{Calzetti}, D., {Armus}, L., {Bohlin}, R.~C., {et~al.} 2000, \apj, 533, 682

\bibitem[{{Cameron}(2011)}]{2011PASA...28..128C}
{Cameron}, E. 2011, \pasa, 28, 128

\bibitem[{{Chabrier}(2003)}]{2003ApJ...586L.133C}
{Chabrier}, G. 2003, \apjl, 586, L133

\bibitem[{{Cluver} {et~al.}(2014){Cluver}, {Jarrett}, {Hopkins}, {Driver},
  {Liske}, {Gunawardhana}, {Taylor}, {Robotham}, {Alpaslan}, {Baldry}, {Brown},
  {Peacock}, {Popescu}, {Tuffs}, {Bauer}, {Bland-Hawthorn}, {Colless},
  {Holwerda}, {Lara-L{\'o}pez}, {Leschinski}, {L{\'o}pez-S{\'a}nchez},
  {Norberg}, {Owers}, {Wang}, \& {Wilkins}}]{2014ApJ...782...90C}
{Cluver}, M.~E., {Jarrett}, T.~H., {Hopkins}, A.~M., {et~al.} 2014, \apj, 782,
  90

\bibitem[{{Cluver} {et~al.}(2020){Cluver}, {Jarrett}, {Taylor}, {Hopkins},
  {Brough}, {Casura}, {Holwerda}, {Liske}, {Pimbblet}, \&
  {Wright}}]{2020ApJ...898...20C}
{Cluver}, M.~E., {Jarrett}, T.~H., {Taylor}, E.~N., {et~al.} 2020, \apj, 898,
  20

\bibitem[{{Condon}(1992)}]{1992ARA&A..30..575C}
{Condon}, J.~J. 1992, \araa, 30, 575

\bibitem[{{Cowie} {et~al.}(1996){Cowie}, {Songaila}, {Hu}, \&
  {Cohen}}]{1996AJ....112..839C}
{Cowie}, L.~L., {Songaila}, A., {Hu}, E.~M., \& {Cohen}, J.~G. 1996, \aj, 112,
  839

\bibitem[{{Croom} {et~al.}(2012){Croom}, {Lawrence}, {Bland-Hawthorn},
  {Bryant}, {Fogarty}, {Richards}, {Goodwin}, {Farrell}, {Miziarski}, {Heald},
  {Jones}, {Lee}, {Colless}, {Brough}, {Hopkins}, {Bauer}, {Birchall}, {Ellis},
  {Horton}, {Leon-Saval}, {Lewis}, {L{\'o}pez-S{\'a}nchez}, {Min}, {Trinh}, \&
  {Trowland}}]{2012MNRAS.421..872C}
{Croom}, S.~M., {Lawrence}, J.~S., {Bland-Hawthorn}, J., {et~al.} 2012, \mnras,
  421, 872

\bibitem[{{Cutri} \& {et al.}(2012)}]{2012yCat.2311....0C}
{Cutri}, R.~M., \& {et al.} 2012, VizieR Online Data Catalog, II/311

\bibitem[{{Daddi} {et~al.}(2007){Daddi}, {Dickinson}, {Morrison}, {Chary},
  {Cimatti}, {Elbaz}, {Frayer}, {Renzini}, {Pope}, {Alexander}, {Bauer},
  {Giavalisco}, {Huynh}, {Kurk}, \& {Mignoli}}]{2007ApJ...670..156D}
{Daddi}, E., {Dickinson}, M., {Morrison}, G., {et~al.} 2007, \apj, 670, 156

\bibitem[{{Davies} {et~al.}(2016){Davies}, {Driver}, {Robotham}, {Grootes},
  {Popescu}, {Tuffs}, {Hopkins}, {Alpaslan}, {Andrews}, {Bland-Hawthorn},
  {Bremer}, {Brough}, {Brown}, {Cluver}, {Croom}, {da Cunha}, {Dunne},
  {Lara-L{\'o}pez}, {Liske}, {Loveday}, {Moffett}, {Owers}, {Phillipps},
  {Sansom}, {Taylor}, {Michalowski}, {Ibar}, {Smith}, \&
  {Bourne}}]{2016MNRAS.461..458D}
{Davies}, L.~J.~M., {Driver}, S.~P., {Robotham}, A.~S.~G., {et~al.} 2016,
  \mnras, 461, 458

\bibitem[{{Davies} {et~al.}(2018){Davies}, {Robotham}, {Driver}, {Lagos},
  {Cortese}, {Mannering}, {Foster}, {Lidman}, {Hashemizadeh}, {Koushan},
  {O'Toole}, {Baldry}, {Bilicki}, {Bland-Hawthorn}, {Bremer}, {Brown},
  {Bryant}, {Catinella}, {Croom}, {Grootes}, {Holwerda}, {Jarvis}, {Maddox},
  {Meyer}, {Moffett}, {Phillipps}, {Taylor}, {Windhorst}, \&
  {Wolf}}]{2018MNRAS.480..768D}
{Davies}, L.~J.~M., {Robotham}, A.~S.~G., {Driver}, S.~P., {et~al.} 2018,
  \mnras, 480, 768

\bibitem[{{de Jong} {et~al.}(2015){de Jong}, {Verdoes Kleijn}, {Boxhoorn},
  {Buddelmeijer}, {Capaccioli}, {Getman}, {Grado}, {Helmich}, {Huang},
  {Irisarri}, {Kuijken}, {La Barbera}, {McFarland}, {Napolitano}, {Radovich},
  {Sikkema}, {Valentijn}, {Begeman}, {Brescia}, {Cavuoti}, {Choi}, {Cordes},
  {Covone}, {Dall'Ora}, {Hildebrandt}, {Longo}, {Nakajima}, {Paolillo},
  {Puddu}, {Rifatto}, {Tortora}, {van Uitert}, {Buddendiek},
  {Harnois-D{\'e}raps}, {Erben}, {Eriksen}, {Heymans}, {Hoekstra}, {Joachimi},
  {Kitching}, {Klaes}, {Koopmans}, {K{\"o}hlinger}, {Roy}, {Sif{\'o}n},
  {Schneider}, {Sutherland}, {Viola}, \& {Vriend}}]{2015A&A...582A..62D}
{de Jong}, J. T.~A., {Verdoes Kleijn}, G.~A., {Boxhoorn}, D.~R., {et~al.} 2015,
  \aap, 582, A62

\bibitem[{{Dey} {et~al.}(2019){Dey}, {Schlegel}, {Lang}, {Blum}, {Burleigh},
  {Fan}, {Findlay}, {Finkbeiner}, {Herrera}, {Juneau}, {Landriau}, {Levi},
  {McGreer}, {Meisner}, {Myers}, {Moustakas}, {Nugent}, {Patej}, {Schlafly},
  {Walker}, {Valdes}, {Weaver}, {Y{\`e}che}, {Zou}, {Zhou}, {Abareshi},
  {Abbott}, {Abolfathi}, {Aguilera}, {Alam}, {Allen}, {Alvarez}, {Annis},
  {Ansarinejad}, {Aubert}, {Beechert}, {Bell}, {BenZvi}, {Beutler}, {Bielby},
  {Bolton}, {Brice{\~n}o}, {Buckley-Geer}, {Butler}, {Calamida}, {Carlberg},
  {Carter}, {Casas}, {Castander}, {Choi}, {Comparat}, {Cukanovaite}, {Delubac},
  {DeVries}, {Dey}, {Dhungana}, {Dickinson}, {Ding}, {Donaldson}, {Duan},
  {Duckworth}, {Eftekharzadeh}, {Eisenstein}, {Etourneau}, {Fagrelius},
  {Farihi}, {Fitzpatrick}, {Font-Ribera}, {Fulmer}, {G{\"a}nsicke},
  {Gaztanaga}, {George}, {Gerdes}, {Gontcho}, {Gorgoni}, {Green}, {Guy},
  {Harmer}, {Hernandez}, {Honscheid}, {Huang}, {James}, {Jannuzi}, {Jiang},
  {Joyce}, {Karcher}, {Karkar}, {Kehoe}, {Kneib}, {Kueter-Young}, {Lan},
  {Lauer}, {Le Guillou}, {Le Van Suu}, {Lee}, {Lesser}, {Perreault Levasseur},
  {Li}, {Mann}, {Marshall}, {Mart{\'\i}nez-V{\'a}zquez}, {Martini}, {du Mas des
  Bourboux}, {McManus}, {Meier}, {M{\'e}nard}, {Metcalfe},
  {Mu{\~n}oz-Guti{\'e}rrez}, {Najita}, {Napier}, {Narayan}, {Newman}, {Nie},
  {Nord}, {Norman}, {Olsen}, {Paat}, {Palanque-Delabrouille}, {Peng},
  {Poppett}, {Poremba}, {Prakash}, {Rabinowitz}, {Raichoor}, {Rezaie},
  {Robertson}, {Roe}, {Ross}, {Ross}, {Rudnick}, {Safonova}, {Saha},
  {S{\'a}nchez}, {Savary}, {Schweiker}, {Scott}, {Seo}, {Shan}, {Silva},
  {Slepian}, {Soto}, {Sprayberry}, {Staten}, {Stillman}, {Stupak}, {Summers},
  {Sien Tie}, {Tirado}, {Vargas-Maga{\~n}a}, {Vivas}, {Wechsler}, {Williams},
  {Yang}, {Yang}, {Yapici}, {Zaritsky}, {Zenteno}, {Zhang}, {Zhang}, {Zhou}, \&
  {Zhou}}]{2019AJ....157..168D}
{Dey}, A., {Schlegel}, D.~J., {Lang}, D., {et~al.} 2019, \aj, 157, 168

\bibitem[{{Draine}(2003)}]{2003ARA&A..41..241D}
{Draine}, B.~T. 2003, \araa, 41, 241

\bibitem[{{Driver} {et~al.}(2011){Driver}, {Hill}, {Kelvin}, {Robotham},
  {Liske}, {Norberg}, {Baldry}, {Bamford}, {Hopkins}, {Loveday}, {Peacock},
  {Andrae}, {Bland-Hawthorn}, {Brough}, {Brown}, {Cameron}, {Ching}, {Colless},
  {Conselice}, {Croom}, {Cross}, {de Propris}, {Dye}, {Drinkwater}, {Ellis},
  {Graham}, {Grootes}, {Gunawardhana}, {Jones}, {van Kampen}, {Maraston},
  {Nichol}, {Parkinson}, {Phillipps}, {Pimbblet}, {Popescu}, {Prescott},
  {Roseboom}, {Sadler}, {Sansom}, {Sharp}, {Smith}, {Taylor}, {Thomas},
  {Tuffs}, {Wijesinghe}, {Dunne}, {Frenk}, {Jarvis}, {Madore}, {Meyer},
  {Seibert}, {Staveley-Smith}, {Sutherland}, \& {Warren}}]{2011MNRAS.413..971D}
{Driver}, S.~P., {Hill}, D.~T., {Kelvin}, L.~S., {et~al.} 2011, \mnras, 413,
  971

\bibitem[{{Driver} {et~al.}(2019){Driver}, {Liske}, {Davies}, {Robotham},
  {Baldry}, {Brown}, {Cluver}, {Kuijken}, {Loveday}, {McMahon}, {Meyer},
  {Norberg}, {Owers}, {Power}, {Taylor}, \& {WAVES Team}}]{2019Msngr.175...46D}
{Driver}, S.~P., {Liske}, J., {Davies}, L.~J.~M., {et~al.} 2019, The Messenger,
  175, 46

\bibitem[{{Driver} {et~al.}(2022){Driver}, {Bellstedt}, {Robotham}, {Baldry},
  {Davies}, {Liske}, {Obreschkow}, {Taylor}, {Wright}, {Alpaslan}, {Bamford},
  {Bauer}, {Bland-Hawthorn}, {Bilicki}, {Bravo}, {Brough}, {Casura}, {Cluver},
  {Colless}, {Conselice}, {Croom}, {de Jong}, {D'Eugenio}, {De Propris},
  {Dogruel}, {Drinkwater}, {Dvornik}, {Farrow}, {Frenk}, {Giblin}, {Graham},
  {Grootes}, {Gunawardhana}, {Hashemizadeh}, {H{\"a}u{\ss}ler}, {Heymans},
  {Hildebrandt}, {Holwerda}, {Hopkins}, {Jarrett}, {Heath Jones}, {Kelvin},
  {Koushan}, {Kuijken}, {Lara-L{\'o}pez}, {Lange}, {L{\'o}pez-S{\'a}nchez},
  {Loveday}, {Mahajan}, {Meyer}, {Moffett}, {Napolitano}, {Norberg}, {Owers},
  {Radovich}, {Raouf}, {Peacock}, {Phillipps}, {Pimbblet}, {Popescu}, {Said},
  {Sansom}, {Seibert}, {Sutherland}, {Thorne}, {Tuffs}, {Turner}, {van der
  Wel}, {van Kampen}, \& {Wilkins}}]{2022MNRAS.513..439D}
{Driver}, S.~P., {Bellstedt}, S., {Robotham}, A. S.~G., {et~al.} 2022, \mnras,
  513, 439

\bibitem[{{Duncan} {et~al.}(2023){Duncan}, {Baker}, {Best}, {Blyth}, {Hatch},
  {Holwerda}, {Jarvis}, {Maddox}, {Smith}, {Arnaudova}, {Chemin}, {Dav{\'e}},
  {Dunlop}, {Frank}, {Gawiser}, {Gloudemans}, {Hale}, {Heywood}, {Kannappan},
  {Kondapally}, {McLure}, {Morabito}, {Nesvadba}, {Pan}, {Ponomareva},
  {Prescott}, {Roberts}, {R{\"o}ttgering}, {Somerville}, {Tudorache},
  {Vaccari}, {Whittam}, {Wu}, \& {Zwaan}}]{2023Msngr.190...25D}
{Duncan}, K., {Baker}, A., {Best}, P., {et~al.} 2023, The Messenger, 190, 25

\bibitem[{{Duncan}(2022)}]{2022MNRAS.512.3662D}
{Duncan}, K.~J. 2022, \mnras, 512, 3662

\bibitem[{{Dwek}(1998)}]{1998ApJ...501..643D}
{Dwek}, E. 1998, \apj, 501, 643

\bibitem[{{Faisst} {et~al.}(2018){Faisst}, {Masters}, {Wang}, {Merson},
  {Capak}, {Malhotra}, \& {Rhoads}}]{2018ApJ...855..132F}
{Faisst}, A.~L., {Masters}, D., {Wang}, Y., {et~al.} 2018, \apj, 855, 132

\bibitem[{{Foster} {et~al.}(2021){Foster}, {Mendel}, {Lagos}, {Wisnioski},
  {Yuan}, {D'Eugenio}, {Barone}, {Harborne}, {Vaughan}, {Schulze}, {Remus},
  {Gupta}, {Collacchioni}, {Khim}, {Taylor}, {Bassett}, {Croom}, {McDermid},
  {Poci}, {Battisti}, {Bland-Hawthorn}, {Bellstedt}, {Colless}, {Davies},
  {Derkenne}, {Driver}, {Ferr{\'e}-Mateu}, {Fisher}, {Gjergo}, {Johnston},
  {Khalid}, {Kobayashi}, {Oh}, {Peng}, {Robotham}, {Sharda}, {Sweet}, {Taylor},
  {Tran}, {Trayford}, {van de Sande}, {Yi}, \& {Zanisi}}]{2021PASA...38...31F}
{Foster}, C., {Mendel}, J.~T., {Lagos}, C.~D.~P., {et~al.} 2021, \pasa, 38,
  e031

\bibitem[{{Franzen} {et~al.}(2015){Franzen}, {Banfield}, {Hales}, {Hopkins},
  {Norris}, {Seymour}, {Chow}, {Herzog}, {Huynh}, {Lenc}, {Mao}, \&
  {Middelberg}}]{2015MNRAS.453.4020F}
{Franzen}, T.~M.~O., {Banfield}, J.~K., {Hales}, C.~A., {et~al.} 2015, \mnras,
  453, 4020

\bibitem[{{Galvin} {et~al.}(2020){Galvin}, {Huynh}, {Norris}, {Wang},
  {Hopkins}, {Polsterer}, {Ralph}, {O'Brien}, \& {Heald}}]{2020MNRAS.497.2730G}
{Galvin}, T.~J., {Huynh}, M.~T., {Norris}, R.~P., {et~al.} 2020, \mnras, 497,
  2730

\bibitem[{{Garn} \& {Best}(2010)}]{2010MNRAS.409..421G}
{Garn}, T., \& {Best}, P.~N. 2010, \mnras, 409, 421

\bibitem[{{Garn} {et~al.}(2010){Garn}, {Sobral}, {Best}, {Geach}, {Smail},
  {Cirasuolo}, {Dalton}, {Dunlop}, {McLure}, \& {Farrah}}]{2010MNRAS.402.2017G}
{Garn}, T., {Sobral}, D., {Best}, P.~N., {et~al.} 2010, \mnras, 402, 2017

\bibitem[{{Gordon} {et~al.}(2017){Gordon}, {Owers}, {Pimbblet}, {Croom},
  {Alpaslan}, {Baldry}, {Brough}, {Brown}, {Cluver}, {Conselice}, {Davies},
  {Holwerda}, {Hopkins}, {Gunawardhana}, {Loveday}, {Taylor}, \&
  {Wang}}]{2017MNRAS.465.2671G}
{Gordon}, Y.~A., {Owers}, M.~S., {Pimbblet}, K.~A., {et~al.} 2017, \mnras, 465,
  2671

\bibitem[{{Groves} {et~al.}(2012){Groves}, {Brinchmann}, \&
  {Walcher}}]{2012MNRAS.419.1402G}
{Groves}, B., {Brinchmann}, J., \& {Walcher}, C.~J. 2012, \mnras, 419, 1402

\bibitem[{{Gunawardhana} {et~al.}(2011){Gunawardhana}, {Hopkins}, {Sharp},
  {Brough}, {Taylor}, {Bland-Hawthorn}, {Maraston}, {Tuffs}, {Popescu},
  {Wijesinghe}, {Jones}, {Croom}, {Sadler}, {Wilkins}, {Driver}, {Liske},
  {Norberg}, {Baldry}, {Bamford}, {Loveday}, {Peacock}, {Robotham}, {Zucker},
  {Parker}, {Conselice}, {Cameron}, {Frenk}, {Hill}, {Kelvin}, {Kuijken},
  {Madore}, {Nichol}, {Parkinson}, {Pimbblet}, {Prescott}, {Sutherland},
  {Thomas}, \& {van Kampen}}]{2011MNRAS.415.1647G}
{Gunawardhana}, M.~L.~P., {Hopkins}, A.~M., {Sharp}, R.~G., {et~al.} 2011,
  \mnras, 415, 1647

\bibitem[{{Gunawardhana} {et~al.}(2013){Gunawardhana}, {Hopkins},
  {Bland-Hawthorn}, {Brough}, {Sharp}, {Loveday}, {Taylor}, {Jones},
  {Lara-L{\'o}pez}, {Bauer}, {Colless}, {Owers}, {Baldry},
  {L{\'o}pez-S{\'a}nchez}, {Foster}, {Bamford}, {Brown}, {Driver},
  {Drinkwater}, {Liske}, {Meyer}, {Norberg}, {Robotham}, {Ching}, {Cluver},
  {Croom}, {Kelvin}, {Prescott}, {Steele}, {Thomas}, \&
  {Wang}}]{2013MNRAS.433.2764G}
{Gunawardhana}, M.~L.~P., {Hopkins}, A.~M., {Bland-Hawthorn}, J., {et~al.}
  2013, \mnras, 433, 2764

\bibitem[{{G{\"u}rkan} {et~al.}(2018){G{\"u}rkan}, {Hardcastle}, {Smith},
  {Best}, {Bourne}, {Calistro-Rivera}, {Heald}, {Jarvis}, {Prandoni},
  {R{\"o}ttgering}, {Sabater}, {Shimwell}, {Tasse}, \&
  {Williams}}]{2018MNRAS.475.3010G}
{G{\"u}rkan}, G., {Hardcastle}, M.~J., {Smith}, D.~J.~B., {et~al.} 2018,
  \mnras, 475, 3010

\bibitem[{{G{\"u}rkan} {et~al.}(2022){G{\"u}rkan}, {Prandoni}, {O'Brien},
  {Raja}, {Marchetti}, {Vaccari}, {Driver}, {Taylor}, {Franzen}, {Brown},
  {Shabala}, {Andernach}, {Hopkins}, {Norris}, {Leahy}, {Bilicki},
  {Farajollahi}, {Galvin}, {Heald}, {Koribalski}, {An}, \&
  {Warhurst}}]{2022MNRAS.512.6104G}
{G{\"u}rkan}, G., {Prandoni}, I., {O'Brien}, A., {et~al.} 2022, \mnras, 512,
  6104

\bibitem[{{Hales} {et~al.}(2014){Hales}, {Norris}, {Gaensler}, {Middelberg},
  {Chow}, {Hopkins}, {Huynh}, {Lenc}, \& {Mao}}]{2014MNRAS.441.2555H}
{Hales}, C.~A., {Norris}, R.~P., {Gaensler}, B.~M., {et~al.} 2014, \mnras, 441,
  2555

\bibitem[{{Heesen} {et~al.}(2022){Heesen}, {Staffehl}, {Basu}, {Beck}, {Stein},
  {Tabatabaei}, {Hardcastle}, {Chy{\.z}y}, {Shimwell}, {Adebahr}, {Beswick},
  {Bomans}, {Botteon}, {Brinks}, {Br{\"u}ggen}, {Dettmar}, {Drabent}, {de
  Gasperin}, {G{\"u}rkan}, {Heald}, {Horellou}, {Nikiel-Wroczynski},
  {Paladino}, {Piotrowska}, {R{\"o}ttgering}, {Smith}, \&
  {Tasse}}]{2022A&A...664A..83H}
{Heesen}, V., {Staffehl}, M., {Basu}, A., {et~al.} 2022, \aap, 664, A83

\bibitem[{{Hindson} {et~al.}(2018){Hindson}, {Kitchener}, {Brinks}, {Heesen},
  {Westcott}, {Hunter}, {Zhang}, {Rupen}, \& {Rau}}]{2018ApJS..234...29H}
{Hindson}, L., {Kitchener}, G., {Brinks}, E., {et~al.} 2018, \apjs, 234, 29

\bibitem[{{Hopkins}(2018)}]{2018PASA...35...39H}
{Hopkins}, A.~M. 2018, \pasa, 35, e039

\bibitem[{{Hopkins} {et~al.}(2003{\natexlab{a}}){Hopkins}, {Afonso}, {Chan},
  {Cram}, {Georgakakis}, \& {Mobasher}}]{2003AJ....125..465H}
{Hopkins}, A.~M., {Afonso}, J., {Chan}, B., {et~al.} 2003{\natexlab{a}}, \aj,
  125, 465

\bibitem[{{Hopkins} {et~al.}(2001){Hopkins}, {Connolly}, {Haarsma}, \&
  {Cram}}]{2001AJ....122..288H}
{Hopkins}, A.~M., {Connolly}, A.~J., {Haarsma}, D.~B., \& {Cram}, L.~E. 2001,
  \aj, 122, 288

\bibitem[{{Hopkins} {et~al.}(2008){Hopkins}, {McClure-Griffiths}, \&
  {Gaensler}}]{2008ApJ...682L..13H}
{Hopkins}, A.~M., {McClure-Griffiths}, N.~M., \& {Gaensler}, B.~M. 2008, \apjl,
  682, L13

\bibitem[{{Hopkins} {et~al.}(1998){Hopkins}, {Mobasher}, {Cram}, \&
  {Rowan-Robinson}}]{1998MNRAS.296..839H}
{Hopkins}, A.~M., {Mobasher}, B., {Cram}, L., \& {Rowan-Robinson}, M. 1998,
  \mnras, 296, 839

\bibitem[{{Hopkins} {et~al.}(2003{\natexlab{b}}){Hopkins}, {Miller}, {Nichol},
  {Connolly}, {Bernardi}, {G{\'o}mez}, {Goto}, {Tremonti}, {Brinkmann},
  {Ivezi{\'c}}, \& {Lamb}}]{2003ApJ...599..971H}
{Hopkins}, A.~M., {Miller}, C.~J., {Nichol}, R.~C., {et~al.}
  2003{\natexlab{b}}, \apj, 599, 971

\bibitem[{{Hopkins} {et~al.}(2013){Hopkins}, {Driver}, {Brough}, {Owers},
  {Bauer}, {Gunawardhana}, {Cluver}, {Colless}, {Foster}, {Lara-L{\'o}pez},
  {Roseboom}, {Sharp}, {Steele}, {Thomas}, {Baldry}, {Brown}, {Liske},
  {Norberg}, {Robotham}, {Bamford}, {Bland-Hawthorn}, {Drinkwater}, {Loveday},
  {Meyer}, {Peacock}, {Tuffs}, {Agius}, {Alpaslan}, {Andrae}, {Cameron},
  {Cole}, {Ching}, {Christodoulou}, {Conselice}, {Croom}, {Cross}, {De
  Propris}, {Delhaize}, {Dunne}, {Eales}, {Ellis}, {Frenk}, {Graham},
  {Grootes}, {H{\"a}u{\ss}ler}, {Heymans}, {Hill}, {Hoyle}, {Hudson}, {Jarvis},
  {Johansson}, {Jones}, {van Kampen}, {Kelvin}, {Kuijken},
  {L{\'o}pez-S{\'a}nchez}, {Maddox}, {Madore}, {Maraston}, {McNaught-Roberts},
  {Nichol}, {Oliver}, {Parkinson}, {Penny}, {Phillipps}, {Pimbblet}, {Ponman},
  {Popescu}, {Prescott}, {Proctor}, {Sadler}, {Sansom}, {Seibert},
  {Staveley-Smith}, {Sutherland}, {Taylor}, {Van Waerbeke}, {V{\'a}zquez-Mata},
  {Warren}, {Wijesinghe}, {Wild}, \& {Wilkins}}]{2013MNRAS.430.2047H}
{Hopkins}, A.~M., {Driver}, S.~P., {Brough}, S., {et~al.} 2013, \mnras, 430,
  2047

\bibitem[{{Hotan} {et~al.}(2021){Hotan}, {Bunton}, {Chippendale}, {Whiting},
  {Tuthill}, {Moss}, {McConnell}, {Amy}, {Huynh}, {Allison}, {Anderson},
  {Bannister}, {Bastholm}, {Beresford}, {Bock}, {Bolton}, {Chapman}, {Chow},
  {Collier}, {Cooray}, {Cornwell}, {Diamond}, {Edwards}, {Feain}, {Franzen},
  {George}, {Gupta}, {Hampson}, {Harvey-Smith}, {Hayman}, {Heywood}, {Jacka},
  {Jackson}, {Jackson}, {Jeganathan}, {Johnston}, {Kesteven}, {Kleiner},
  {Koribalski}, {Lee-Waddell}, {Lenc}, {Lensson}, {Mackay}, {Mahony},
  {McClure-Griffiths}, {McConigley}, {Mirtschin}, {Ng}, {Norris}, {Pearce},
  {Phillips}, {Pilawa}, {Raja}, {Reynolds}, {Roberts}, {Roxby}, {Sadler},
  {Shields}, {Schinckel}, {Serra}, {Shaw}, {Sweetnam}, {Troup}, {Tzioumis},
  {Voronkov}, \& {Westmeier}}]{2021PASA...38....9H}
{Hotan}, A.~W., {Bunton}, J.~D., {Chippendale}, A.~P., {et~al.} 2021, \pasa,
  38, e009

\bibitem[{{Jarrett} {et~al.}(2019){Jarrett}, {Cluver}, {Brown}, {Dale}, {Tsai},
  \& {Masci}}]{2019ApJS..245...25J}
{Jarrett}, T.~H., {Cluver}, M.~E., {Brown}, M.~J.~I., {et~al.} 2019, \apjs,
  245, 25

\bibitem[{{Jarrett} {et~al.}(2011){Jarrett}, {Cohen}, {Masci}, {Wright},
  {Stern}, {Benford}, {Blain}, {Carey}, {Cutri}, {Eisenhardt}, {Lonsdale},
  {Mainzer}, {Marsh}, {Padgett}, {Petty}, {Ressler}, {Skrutskie}, {Stanford},
  {Surace}, {Tsai}, {Wheelock}, \& {Yan}}]{2011ApJ...735..112J}
{Jarrett}, T.~H., {Cohen}, M., {Masci}, F., {et~al.} 2011, \apj, 735, 112

\bibitem[{{Jarrett} {et~al.}(2017){Jarrett}, {Cluver}, {Magoulas}, {Bilicki},
  {Alpaslan}, {Bland-Hawthorn}, {Brough}, {Brown}, {Croom}, {Driver},
  {Holwerda}, {Hopkins}, {Loveday}, {Norberg}, {Peacock}, {Popescu}, {Sadler},
  {Taylor}, {Tuffs}, \& {Wang}}]{2017ApJ...836..182J}
{Jarrett}, T.~H., {Cluver}, M.~E., {Magoulas}, C., {et~al.} 2017, \apj, 836,
  182

\bibitem[{{Kauffmann} {et~al.}(2003){Kauffmann}, {Heckman}, {Tremonti},
  {Brinchmann}, {Charlot}, {White}, {Ridgway}, {Brinkmann}, {Fukugita}, {Hall},
  {Ivezi{\'c}}, {Richards}, \& {Schneider}}]{2003MNRAS.346.1055K}
{Kauffmann}, G., {Heckman}, T.~M., {Tremonti}, C., {et~al.} 2003, \mnras, 346,
  1055

\bibitem[{{Kennicutt}(1998{\natexlab{a}})}]{1998ARA&A..36..189K}
{Kennicutt}, Robert~C., J. 1998{\natexlab{a}}, \araa, 36, 189

\bibitem[{{Kennicutt}(1998{\natexlab{b}})}]{1998ApJ...498..541K}
---. 1998{\natexlab{b}}, \apj, 498, 541

\bibitem[{{Kewley} {et~al.}(2001){Kewley}, {Dopita}, {Sutherland}, {Heisler},
  \& {Trevena}}]{2001ApJ...556..121K}
{Kewley}, L.~J., {Dopita}, M.~A., {Sutherland}, R.~S., {Heisler}, C.~A., \&
  {Trevena}, J. 2001, \apj, 556, 121

\bibitem[{{Kondapally} {et~al.}(2021){Kondapally}, {Best}, {Hardcastle},
  {Nisbet}, {Bonato}, {Sabater}, {Duncan}, {McCheyne}, {Cochrane}, {Bowler},
  {Williams}, {Shimwell}, {Tasse}, {Croston}, {Goyal}, {Jamrozy}, {Jarvis},
  {Mahatma}, {R{\"o}ttgering}, {Smith}, {Wo{\l}owska}, {Bondi}, {Brienza},
  {Brown}, {Br{\"u}ggen}, {Chambers}, {Garrett}, {G{\"u}rkan}, {Huber},
  {Kunert-Bajraszewska}, {Magnier}, {Mingo}, {Mostert},
  {Nikiel-Wroczy{\'n}ski}, {O'Sullivan}, {Paladino}, {Ploeckinger}, {Prandoni},
  {Rosenthal}, {Schwarz}, {Shulevski}, {Wagenveld}, \&
  {Wang}}]{2021A&A...648A...3K}
{Kondapally}, R., {Best}, P.~N., {Hardcastle}, M.~J., {et~al.} 2021, \aap, 648,
  A3

\bibitem[{{Lara-L{\'o}pez} {et~al.}(2013){Lara-L{\'o}pez}, {Hopkins},
  {L{\'o}pez-S{\'a}nchez}, {Brough}, {Gunawardhana}, {Colless}, {Robotham},
  {Bauer}, {Bland-Hawthorn}, {Cluver}, {Driver}, {Foster}, {Kelvin}, {Liske},
  {Loveday}, {Owers}, {Ponman}, {Sharp}, {Steele}, {Taylor}, \&
  {Thomas}}]{2013MNRAS.434..451L}
{Lara-L{\'o}pez}, M.~A., {Hopkins}, A.~M., {L{\'o}pez-S{\'a}nchez}, A.~R.,
  {et~al.} 2013, \mnras, 434, 451

\bibitem[{{Leahy} {et~al.}(2019){Leahy}, {Hopkins}, {Norris}, {Marvil},
  {Collier}, {Taylor}, {Allison}, {Anderson}, {Bell}, {Bilicki},
  {Bland-Hawthorn}, {Brough}, {Brown}, {Driver}, {Gurkan}, {Harvey-Smith},
  {Heywood}, {Holwerda}, {Liske}, {Lopez-Sanchez}, {McConnell}, {Moffett},
  {Owers}, {Pimbblet}, {Raja}, {Seymour}, {Voronkov}, \&
  {Wang}}]{2019PASA...36...24L}
{Leahy}, D.~A., {Hopkins}, A.~M., {Norris}, R.~P., {et~al.} 2019, \pasa, 36,
  e024

\bibitem[{{Lee} {et~al.}(2007){Lee}, {Kennicutt}, {Funes}, {Sakai}, \&
  {Akiyama}}]{2007ApJ...671L.113L}
{Lee}, J.~C., {Kennicutt}, R.~C., {Funes}, S.~J., J.~G., {Sakai}, S., \&
  {Akiyama}, S. 2007, \apjl, 671, L113

\bibitem[{{Lee} {et~al.}(2009){Lee}, {Gil de Paz}, {Tremonti}, {Kennicutt},
  {Salim}, {Bothwell}, {Calzetti}, {Dalcanton}, {Dale}, {Engelbracht}, {Funes},
  {Johnson}, {Sakai}, {Skillman}, {van Zee}, {Walter}, \&
  {Weisz}}]{2009ApJ...706..599L}
{Lee}, J.~C., {Gil de Paz}, A., {Tremonti}, C., {et~al.} 2009, \apj, 706, 599

\bibitem[{{Liske} {et~al.}(2015){Liske}, {Baldry}, {Driver}, {Tuffs},
  {Alpaslan}, {Andrae}, {Brough}, {Cluver}, {Grootes}, {Gunawardhana},
  {Kelvin}, {Loveday}, {Robotham}, {Taylor}, {Bamford}, {Bland-Hawthorn},
  {Brown}, {Drinkwater}, {Hopkins}, {Meyer}, {Norberg}, {Peacock}, {Agius},
  {Andrews}, {Bauer}, {Ching}, {Colless}, {Conselice}, {Croom}, {Davies}, {De
  Propris}, {Dunne}, {Eardley}, {Ellis}, {Foster}, {Frenk}, {H{\"a}u{\ss}ler},
  {Holwerda}, {Howlett}, {Ibarra}, {Jarvis}, {Jones}, {Kafle}, {Lacey},
  {Lange}, {Lara-L{\'o}pez}, {L{\'o}pez-S{\'a}nchez}, {Maddox}, {Madore},
  {McNaught-Roberts}, {Moffett}, {Nichol}, {Owers}, {Palamara}, {Penny},
  {Phillipps}, {Pimbblet}, {Popescu}, {Prescott}, {Proctor}, {Sadler},
  {Sansom}, {Seibert}, {Sharp}, {Sutherland}, {V{\'a}zquez-Mata}, {van Kampen},
  {Wilkins}, {Williams}, \& {Wright}}]{2015MNRAS.452.2087L}
{Liske}, J., {Baldry}, I.~K., {Driver}, S.~P., {et~al.} 2015, \mnras, 452, 2087

\bibitem[{{L{\'o}pez-S{\'a}nchez} \& {Esteban}(2009)}]{2009A&A...508..615L}
{L{\'o}pez-S{\'a}nchez}, A.~R., \& {Esteban}, C. 2009, \aap, 508, 615

\bibitem[{{L{\'o}pez-S{\'a}nchez} {et~al.}(2015){L{\'o}pez-S{\'a}nchez},
  {Westmeier}, {Esteban}, \& {Koribalski}}]{2015MNRAS.450.3381L}
{L{\'o}pez-S{\'a}nchez}, {\'A}.~R., {Westmeier}, T., {Esteban}, C., \&
  {Koribalski}, B.~S. 2015, \mnras, 450, 3381

\bibitem[{{Mao} {et~al.}(2012){Mao}, {Sharp}, {Norris}, {Hopkins}, {Seymour},
  {Lovell}, {Middelberg}, {Randall}, {Sadler}, {Saikia}, {Shabala}, \&
  {Zinn}}]{2012MNRAS.426.3334M}
{Mao}, M.~Y., {Sharp}, R., {Norris}, R.~P., {et~al.} 2012, \mnras, 426, 3334

\bibitem[{{Masters} {et~al.}(2016){Masters}, {Faisst}, \&
  {Capak}}]{2016ApJ...828...18M}
{Masters}, D., {Faisst}, A., \& {Capak}, P. 2016, \apj, 828, 18

\bibitem[{{McConnell} {et~al.}(2020){McConnell}, {Hale}, {Lenc}, {Banfield},
  {Heald}, {Hotan}, {Leung}, {Moss}, {Murphy}, {O'Brien}, {Pritchard}, {Raja},
  {Sadler}, {Stewart}, {Thomson}, {Whiting}, {Allison}, {Amy}, {Anderson},
  {Ball}, {Bannister}, {Bell}, {Bock}, {Bolton}, {Bunton}, {Chippendale},
  {Collier}, {Cooray}, {Cornwell}, {Diamond}, {Edwards}, {Gupta}, {Hayman},
  {Heywood}, {Jackson}, {Koribalski}, {Lee-Waddell}, {McClure-Griffiths}, {Ng},
  {Norris}, {Phillips}, {Reynolds}, {Roxby}, {Schinckel}, {Shields},
  {Tremblay}, {Tzioumis}, {Voronkov}, \& {Westmeier}}]{2020PASA...37...48M}
{McConnell}, D., {Hale}, C.~L., {Lenc}, E., {et~al.} 2020, \pasa, 37, e048

\bibitem[{{Meurer} {et~al.}(2009){Meurer}, {Wong}, {Kim}, {Hanish}, {Heckman},
  {Werk}, {Bland-Hawthorn}, {Dopita}, {Zwaan}, {Koribalski}, {Seibert},
  {Thilker}, {Ferguson}, {Webster}, {Putman}, {Knezek}, {Doyle}, {Drinkwater},
  {Hoopes}, {Kilborn}, {Meyer}, {Ryan-Weber}, {Smith}, \&
  {Staveley-Smith}}]{2009ApJ...695..765M}
{Meurer}, G.~R., {Wong}, O.~I., {Kim}, J.~H., {et~al.} 2009, \apj, 695, 765

\bibitem[{{Nanayakkara} {et~al.}(2017){Nanayakkara}, {Glazebrook}, {Kacprzak},
  {Yuan}, {Fisher}, {Tran}, {Kewley}, {Spitler}, {Alcorn}, {Cowley}, {Labbe},
  {Straatman}, \& {Tomczak}}]{2017MNRAS.468.3071N}
{Nanayakkara}, T., {Glazebrook}, K., {Kacprzak}, G.~G., {et~al.} 2017, \mnras,
  468, 3071

\bibitem[{{Nikutta} {et~al.}(2014){Nikutta}, {Hunt-Walker}, {Nenkova},
  {Ivezi{\'c}}, \& {Elitzur}}]{2014MNRAS.442.3361N}
{Nikutta}, R., {Hunt-Walker}, N., {Nenkova}, M., {Ivezi{\'c}}, {\v{Z}}., \&
  {Elitzur}, M. 2014, \mnras, 442, 3361

\bibitem[{{Noeske} {et~al.}(2007{\natexlab{a}}){Noeske}, {Faber}, {Weiner},
  {Koo}, {Primack}, {Dekel}, {Papovich}, {Conselice}, {Le Floc'h}, {Rieke},
  {Coil}, {Lotz}, {Somerville}, \& {Bundy}}]{2007ApJ...660L..47N}
{Noeske}, K.~G., {Faber}, S.~M., {Weiner}, B.~J., {et~al.} 2007{\natexlab{a}},
  \apjl, 660, L47

\bibitem[{{Noeske} {et~al.}(2007{\natexlab{b}}){Noeske}, {Weiner}, {Faber},
  {Papovich}, {Koo}, {Somerville}, {Bundy}, {Conselice}, {Newman},
  {Schiminovich}, {Le Floc'h}, {Coil}, {Rieke}, {Lotz}, {Primack}, {Barmby},
  {Cooper}, {Davis}, {Ellis}, {Fazio}, {Guhathakurta}, {Huang}, {Kassin},
  {Martin}, {Phillips}, {Rich}, {Small}, {Willmer}, \&
  {Wilson}}]{2007ApJ...660L..43N}
{Noeske}, K.~G., {Weiner}, B.~J., {Faber}, S.~M., {et~al.} 2007{\natexlab{b}},
  \apjl, 660, L43

\bibitem[{{Norris} {et~al.}(2006){Norris}, {Afonso}, {Appleton}, {Boyle},
  {Ciliegi}, {Croom}, {Huynh}, {Jackson}, {Koekemoer}, {Lonsdale},
  {Middelberg}, {Mobasher}, {Oliver}, {Polletta}, {Siana}, {Smail}, \&
  {Voronkov}}]{2006AJ....132.2409N}
{Norris}, R.~P., {Afonso}, J., {Appleton}, P.~N., {et~al.} 2006, \aj, 132, 2409

\bibitem[{{Norris} {et~al.}(2011){Norris}, {Hopkins}, {Afonso}, {Brown},
  {Condon}, {Dunne}, {Feain}, {Hollow}, {Jarvis}, {Johnston-Hollitt}, {Lenc},
  {Middelberg}, {Padovani}, {Prandoni}, {Rudnick}, {Seymour}, {Umana},
  {Andernach}, {Alexander}, {Appleton}, {Bacon}, {Banfield}, {Becker}, {Brown},
  {Ciliegi}, {Jackson}, {Eales}, {Edge}, {Gaensler}, {Giovannini}, {Hales},
  {Hancock}, {Huynh}, {Ibar}, {Ivison}, {Kennicutt}, {Kimball}, {Koekemoer},
  {Koribalski}, {L{\'o}pez-S{\'a}nchez}, {Mao}, {Murphy}, {Messias},
  {Pimbblet}, {Raccanelli}, {Randall}, {Reiprich}, {Roseboom},
  {R{\"o}ttgering}, {Saikia}, {Sharp}, {Slee}, {Smail}, {Thompson}, {Urquhart},
  {Wall}, \& {Zhao}}]{2011PASA...28..215N}
{Norris}, R.~P., {Hopkins}, A.~M., {Afonso}, J., {et~al.} 2011, \pasa, 28, 215

\bibitem[{{Norris} {et~al.}(2013){Norris}, {Afonso}, {Bacon}, {Beck}, {Bell},
  {Beswick}, {Best}, {Bhatnagar}, {Bonafede}, {Brunetti}, {Budav{\'a}ri},
  {Cassano}, {Condon}, {Cress}, {Dabbech}, {Feain}, {Fender}, {Ferrari},
  {Gaensler}, {Giovannini}, {Haverkorn}, {Heald}, {Van der Heyden}, {Hopkins},
  {Jarvis}, {Johnston-Hollitt}, {Kothes}, {Van Langevelde}, {Lazio}, {Mao},
  {Mart{\'\i}nez-Sansigre}, {Mary}, {Mcalpine}, {Middelberg}, {Murphy},
  {Padovani}, {Paragi}, {Prandoni}, {Raccanelli}, {Rigby}, {Roseboom},
  {R{\"o}ttgering}, {Sabater}, {Salvato}, {Scaife}, {Schilizzi}, {Seymour},
  {Smith}, {Umana}, {Zhao}, \& {Zinn}}]{2013PASA...30...20N}
{Norris}, R.~P., {Afonso}, J., {Bacon}, D., {et~al.} 2013, \pasa, 30, e020

\bibitem[{{Norris} {et~al.}(2021){Norris}, {Marvil}, {Collier}, {Kapi{\'n}ska},
  {O'Brien}, {Rudnick}, {Andernach}, {Asorey}, {Brown}, {Br{\"u}ggen},
  {Crawford}, {English}, {Rahman}, {Filipovi{\'c}}, {Gordon}, {G{\"u}rkan},
  {Hale}, {Hopkins}, {Huynh}, {HyeongHan}, {James Jee}, {Koribalski}, {Lenc},
  {Luken}, {Parkinson}, {Prandoni}, {Raja}, {Reiprich}, {Riseley}, {Shabala},
  {Sheil}, {Vernstrom}, {Whiting}, {Allison}, {Anderson}, {Ball}, {Bell},
  {Bunton}, {Galvin}, {Gupta}, {Hotan}, {Jacka}, {Macgregor}, {Mahony}, {Maio},
  {Moss}, {Pandey-Pommier}, \& {Voronkov}}]{2021PASA...38...46N}
{Norris}, R.~P., {Marvil}, J., {Collier}, J.~D., {et~al.} 2021, \pasa, 38, e046

\bibitem[{{Novak} {et~al.}(2017){Novak}, {Smol{\v{c}}i{\'c}}, {Delhaize},
  {Delvecchio}, {Zamorani}, {Baran}, {Bondi}, {Capak}, {Carilli}, {Ciliegi},
  {Civano}, {Ilbert}, {Karim}, {Laigle}, {Le F{\`e}vre}, {Marchesi},
  {McCracken}, {Miettinen}, {Salvato}, {Sargent}, {Schinnerer}, \&
  {Tasca}}]{2017A&A...602A...5N}
{Novak}, M., {Smol{\v{c}}i{\'c}}, V., {Delhaize}, J., {et~al.} 2017, \aap, 602,
  A5

\bibitem[{{Osterbrock}(1989)}]{1989agna.book.....O}
{Osterbrock}, D.~E. 1989, {Astrophysics of gaseous nebulae and active galactic
  nuclei}

\bibitem[{{Phillipps} {et~al.}(2023){Phillipps}, {Bellstedt}, {Bremer}, {De
  Propris}, {James}, {Casura}, {Liske}, \& {Holwerda}}]{2023MNRAS.518.5475P}
{Phillipps}, S., {Bellstedt}, S., {Bremer}, M.~N., {et~al.} 2023, \mnras, 518,
  5475

\bibitem[{{Robotham} {et~al.}(2010){Robotham}, {Driver}, {Norberg}, {Baldry},
  {Bamford}, {Hopkins}, {Liske}, {Loveday}, {Peacock}, {Cameron}, {Croom},
  {Doyle}, {Frenk}, {Hill}, {Jones}, {van Kampen}, {Kelvin}, {Kuijken},
  {Nichol}, {Parkinson}, {Popescu}, {Prescott}, {Sharp}, {Sutherland},
  {Thomas}, \& {Tuffs}}]{2010PASA...27...76R}
{Robotham}, A., {Driver}, S.~P., {Norberg}, P., {et~al.} 2010, \pasa, 27, 76

\bibitem[{{Rodighiero} {et~al.}(2011){Rodighiero}, {Daddi}, {Baronchelli},
  {Cimatti}, {Renzini}, {Aussel}, {Popesso}, {Lutz}, {Andreani}, {Berta},
  {Cava}, {Elbaz}, {Feltre}, {Fontana}, {F{\"o}rster Schreiber},
  {Franceschini}, {Genzel}, {Grazian}, {Gruppioni}, {Ilbert}, {Le Floch},
  {Magdis}, {Magliocchetti}, {Magnelli}, {Maiolino}, {McCracken}, {Nordon},
  {Poglitsch}, {Santini}, {Pozzi}, {Riguccini}, {Tacconi}, {Wuyts}, \&
  {Zamorani}}]{2011ApJ...739L..40R}
{Rodighiero}, G., {Daddi}, E., {Baronchelli}, I., {et~al.} 2011, \apjl, 739,
  L40

\bibitem[{{Salim} {et~al.}(2007){Salim}, {Rich}, {Charlot}, {Brinchmann},
  {Johnson}, {Schiminovich}, {Seibert}, {Mallery}, {Heckman}, {Forster},
  {Friedman}, {Martin}, {Morrissey}, {Neff}, {Small}, {Wyder}, {Bianchi},
  {Donas}, {Lee}, {Madore}, {Milliard}, {Szalay}, {Welsh}, \&
  {Yi}}]{2007ApJS..173..267S}
{Salim}, S., {Rich}, R.~M., {Charlot}, S., {et~al.} 2007, \apjs, 173, 267

\bibitem[{{S{\'a}nchez} {et~al.}(2012){S{\'a}nchez}, {Kennicutt}, {Gil de Paz},
  {van de Ven}, {V{\'\i}lchez}, {Wisotzki}, {Walcher}, {Mast}, {Aguerri},
  {Albiol-P{\'e}rez}, {Alonso-Herrero}, {Alves}, {Bakos}, {Bart{\'a}kov{\'a}},
  {Bland-Hawthorn}, {Boselli}, {Bomans}, {Castillo-Morales}, {Cortijo-Ferrero},
  {de Lorenzo-C{\'a}ceres}, {Del Olmo}, {Dettmar}, {D{\'\i}az}, {Ellis},
  {Falc{\'o}n-Barroso}, {Flores}, {Gallazzi}, {Garc{\'\i}a-Lorenzo},
  {Gonz{\'a}lez Delgado}, {Gruel}, {Haines}, {Hao}, {Husemann},
  {Igl{\'e}sias-P{\'a}ramo}, {Jahnke}, {Johnson}, {Jungwiert}, {Kalinova},
  {Kehrig}, {Kupko}, {L{\'o}pez-S{\'a}nchez}, {Lyubenova}, {Marino},
  {M{\'a}rmol-Queralt{\'o}}, {M{\'a}rquez}, {Masegosa}, {Meidt},
  {Mendez-Abreu}, {Monreal-Ibero}, {Montijo}, {Mour{\~a}o}, {Palacios-Navarro},
  {Papaderos}, {Pasquali}, {Peletier}, {P{\'e}rez}, {P{\'e}rez}, {Quirrenbach},
  {Rela{\~n}o}, {Rosales-Ortega}, {Roth}, {Ruiz-Lara},
  {S{\'a}nchez-Bl{\'a}zquez}, {Sengupta}, {Singh}, {Stanishev}, {Trager},
  {Vazdekis}, {Viironen}, {Wild}, {Zibetti}, \&
  {Ziegler}}]{2012A&A...538A...8S}
{S{\'a}nchez}, S.~F., {Kennicutt}, R.~C., {Gil de Paz}, A., {et~al.} 2012,
  \aap, 538, A8

\bibitem[{{S{\'a}nchez} {et~al.}(2015){S{\'a}nchez}, {P{\'e}rez},
  {Rosales-Ortega}, {Miralles-Caballero}, {L{\'o}pez-S{\'a}nchez},
  {Iglesias-P{\'a}ramo}, {Marino}, {S{\'a}nchez-Menguiano},
  {Garc{\'\i}a-Benito}, {Mast}, {Mendoza}, {Papaderos}, {Ellis}, {Galbany},
  {Kehrig}, {Monreal-Ibero}, {Gonz{\'a}lez Delgado}, {Moll{\'a}}, {Ziegler},
  {de Lorenzo-C{\'a}ceres}, {Mendez-Abreu}, {Bland-Hawthorn},
  {Bekerait{\.{e}}}, {Roth}, {Pasquali}, {D{\'\i}az}, {Bomans}, {van de Ven},
  \& {Wisotzki}}]{2015A&A...574A..47S}
{S{\'a}nchez}, S.~F., {P{\'e}rez}, E., {Rosales-Ortega}, F.~F., {et~al.} 2015,
  \aap, 574, A47

\bibitem[{{Schreiber} {et~al.}(2015){Schreiber}, {Pannella}, {Elbaz},
  {B{\'e}thermin}, {Inami}, {Dickinson}, {Magnelli}, {Wang}, {Aussel}, {Daddi},
  {Juneau}, {Shu}, {Sargent}, {Buat}, {Faber}, {Ferguson}, {Giavalisco},
  {Koekemoer}, {Magdis}, {Morrison}, {Papovich}, {Santini}, \&
  {Scott}}]{2015A&A...575A..74S}
{Schreiber}, C., {Pannella}, M., {Elbaz}, D., {et~al.} 2015, \aap, 575, A74

\bibitem[{{Seymour} {et~al.}(2008){Seymour}, {Dwelly}, {Moss}, {McHardy},
  {Zoghbi}, {Rieke}, {Page}, {Hopkins}, \& {Loaring}}]{2008MNRAS.386.1695S}
{Seymour}, N., {Dwelly}, T., {Moss}, D., {et~al.} 2008, \mnras, 386, 1695

\bibitem[{{Smail} {et~al.}(1999){Smail}, {Morrison}, {Gray}, {Owen}, {Ivison},
  {Kneib}, \& {Ellis}}]{1999ApJ...525..609S}
{Smail}, I., {Morrison}, G., {Gray}, M.~E., {et~al.} 1999, \apj, 525, 609

\bibitem[{{Smith} {et~al.}(2016){Smith}, {Best}, {Duncan}, {Hatch}, {Jarvis},
  {R{\"o}ttgering}, {Simpson}, {Stott}, {Cochrane}, {Coppin}, {Dannerbauer},
  {Davis}, {Geach}, {Hale}, {Hardcastle}, {Hatfield}, {Houghton}, {Maddox},
  {McGee}, {Morabito}, {Nisbet}, {Pandey-Pommier}, {Prandoni}, {Saxena},
  {Shimwell}, {Tarr}, {van Bemmel}, {Verma}, {White}, \&
  {Williams}}]{2016sf2a.conf..271S}
{Smith}, D.~J.~B., {Best}, P.~N., {Duncan}, K.~J., {et~al.} 2016, in SF2A-2016:
  Proceedings of the Annual meeting of the French Society of Astronomy and
  Astrophysics, ed. C.~{Reyl{\'e}}, J.~{Richard}, L.~{Cambr{\'e}sy},
  M.~{Deleuil}, E.~{P{\'e}contal}, L.~{Tresse}, \& I.~{Vauglin}, 271--280

\bibitem[{{Smol{\v{c}}i{\'c}} {et~al.}(2017){Smol{\v{c}}i{\'c}}, {Delvecchio},
  {Zamorani}, {Baran}, {Novak}, {Delhaize}, {Schinnerer}, {Berta}, {Bondi},
  {Ciliegi}, {Capak}, {Civano}, {Karim}, {Le Fevre}, {Ilbert}, {Laigle},
  {Marchesi}, {McCracken}, {Tasca}, {Salvato}, \&
  {Vardoulaki}}]{2017A&A...602A...2S}
{Smol{\v{c}}i{\'c}}, V., {Delvecchio}, I., {Zamorani}, G., {et~al.} 2017, \aap,
  602, A2

\bibitem[{{Speagle} {et~al.}(2014){Speagle}, {Steinhardt}, {Capak}, \&
  {Silverman}}]{2014ApJS..214...15S}
{Speagle}, J.~S., {Steinhardt}, C.~L., {Capak}, P.~L., \& {Silverman}, J.~D.
  2014, \apjs, 214, 15

\bibitem[{{Strazzullo} {et~al.}(2010){Strazzullo}, {Pannella}, {Owen},
  {Bender}, {Morrison}, {Wang}, \& {Shupe}}]{2010ApJ...714.1305S}
{Strazzullo}, V., {Pannella}, M., {Owen}, F.~N., {et~al.} 2010, \apj, 714, 1305

\bibitem[{{Sutherland}(2012)}]{2012sngi.confE..40S}
{Sutherland}, W. 2012, in Science from the Next Generation Imaging and
  Spectroscopic Surveys, 40

\bibitem[{{Taylor} {et~al.}(2011){Taylor}, {Hopkins}, {Baldry}, {Brown},
  {Driver}, {Kelvin}, {Hill}, {Robotham}, {Bland-Hawthorn}, {Jones}, {Sharp},
  {Thomas}, {Liske}, {Loveday}, {Norberg}, {Peacock}, {Bamford}, {Brough},
  {Colless}, {Cameron}, {Conselice}, {Croom}, {Frenk}, {Gunawardhana},
  {Kuijken}, {Nichol}, {Parkinson}, {Phillipps}, {Pimbblet}, {Popescu},
  {Prescott}, {Sutherland}, {Tuffs}, {van Kampen}, \&
  {Wijesinghe}}]{2011MNRAS.418.1587T}
{Taylor}, E.~N., {Hopkins}, A.~M., {Baldry}, I.~K., {et~al.} 2011, \mnras, 418,
  1587

\bibitem[{{Tremonti} {et~al.}(2004){Tremonti}, {Heckman}, {Kauffmann},
  {Brinchmann}, {Charlot}, {White}, {Seibert}, {Peng}, {Schlegel}, {Uomoto},
  {Fukugita}, \& {Brinkmann}}]{2004ApJ...613..898T}
{Tremonti}, C.~A., {Heckman}, T.~M., {Kauffmann}, G., {et~al.} 2004, \apj, 613,
  898

\bibitem[{{Veilleux} \& {Osterbrock}(1987)}]{1987ApJS...63..295V}
{Veilleux}, S., \& {Osterbrock}, D.~E. 1987, \apjs, 63, 295

\bibitem[{{Wang} \& {Heckman}(1996)}]{1996ApJ...457..645W}
{Wang}, B., \& {Heckman}, T.~M. 1996, \apj, 457, 645

\bibitem[{{Whittam} {et~al.}(2022){Whittam}, {Jarvis}, {Hale}, {Prescott},
  {Morabito}, {Heywood}, {Adams}, {Afonso}, {An}, {Ao}, {Bowler}, {Collier},
  {Deane}, {Delhaize}, {Frank}, {Glowacki}, {Hatfield}, {Maddox}, {Marchetti},
  {Matthews}, {Prandoni}, {Randriamampandry}, {Randriamanakoto}, {Smith},
  {Taylor}, {Thomas}, \& {Vaccari}}]{2022MNRAS.516..245W}
{Whittam}, I.~H., {Jarvis}, M.~J., {Hale}, C.~L., {et~al.} 2022, \mnras, 516,
  245

\bibitem[{{Wilman} {et~al.}(2010){Wilman}, {Jarvis}, {Mauch}, {Rawlings}, \&
  {Hickey}}]{2010MNRAS.405..447W}
{Wilman}, R.~J., {Jarvis}, M.~J., {Mauch}, T., {Rawlings}, S., \& {Hickey}, S.
  2010, \mnras, 405, 447

\bibitem[{{Windhorst}(2003)}]{2003NewAR..47..357W}
{Windhorst}, R.~A. 2003, \nar, 47, 357

\bibitem[{{Windhorst} {et~al.}(1999){Windhorst}, {Hopkins}, {Richards}, \&
  {Waddington}}]{1999ASPC..193...55W}
{Windhorst}, R.~A., {Hopkins}, A., {Richards}, E.~A., \& {Waddington}, I. 1999,
  in Astronomical Society of the Pacific Conference Series, Vol. 193, The
  Hy-Redshift Universe: Galaxy Formation and Evolution at High Redshift, ed.
  A.~J. {Bunker} \& W.~J.~M. {van Breugel}, 55

\bibitem[{{Wright} {et~al.}(2010){Wright}, {Eisenhardt}, {Mainzer}, {Ressler},
  {Cutri}, {Jarrett}, {Kirkpatrick}, {Padgett}, {McMillan}, {Skrutskie},
  {Stanford}, {Cohen}, {Walker}, {Mather}, {Leisawitz}, {Gautier}, {McLean},
  {Benford}, {Lonsdale}, {Blain}, {Mendez}, {Irace}, {Duval}, {Liu}, {Royer},
  {Heinrichsen}, {Howard}, {Shannon}, {Kendall}, {Walsh}, {Larsen}, {Cardon},
  {Schick}, {Schwalm}, {Abid}, {Fabinsky}, {Naes}, \&
  {Tsai}}]{2010AJ....140.1868W}
{Wright}, E.~L., {Eisenhardt}, P. R.~M., {Mainzer}, A.~K., {et~al.} 2010, \aj,
  140, 1868

\bibitem[{{Yao} {et~al.}(2020){Yao}, {Jarrett}, {Cluver}, {Marchetti},
  {Taylor}, {Santos}, {Owers}, {Lopez-Sanchez}, {Gordon}, {Brown}, {Brough},
  {Phillipps}, {Holwerda}, {Hopkins}, \& {Wang}}]{2020ApJ...903...91Y}
{Yao}, H.~F.~M., {Jarrett}, T.~H., {Cluver}, M.~E., {et~al.} 2020, \apj, 903,
  91

\bibitem[{{Zahid} {et~al.}(2017){Zahid}, {Kudritzki}, {Conroy}, {Andrews}, \&
  {Ho}}]{2017ApJ...847...18Z}
{Zahid}, H.~J., {Kudritzki}, R.-P., {Conroy}, C., {Andrews}, B., \& {Ho}, I.~T.
  2017, \apj, 847, 18

\end{thebibliography}

\appendix

\end{document}